\documentclass{article}  

\usepackage{graphicx}
\usepackage{color}
\usepackage{subfigure}
\usepackage{mathptmx} 
\usepackage{times} 
\usepackage{amsmath}
\usepackage{amssymb} 
\usepackage{wrapfig}
\usepackage{subfigure}

\usepackage[T1]{fontenc}
\usepackage[utf8]{inputenc}
\usepackage[font=small,labelfont=bf]{caption}

\usepackage{cite}
\usepackage{pstricks}
\usepackage{amsfonts}
\usepackage[amsmath,thmmarks]{ntheorem}

\newrgbcolor{aasacolor}{.7 .1 .1}
\newrgbcolor{megancolor}{.1 .7 .1}

\newboolean{DEBUG}
\setboolean{DEBUG}{true}

\ifthenelse {\boolean{DEBUG}}
{\newcommand{\aasa}[1]{{\bf \aasacolor{[Aasa says: #1]}}}}
{\newcommand{\aasa}[1]{}}

\ifthenelse {\boolean{DEBUG}}
{\newcommand{\megan}[1]{{\bf \megancolor{[Megan says: #1]}}}}
{\newcommand{\megan}[1]{}}

\title{\LARGE \bf Quantification and visualization of variation in anatomical trees}

\author{Nina Amenta$^{1}$ and 
Manasi Datar$^{2}$ and 
Asger Dirksen$^{3}$\\ and 
Marleen de Bruijne$^{4,11}$ and 
Aasa Feragen$^{4}$ $^\sharp$ and 
Xiaoyin Ge$^{5}$\\ 
and Jesper Holst Pedersen$^{6}$ and 
Marylesa Howard$^{7}$ and 
Megan Owen$^{8}$ $^\sharp$\\ and 
Jens Petersen$^{4}$ and 
Jie Shi$^{9}$ and 
Qiuping Xu$^{10}$
\thanks{$^{1}$University of California at Davis; 
$^{2}$Scientific Computing and Imaging Institute, University of Utah; 
$^{3}$Lungemedicinsk Afdeling, Gentofte Hospital, Denmark; 
$^{4}$Department of Computer Science, University of Copenhagen, Denmark; 
$^{5}$Department of Computer Science and Engineering, Ohio State University; 
$^{6}$Department of Cardiothoracic Surgery, Rigshospitalet, University of Copenhagen; 
$^{7}$National Security Technologies, LLC (A Department of Energy Contractor); 
$^{8}$ Department of Mathematics and Computer Science, Lehman College, City University of New York, USA; 
$^{9}$ School of Computing, Informatics, and Decision Systems Engineering, Arizona State University; 
$^{10}$Department of Mathematics, Florida State University; 
$^{11}$Erasmus MC Rotterdam, The Netherlands; 
$^\sharp$Corresponding authors: Megan Owen ({\tt\small megan.owen@lehman.cuny.edu}) and Aasa Feragen ({\tt\small aasa@diku.dk}).}        
}

\begin{document}

\maketitle
\thispagestyle{plain}
\pagestyle{plain}

\begin{abstract}
This paper presents two approaches to quantifying and visualizing variation in datasets of trees. The first approach localizes subtrees in which significant population differences are found through hypothesis testing and sparse classifiers on subtree features. The second approach visualizes the global metric structure of datasets through low-distortion embedding into hyperbolic planes in the style of multidimensional scaling. A case study is made on a dataset of airway trees in relation to Chronic Obstructive Pulmonary Disease.
\end{abstract}

\section{Introduction}

Tree-structured data appears in many medical imaging applications, e.g., airway trees~\cite{feragen_ipmi2013_stats}, blood vessel trees~\cite{Geers01032011}, dendrites~\cite{dendrite2} and galactograms~\cite{megaloo}. Typically, these anatomical trees vary both in tree topology and associated branch features such as branch length or shape, and as a result there is no straight-forward way to analyze the trees using standard Euclidean statistics. One way to integrate both tree topology and branch features in a single parametric framework is by modeling trees as residing in a non-linear, non-smooth \emph{tree-space}~\cite{feragen_ipmi2013_stats,BHV}. The non-linear, non-smooth nature of tree-space creates several problems for data analysis. First, statistics have to be redefined, as the standard statistical procedures such as finding an average or a principal component, or performing classification, do not translate directly to the tree-space setting. Second, even if we define classification algorithms in tree-space, we do not know which parts of the anatomical tree are responsible for causing class differences, for example, because each tree-space point represents an entire tree structure. Third, due to the lack of statistical tools such as principal component analysis, it is hard to visualize how distributions of trees vary in  tree-space. While recent work has resulted in basic statistical tools~\cite{sturmmean,bacak, feragen_ipmi2013_stats,nye}, the two latter problems are still unsolved. In this paper we investigate two approaches to these two problems: First, we study the influence of local subtrees on the results of hypothesis testing and classification, and the identification of subtrees which are responsible for significant differences between two populations of trees. Second, we use \emph{hyperbolic low-distortion embedding} to visualize the global metric structure of data living in tree-space. As a case study, we demonstrate the use of these techniques on a population of airway trees from a lung cancer screening study.

This paper presents 
results from the one-week collaboration workshop \emph{Women in Shape: Modeling Boundaries of Objects in $2$- and $3$-Dimensions} held at the Institute of Pure and Applied Mathematics at UCLA, July 15-19 2013. At this workshop, most of the authors of this paper spent a week working together on two projects related to quantifying and visualizing variance in populations of trees, which are described in Sections~\ref{quantification} and~\ref{visualization}, respectively.

\subsection{Tree-space}
A tree-space is any geometric space in which points represent trees.  The tree-space used in this paper, described in \cite{feragen_miccai,feragen_ipmi2013_stats}, is a generalization of the phylogenetic tree-space proposed by Billera et al.~\cite{BHV}.  This tree-space, denoted $\mathcal{T}_n$, contains all rooted trees with $n$ labeled leaves with vertices of degree at least $3$, where the $n$ leaf labels are given by a fixed set of cardinality $n$.  In this paper, the root of a tree is not considered to be among the leaves. Furthermore, for any tree in this tree-space, each edge has a $k$-dimensional vector associated with it. An example of such an edge vector is a non-negative real number representing the edge length (i.e. $k = 1$); a second example is the vector of $l$ $3$-dimensional landmark points sampled along a branch centerline, giving $k = 3 \cdot l$. For each edge, the landmark points are translated so that the edge starts at the origin. We refer to the latter edge vector as the \emph{shape} of the edge. The trees in such a tree-space can, for instance, be used to model airways in the lung. In this space, we will use edge shape with $l = 5$ to describe edges unless otherwise stated.

\begin{figure}
  \begin{center}
       \subfigure[]{\label{fig:tree-space_2orthants}\includegraphics[scale=0.22]{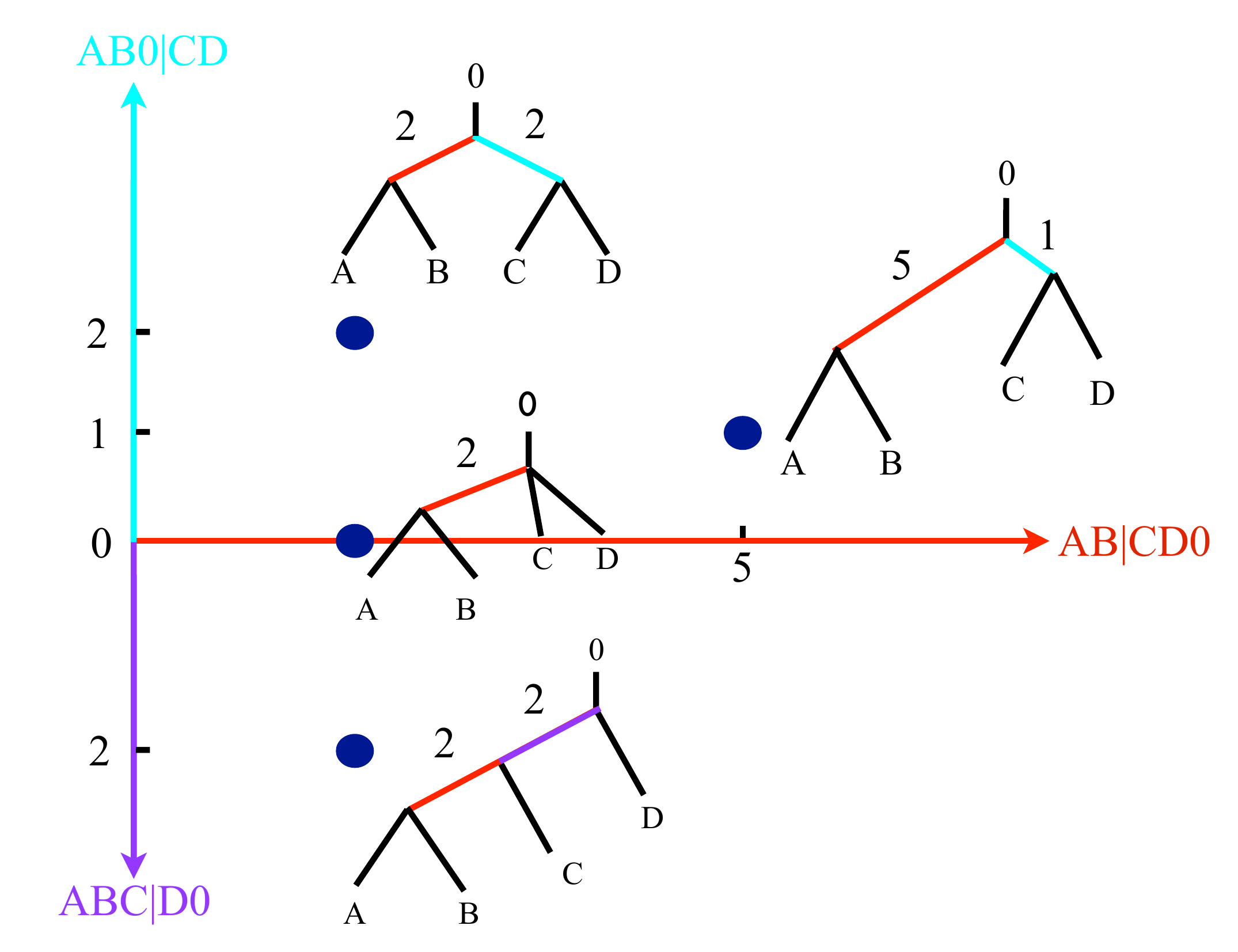}} \hfil
       \subfigure[]{\label{fig:corner_ex}\includegraphics[scale=0.18]{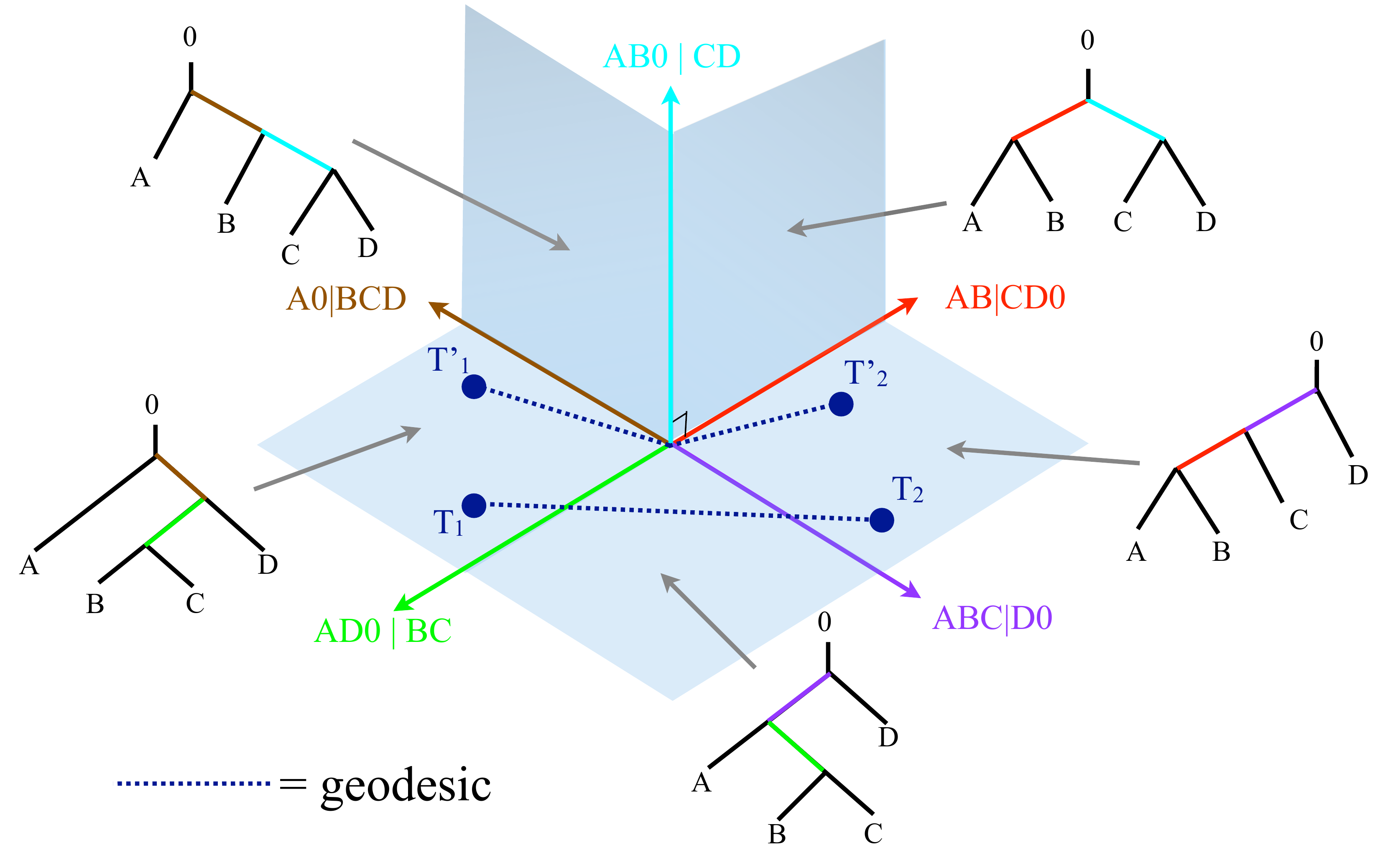}} \hfil
   \end{center}
   \vspace{-0.4cm}
  \caption{(a) Two adjacent quadrants in the tree space, $\mathcal{T}_4$, of trees with 4 leaves and edge vectors of 1 dimension.  Here, the edge vector, or length, is restricted to being non-negative and the pendant edges are ignored, so the Euclidean subspaces are represented as quadrants.  A quadrant contains all trees with a given topology, and each tree with that topology is represented by the coordinates corresponding to its internal edge lengths.  (b) Representation of 5 of the 15 quadrants in the tree space $\mathcal{T}_4$, with edge vectors of 1 dimension.}
  \label{fig:tree-space}
\end{figure}

We now give a description of our tree-space, $\mathcal{T}_n$, for a fixed $n$, which is illustrated for $n= 4$ in Figure~\ref{fig:tree-space}.  All the trees in $\mathcal{T}_n$ with the same \emph{tree topology}, or branching order, form a lower-dimension Euclidean subspace in the tree-space.  The dimension of this subspace is $mk$, where $m$ is the number of edges in the tree topology and $k$ is the dimension of the vector associated with each tree edge.  Each tree edge in the tree topology is put into correspondence with $k$ of the subspace's dimensions, and a particular tree with that topology can be written as a $km$-dimensional vector in that subspace, with the coordinates being the consecutive $k$-dimensional edge vectors.  That is, if a tree has edges $e_1, e_2, ..., e_m$, with corresponding edge vectors $\mathbf{\ell_1} = (\ell_1^1, \ell_2^1, ..., \ell_5^1), ... \mathbf{\ell_m} = (\ell_m^1, \ell_m^2, ..., \ell_m^5)$, then that tree corresponds to the point $(\mathbf{\ell_1}, ..., \mathbf{\ell_m})$.  All trees within the subspace must map their $k$-dimensional edge vectors to the $km$-dimensional vector in the same order, but what this order is does not matter.  

The Euclidean subspaces for each tree topology are glued together in the following way.  Consider a tree containing the edges $e_1, ..., e_m$, where each edge can be identified by the unique partition of the leaves it makes when it is removed from the tree (i.e. removing the edge forms a forest of two trees, each of whose leaves, including the root, forms one half of the partition).  Let exactly one of the edges, say $e_1$, have an all $0$ edge vector.  Then this tree lies on the boundary of the Euclidean subspace corresponding to its tree topology, and furthermore, it actually lies in a lower dimensional Euclidean subspace $E$ corresponding to trees with only the edges $e_2, ..., e_m$ in their topologies.  This lower dimensional subspace $E$ is also on the boundary of two other Euclidean subspaces, and we identify all such common subspaces in all the Euclidean subspaces corresponding to tree topologies to form $\mathcal{T}_n$.  See \cite{feragen_miccai} and \cite{BHV} for a more detailed description of the tree space.

The metric on $\mathcal{T}_n$ is induced by the Euclidean metric on each of its constituent subspaces.  Specifically, the distance between two trees with the same topology is the Euclidean distance between the two points representing those trees in the subspace for that tree topology. The distance between two trees with different topologies is the length of the shortest path joining their points in tree space.  Such a path will consist of a sequence of line segments, each contained in exactly exactly one of the subspaces, and thus the path length is just the sum of the Euclidean lengths of each segment.

Most importantly for this paper, tree space is a non-positively curved metric space~\cite{BHV}, which implies that there is a unique shortest path within the space between any two trees, called the \emph{geodesic}.  The \emph{geodesic distance} between two trees is the length of the geodesic between them, and it can be computed in polynomial time \cite{OwenProvan11}.  Certain statistics, such as means and first principal components, can also be computed on trees in this space~\cite{sturmmean,bacak,feragen_ipmi2013_stats,nye}.

\begin{figure}
\centering 
\includegraphics[scale=0.3]{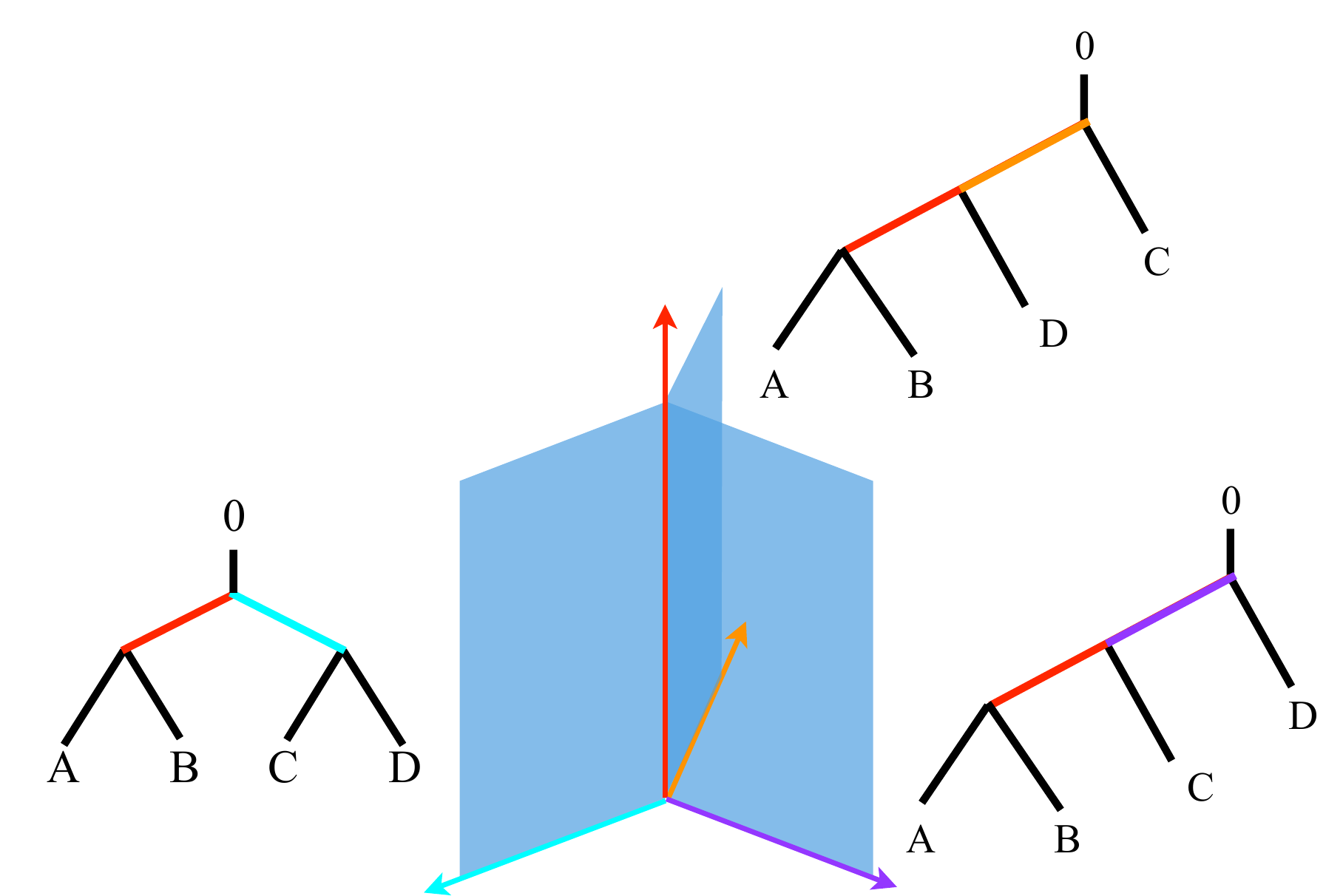}
\caption{An occurrence of a half open book with 3 2-dimensional sheets in tree-space.}
\label{fig:open book}
\end{figure}

While sections of tree-space are identical to higher dimensional Euclidean spaces, tree-space itself is not a manifold.  In particular, it has several singularities, which have infinite negative curvature. One such singularity is at the origin, which corresponds to the tree which has all 0 edge lengths or vectors.  A simpler model of this singularity in $\mathcal{T}_4$ is a \emph{corner}:  five Euclidean quadrants, glued together around an single origin, see Fig.~\ref{fig:corner_ex}.  Singular points in tree-space also occur where the the higher dimensional Euclidean subspaces join together to form a space that locally resembles an \emph{open book}.  An open book is a set of Euclidean half planes, or \emph{sheets}, which are identified along their boundary hyperplanes, which form the \emph{spine}, see Fig.~\ref{fig:open book}. 

\subsubsection*{The Fr\'echet mean in tree-space} \label{meandef}

In Euclidean space, there are a number of equivalent definitions of the mean.  Some of these definitions cannot be carried over to tree-space, while those that can be carried over are no longer equivalent.  One of the definitions of mean in Euclidean space is the Fr\'echet mean, or barycenter, which minimizes the sum of square distances to the input set.  If $\{T_1, T_2, ..., T_r\}$ are a set of input trees, then their Fr\'echet mean in tree-space is the tree $t$ which minimizes $\sum_{i=1}^r d(t,T_i)^2$, where $d$ is the geodesic distance.  The Fr\'echet mean was introduced for tree-space independently by \cite{bacak} and \cite{sturmmean}, both of whom also gave an algorithm to approximate it based on a Law of Large Numbers holding in non-positively curved spaces \cite{sturm}.

\section{Related work}

This paper studies two problems related to understanding variance in datasets of trees: (i) detection of local subtree differences, and (ii) visualization of global population-level geometry.

\subsubsection*{Local significant differences}

Many data types represent entities which can be decomposed into parts or regions. Examples are graph-structured data~\cite{feragen_nips2013,jenatton}, anatomical data which can be segmented into different organs~\cite{Gorczowski,gasparovic} or even single anatomical organs where additional spatial information is relevant; for instance, in the framework of shape analysis~\cite{cates,busch} where local analysis is made on correspondence points on biomedical shape surfaces. A typical problem when studying such data is \emph{interpretability}: A classifier will often only predict a certain diagnosis or class, but in order to understand the cause of the result (and, e.g.~in diagnostic settings, react on it) one also desires to know which parts of the collection caused a certain classification outcome.

While a large body of work has been done on classifying structured data, less is known about how to identify which parts of a structure are relevant for the classification problem. Most such work has been done in settings where there is a correspondence between the parts constituting the data object: In analysis of brain connectivity~\cite{jenatton,ghanbari}, one usually has a matching between the nodes in the dataset, while in voxel-based morphometry~\cite{ashburner} or shape analysis~\cite{cates}, registration is used to match different images to a template. A popular approach to such problems is \emph{structured sparsity}~\cite{huang,jenatton,azencott}, which detects discriminative substructures in data described by fixed-length Euclidean vectors with a known underlying structure relating the vector coordinates. However, anatomical trees usually cannot be described by fixed-length vectors without discarding parts of the tree. Thus, these methods are not directly applicable.

\subsubsection*{Low-distortion embeddings}   

The standard technique for visualising population structure in high-dimensional or non-Euclidean datasets is to extract the pairwise distances between data points, and then use \emph{multidimensional scaling} (MDS), which attempts to embed the points into a lower dimensional Euclidean space such that the given distances between the points are preserved. This is expressed mathematically as minimizing the sum of the differences between original and embedded pairwise distances~\cite{MDSbook}. In a sequence of work \cite{amenta2002case,montealegre2002visualizing,hillis2005analysis} Amenta, St. John et al.~investigate visualization of sets of phylogenetic, or evolutionary, trees using multidimensional scaling. In this work, inter-tree distances are given by the Robinson-Foulds distance \cite{RF81}, which only measures topological differences in the trees. More recently, Wilgenbusch et al. \cite{wilgenbusch2010MDSevaluation} compare 
several non-linear versions of MDS on phylogenetic trees, and find that a metric that 
places less weight on large distances gives more meaningful visualizations.
Chakerian and Holmes \cite{chakerian2012} use MDS with the geodesic distance between trees~\cite{BHV}. A different approach is that of Sundberg et al.~\cite{sundberg2010cartographic}, who visualize phylogenetic trees by projecting them onto a hypersphere; this approach does not consider branch lengths, only tree topology.

All of these methods approach visualization through embedding into a Euclidean space in a low-distortion way. However, embedding spaces need not be restricted to only Euclidean spaces. For instance, low-distortion embedding of a general metric into a tree has been considered for various measures of distortion~\cite{badoiu,agarwala}.  Low-distortion embedding of general metrics into hyperbolic spaces has also been considered by Walter et al.~\cite{walter2004h,walter2002interactive} and Cvetkovski and Crovella \cite{Cvetkovski}. In this paper, we use hyperbolic MDS for more truthful visualizations of tree variation.


\section{Quantification and visualization of local tree-shape differences} \label{quantification}

While previous work~\cite{feragen_ipmi2013_stats} developed methods for finding significant differences between populations of trees, this work did not address the question of where these changes came from. In this section we investigate different methods for detecting \emph{where} in a tree significant differences appear. In Sec.~\ref{subtree_perm} we perform hypothesis testing on nested subtrees in order to detect how significant changes take place in particular subtrees. In Sec.~\ref{classification} we develop a structured sparsity framework which takes advantage of the tree-space geometry in order to handle the fact that subtrees have variable topological structure. In both of these sections, we obtain results on which subtrees induce significant differences. A disadvantage of the methods developed in Sections~\ref{subtree_perm} and \ref{classification} is that they do not take correlation between different subtrees into account. In Sec.~\ref{corr} we therefore develop a method that allows us to study how subtree differences correlate with each other.

{\bf Case study.} We apply the developed methods to a case study of airway trees from subjects with and without Chronic Obstructive Pulmonary Disease (COPD). The $600$ airway trees are from randomly selected subjects from the Danish Lung Cancer Screening Trial~\cite{dlcst}, of which $300$ were diagnosed with COPD at scan time and $300$ were symptom free. The hypothesis testing and classification experiments performed in this chapter all have the common goal of separating the class of COPD patients from the class of symptom free subjects.

The airway trees were extracted from low-dose (120 kV and 40 mAs) pulmonary CT scans. To extract the tree, the airway lumen surface was extracted from the images using the locally optimal path approach of \cite{pechin_miccai} and then refined using the optimal surface approach of \cite{petersen2014}. Afterwards centerlines were computed by front propagation within the refined lumen surface as described in \cite{exact09b}. The resulting centerlines were disconnected in bifurcation regions and so Dijkstra's algorithm was used to connect them along shortest paths within an inverted distance transform of the refined lumen surface. These centerlines were then represented by $6$ equidistantly sampled landmark points. The airway trees were normalized by patient height as an affine scaling parameter.

Airway trees are somewhat regular in the sense that some of the branches have anatomical names and can be found in most human lungs. The subtrees rooted at these branches feed different subdivisions of the lung at different hierarchical levels, as schematically illustrated in Fig.~\ref{fig:labels}. The Trachea is the root branch that feeds both lungs. The left and right main bronchi (LMB, RMB) feed the left and right lungs. The left upper lobe and lower lobe branches (LUL and LLB) feed the left upper and lower lobes. The lower lobe splits into the branches L7-L10. The left upper lobe branches into two subsections; the first feeds the three segments L1, L2 and L3, and the branch feeding all of these is called L1+2+3. The second subsection feeds the segments L4 and L5, and their parent is called L4+5. The right lung us subdivided into the upper lobe, fed by the right upper lobe branch (RUL), and the middle- and lower lobes, both fed by the bronchus intermedius (BronchInt). The middle lobe consists of the segments R4-R5, fed by the the parent R4+5, and the lower lobe is fed by the right lower lobe branch (RLL), and splits into the segments R7-R10.

Due to variation in airway tree topology, these branches are not always all present. In our dataset, however, which has been automatically labeled using the algorithm presented in \cite{feragen_miccai}, the following branches are consistently present:
\begin{equation} \label{cut_branches}
\begin{array}{c}
\textrm{Trachea, LMB, RMB, LUL, RUL, L1+2+3, LLB, BronchInt, and RLL}.
\end{array}
\end{equation}

\begin{minipage}[b]{\textwidth}
\vspace{0.2cm}
\begin{minipage}[b]{0.49\textwidth}
\centering
\includegraphics[width=\textwidth]{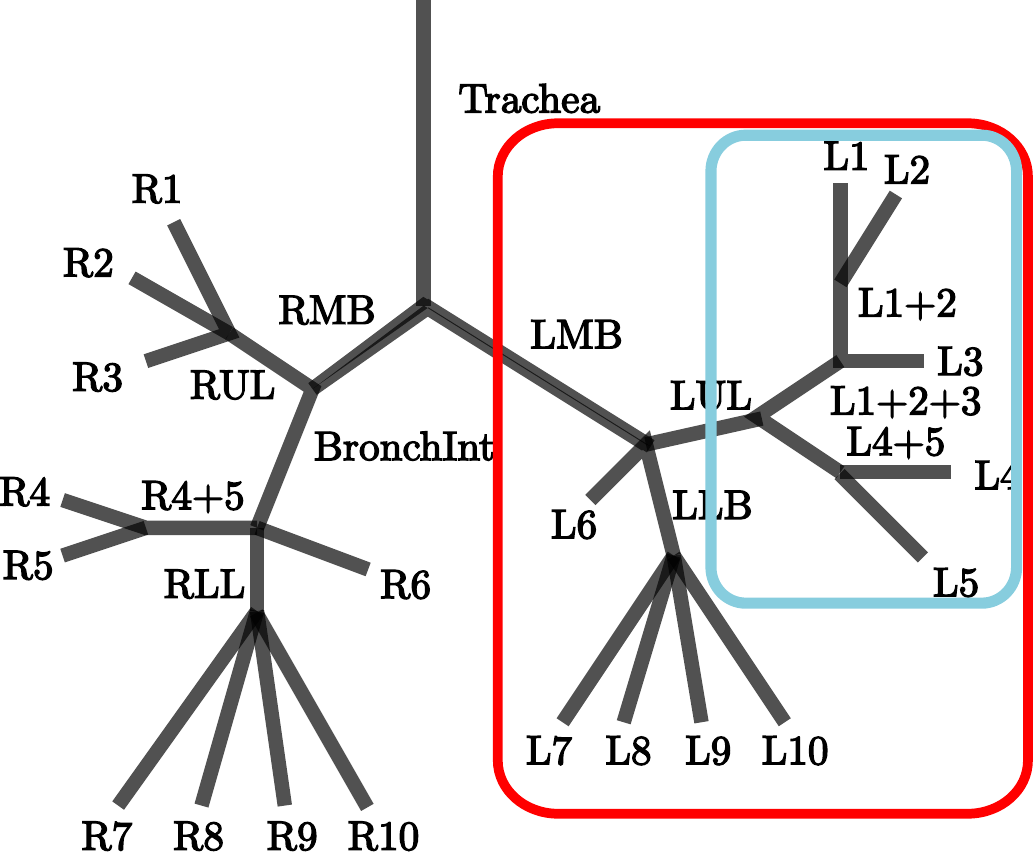}
\captionof{figure}{Airway tree (black) and sub-trees (LMB, red) and (LUL, blue).\\ \\ \\ \\} 
\label{fig:labels}
\end{minipage}
\hfill
\begin{minipage}[b]{0.49\textwidth}
\centering
\begin{tabular}{ c  c  c }
\hline
LABEL & P-VALUE & P-VALUE \\
& mean & variance\\
\hline
full & \bf 0.0010 & \bf 0.0060 \\
RMB & \bf 0.0020 & 0.0939\\
RUL & 0.2298 & 0.1668\\
BrInt & \bf 0.0050 & 0.1249 \\
RLL & \bf 0.0300 & 0.0959 \\
LMB & 0.0859 & \bf 0.0210\\
LUL & \bf 0.0320 & \bf 0.0390\\
L123 & \bf 0.0260 & \bf 0.0410\\
LLB & 0.5524 & 0.1588\\
\hline
\end{tabular}
\captionof{table}{{\bf Case study:} Group comparison showing results of permutation tests on subtrees of the full airway trees rooted at the branches listed in~\eqref{cut_branches}. The permutation test compares the populations of airway trees from COPD patients and symptom free subjects.}
\label{tab:permTest}
\end{minipage}
\vspace{0.2cm}
\end{minipage}

\subsection{Permutation tests for subtree statistics} \label{subtree_perm}

  
In this section we perform subtree hypothesis testing using the tree-shape permutation tests for equality of means and variances developed in~\cite{feragen_ipmi2013_stats} on the nested subtrees defined by the subtree root branches~\eqref{cut_branches}. These tests are standard permutation tests which, for samples $G_1$ and $G_2$ drawn from two different classes of trees (in our experiments: healthy subjects and COPD patients), use test statistics used for means and variances between classes defined as
\[
t_m = d(\mu(G_1), \mu(G_2)) \textrm{ and } t_v = \vert v(G_1)-v(G_2)\vert,
\]
respectively, where $\mu(G_i)$ is the Fr\'{e}chet mean of the trees in the $i^{th}$ class as defined on p.~\pageref{meandef}, and $v(G_i)$ denotes the variance of the $i^{th}$ class $\frac{1}{(N-1)}\sum_{t \in G_i} d^2(t,\mu(G_i))$. 

Under the null hypothesis, namely that there is no difference between the two classes, the samples $G_1$ and $G_2$ are drawn from the same distribution on $\mathcal{T}_n$, and randomly permuting the elements of $G_1$ and $G_2$ should not affect the value of the test statistic $t_\ast$.

Form the two-class data set $G = G_1 \cup G_2$ and consider partitions of $G$ into subsets of size $N_1 = |G_1|$ and $N_2 = |G_2|$. Due to the size of $G$ we cannot check all possible permutations, but instead compute the test statistics $(t_\ast)_m$ for means and variances for the new subsets, $m = 1, \ldots, M$, for $M$ random partitions of $G$ into sets of size $N_1$ and $N_2$. Comparing the $(t_\ast)_m$ to the original statistic value $t_\ast$ for the samples $G_1$ and $G_2$, we obtain a $p$-value approximating the probability of observing $t_\ast$ under the null hypothesis:
\[
p = \frac{1 + \sum_{(t_\ast)_m \ge t_\ast, m \in \{1, \ldots, M\}} 1}{M + 1},
\]
where the additional $1$ is added to avoid $p = 0$.

Permutation tests for the two statistics were performed on each subtree with $M = 1000$ permutations. The results are summarized in Table~\ref{tab:permTest}, and show significant differences in several subtrees. These subtrees are identified by their root branches, which are illustrated schematically in Fig.~\ref{fig:labels}. In comparison, the same hypothesis test was made on the individual branches~\eqref{cut_branches} along with the segment branches R1-R10, L1-L10, with results shown in Table~\ref{tab:permTest_branches}. The tests on individually identified branches show relatively fewer significant differences between the two populations, emphasizing a need for considering the airway subtrees as entities rather than collections of independent branches.

This suggests that the permutation test described here can be applied to study local group differences between subtrees in a hierarchical manner.

\begin{table}[t]
  \centering
    \caption{{\bf Case study benchmark:} Permutation tests for shape differences in individual branches between populations of airway trees from healthy individuals and COPD patients. P-values below a threshold of $0.05$ are shown in bold.}  
  \begin{tabular}{c c c | c c c}
    \hline
    LABEL & P-VALUE & P-VALUE & LABEL & P-VALUE & P-VALUE \\
     & mean & variance & &  mean & variance\\
    \hline
    RMB & 0.163 & 0.621 & LMB & \bf 0.020 & 0.786\\
    RUL & 0.134 & 0.416 & LUL & 0.297 & 0.118\\
    R1 & 0.410 & 0.363 & L1 & 0.163 & 0.391\\
    R2 & 0.116 & 0.255 & L2 & 0.324 & \bf 0.017\\
    R3 & 0.329 & 0.854 & L3 & 0.968 & 0.800\\
    BronchInt & \bf 0.001 & 0.764 & L45 & 0.078 & 0.312\\
    R4 & 0.134 & 0.190 & L4 & 0.372 & 0.570\\
    R5 & \bf 0.027 & 0.175 & L5 & \bf 0.023 & \bf 0.050\\
    R6 & 0.992 & 0.135 & L6 & 0.260 & 0.833\\
    RLL & \bf 0.001 & 0.865 & LLB & 0.177 & 0.112\\
    R7 & 0.058 & 0.325 & L7 & 0.496 & 0.611\\
    R8 & \bf 0.014 & 0.207 & L8 & 0.466 & 0.900\\
    R9 & \bf 0.037 & 0.127 & L9 & 0.146 & \bf 0.026\\
    R10 & 0.308 & 0.652 & L10 & 0.855 & 0.162\\
    L123 & 0.393 & 0.361 & & & \\
    \hline
  \end{tabular}
  \label{tab:permTest_branches}
\end{table}

\subsection{Subtree classification} \label{classification}

In the previous section we saw how hypothesis testing on subtrees allowed us to learn about which subtrees differed significantly between two populations of trees. While significant differences are interesting in their own right, we are often particularly interested in finding \emph{predictive} differences. In particular, we want to find subtrees such that restricting prediction to these subtrees results in good predictive performance, giving interpretable classifiers in the sense that we can detect \emph{which} tree changes are predictive.

\begin{table}
\caption{Mean $\pm$ standard deviation of COPD classification accuracy using branch length (left) and branch shape (right).}
\vspace{-0.7cm}
\begin{center}\begin{tabular}{ccc}
\hline\noalign{\smallskip}
Method & Accuracy & Accuracy\\
 & length & shape\\
\noalign{\smallskip}\hline\noalign{\smallskip}
 LDA & $0.56 \pm 0.06$ & $0.52 \pm 0.06$\\
 QDA & $0.55 \pm 0.05$ & N/A\\
 Mahalanobis & $0.54 \pm 0.05$ & N/A \\
 $k$NN & $0.53 \pm 0.06$ & $0.53 \pm 0.06$\\
 SVM & $0.56 \pm 0.06$ & $0.56 \pm 0.06$\\
\noalign{\smallskip}\hline
\end{tabular}\end{center}
\label{tab:class_results}
\end{table}

\subsubsection{Classification on known branches} \label{baseline}

A straight-forward approach to tree classification and identification of discriminative substructures of trees is to use standard classification methods on vectors whose coordinates correspond to a fixed set of identified branches. In our case, these branches will be identified by their anatomical names, which create a natural matching between the branches of different trees. Classification of such vectors can return information about which branches are more discriminative, because classifiers such as the support vector machine (SVM), include coordinate weights that intuitively correspond to the relevance of the individual coordinate feature for the classification problem. This method is simple, but has the disadvantage that it can only use branches that are present in every single tree in the dataset. This method will form a baseline to which our proposed methods are compared.

Since we could only use branches that were present in every single dataset tree, we used the list of branches~\eqref{cut_branches} along with the leaves $\{$R1-R10, L1-L10$\}$, which are guaranteed to be present. We performed classification with linear discriminant analysis (LDA), quadratic discriminant analysis (QDA), Mahalanobis distance, $k$-nearest neighbor ($k$NN) using $k = 5$, and support vector machine (SVM) using $10$ repetitions of $10$-cross validation. The corresponding classification accuracies are reported in Table~\ref{tab:class_results}. Note that the QDA and Mahalanobis distance are missing for the shape branch features; this is because these both require a positive definite covariance matrix, for which the data set was too small for the higher-dimensional shape vectors.

While the mean classification accuracy is above chance for all classifiers, none of them are significantly above chance. A common heuristic to find features which are important in classification is to study the magnitudes of the coordinates of the weight vector produced by the SVM algorithm, as shown in Table \ref{tab:SVMweights}. Note that in addition to the classification accuracy being very low, the weights of high magnitude are scattered around the airway tree, not adding much in terms of interpretation.

The poor performance of classifiers on the set of all branches could be explained by the fact that many branches are highly correlated. In such cases, the weights might not carry much information. The poor classification accuracy may be explained by the dimensionality of the data. This motivates our search for a more predictive and interpretable classification algorithm by including subtree information in the classifier.

\begin{table}[ht!]
\caption{The mean and standard deviations of the SVM weight vectors on the COPD/healthy classification. The largest weight vectors as well as those falling within one standard deviation are highlighted.}
\resizebox{\linewidth}{!}{
\begin{tabular}{cc|cc|cc|cc}
\hline\noalign{\smallskip}
Branch & SVM weight & Branch & SVM weight  & Branch & SVM weight  & Branch & SVM weight\\
\noalign{\smallskip}\hline\noalign{\smallskip}
RMB & $2.0 \pm 1.45$ & R5 & $2.8 \pm 1.6$ & LMB & $-2.6 \pm 1.6$  & L4 & $-1.4 \pm 1.2$\\
RUL &  $\bf 4.3 \pm 1.6$ & R6 & $-1.7 \pm 1.6$ & LUL & $1.7 \pm 2.1$ & L5 & $2.3 \pm 1.6$\\
R1 & $-0.7 \pm 1.2$ & L6 & $1.5 \pm 1.6$ & L123 & $-0.8 \pm 1.4$ & RLL & $2.6 \pm 1.6$ \\
R2 & $\bf 4.9 \pm 1.2$ & R7 & $1.4 \pm 1.5$ & L1 & $-1.6 \pm 1.2$ & LLB & $\bf -3.6 \pm 1.5$\\
R3 & $\bf -3.6 \pm 1.4$ & R8 & $3.3 \pm 2.2$ & L2 & $\bf -3.2 \pm 1.3$ & L7 & $-2.8 \pm 1.3$\\
BrInt & $\bf -5.0 \pm 1.6$ & R9 & $\bf 7.3 \pm 1.5$  & L3 & $1.4 \pm 1.3$ & L8 & $2.2 \pm 1.5$\\
R4 & $-0.4 \pm 1.6$ & R10 & $2.9 \pm 1.6$ &  L45 & $\bf 3.6 \pm 1.8$ & L9 & $\bf 3.2 \pm 1.5$\\
& & & &  & & L10 & $\bf 5.2 \pm 1.5$\\
 \noalign{\smallskip}\hline
 \end{tabular}
}
\label{tab:SVMweights}
\end{table}

\subsubsection{Structured sparse feature selection through regularized logistic regression on subtree similarity} \label{logreg}

Under a hypothesis that significant differences are found in local subtrees, we incorporate local subtree structure into classification through a sparse classifier taking subtree similarity as input. Logistic regression measures the relationship between a categorical dependent variable (class label) and one or more independent variables by using conditional probabilities as predicted values of the dependent variable. The \(L_1\) regularized logistic regression, or the so-called sparse logistic regression \cite{Tibshirani96}, regularizes the classifier by forcing the weight vector of the classifier to have a small number of nonzero values. This results in implicit feature selection and robustness to noise, as well as interpretability through the selected subtree features. In addition to its solid theoretical foundation, this model is computationally efficient~\cite{Friedman10}.

Consider a set of \(n\) training examples \(T=\left\{(x_1,y_1),(x_2,y_2),\cdots,(x_n,y_n)\right\}\) from which a tree classifier \(y=f(x)\) will be learned. The \(i^{th}\) tree is represented by a \(D\)-dimensional feature vector \(x_i=\left[x_{i1},x_{i2}, \cdots, x_{id}\right]^{T}\) where $x_{ij} = d(S^i_j,\mu_j)$, where $S^i_j$ is the $j^{th}$ subtree of the $i^{th}$ tree, $\mu_j$ is the Fr\'echet mean tree of all $j^{th}$ subtrees in the training set, and $d$ denotes geodesic distance between trees. The $D$ subtrees are rooted at the branches listed in~\eqref{cut_branches}. The values \(y_i \in \left\{0,1\right\} \) indicate the class labels of the two groups, modeling the conditional probability distribution of the class label \(y\) given a feature vector \(x\) as:
\begin{equation*}
p(y=1|x,\beta)=\frac{1}{1+\exp(-\beta^{T} x)},
\label{eq:logistic}
\end{equation*}
where \(\beta \in \mathrm{R}^{D}\) are the parameters of logistic model. The estimation of the parameters \(\beta\) is done by likelihood maximization, equivalent to minimizing the negative log-likelihood
\begin{equation*}
\hat{\beta}=\arg\min \sum_{i=1}^{n}-\log p(y_{i}|x_{i},\beta).
\end{equation*}
Applying a sparse regularizer we obtain feature selection, interpretability and reduced overfitting. This is done by adding a so-called \emph{lasso} regularization term:
\begin{equation*}
\hat{\beta}=\arg\min \sum_{i=1}^{n}-\log p(y_{i}|x_{i},\beta)+\lambda||\beta||_{1},
\end{equation*}
where \(\lambda\) is a parameter controlling the sparsity of \(\beta\), in the sense that fewer nonzero coefficients of \(\beta\) remain as \(\lambda\) increases. The optimal \(\lambda\) is chosen to optimize classification accuracy by $5$-fold cross validation.

The nested subtrees used will be correlated by definition. One way of handling this is by adding an $l_2$ norm regularization term as well, known as \emph{elastic net regularization}. This leads to an objective function
\begin{equation*}
\hat{\beta}=\arg\min \sum_{i=1}^{n}-\log p(y_{i}|x_{i},\beta)+\lambda (\alpha||\beta||_{1} + \frac{1 - \alpha}{2} ||\beta||_2^2).
\end{equation*}
The results of the sparse classifiers for different values of $\alpha$ are shown in Table~\ref{tab:logreg}. Note that the classification performance is significantly better than that of the standard classifiers on identified branches seen in Table.~\ref{tab:class_results}. Moreover, note the discriminative subtrees selected by the classifier. The fact that the lasso regularizer results in only the full tree being selected is most likely a result of the correlation between subtrees. As an $l_2$ regularizer is also added, we obtain a tradeoff between sparsity and including correlated subtrees.

\begin{table}
\caption{Results of the structured logistic classifier with lasso and elastic net regularization. Classification results are averaged over $10$ randomized folds, and significant features are those where the distances to both class means were kept as features in the classifier in all folds.}
\begin{tabular}{l c c}
\hline
Value of $\alpha $ & Classification result & Significant features\\
\hline
$1$ (lasso) & $65 \pm 2.7 \%$ & full\\
$0.75$ & $64.3 \pm 2.1 \%$ & RMB, full\\
$0.5$ & $62.5 \pm 2.4 \%$  & RMB, BronchInt, full \\
$0.25$ & $62.5 \pm 2.3 \%$ & RMB, LMB, LLB, BronchInt, RLL, full\\
\hline
\end{tabular}
\label{tab:logreg}
\end{table}

\subsection{Subtree variance correlation testing} \label{corr}

As the localized methods use features extracted from nested subtrees, we expect a high degree of correlation between overlapping subtrees. Most of the previously described methods do not take such correlations into account, and this may, in particular, be a problem for the interpretability through selected features. Moreover, it is interesting to know whether variation in non-overlapping subtrees is correlated. In this section we provide a method for testing the correlation between variance in the subtrees.  We use the notation from Section~\ref{logreg}.

To compare the variance between subtrees in the same airway tree, we use the distance from the $j$-th subtree $S_j^i$ in tree $i$ to the population mean $\mu_j$ of all $j$-th subtrees in some class as a measure of the amount of variation in that subtree $S_j^i$.  We compute the correlation between these distances, represented by the random variable $X$ and $Y$, for each pair of subtrees $(j,k)$, and measure whether deviation from the mean subtree $\mu_j$ is correlated with deviation from the mean subtree $\mu_k$. To measure the correlation, we use Pearson's sample correlation coefficient, $$r_{xy} = \frac{\sum_{i=1}^n (x_i - \bar{x})(y_i - \bar{y})}{(n-1)s_xs_y},$$ where $\bar{x}$ and $\bar{y}$ are the same means of the two distance variables $X$ and $Y$, and $s_x$ and $s_y$ are the sample standard deviations of the $X$ and $Y$.  This is equivalent to the sample covariance divided by the sample standard deviations.

\begin{figure}[!ht]
  \centering
    \includegraphics[width=5in]{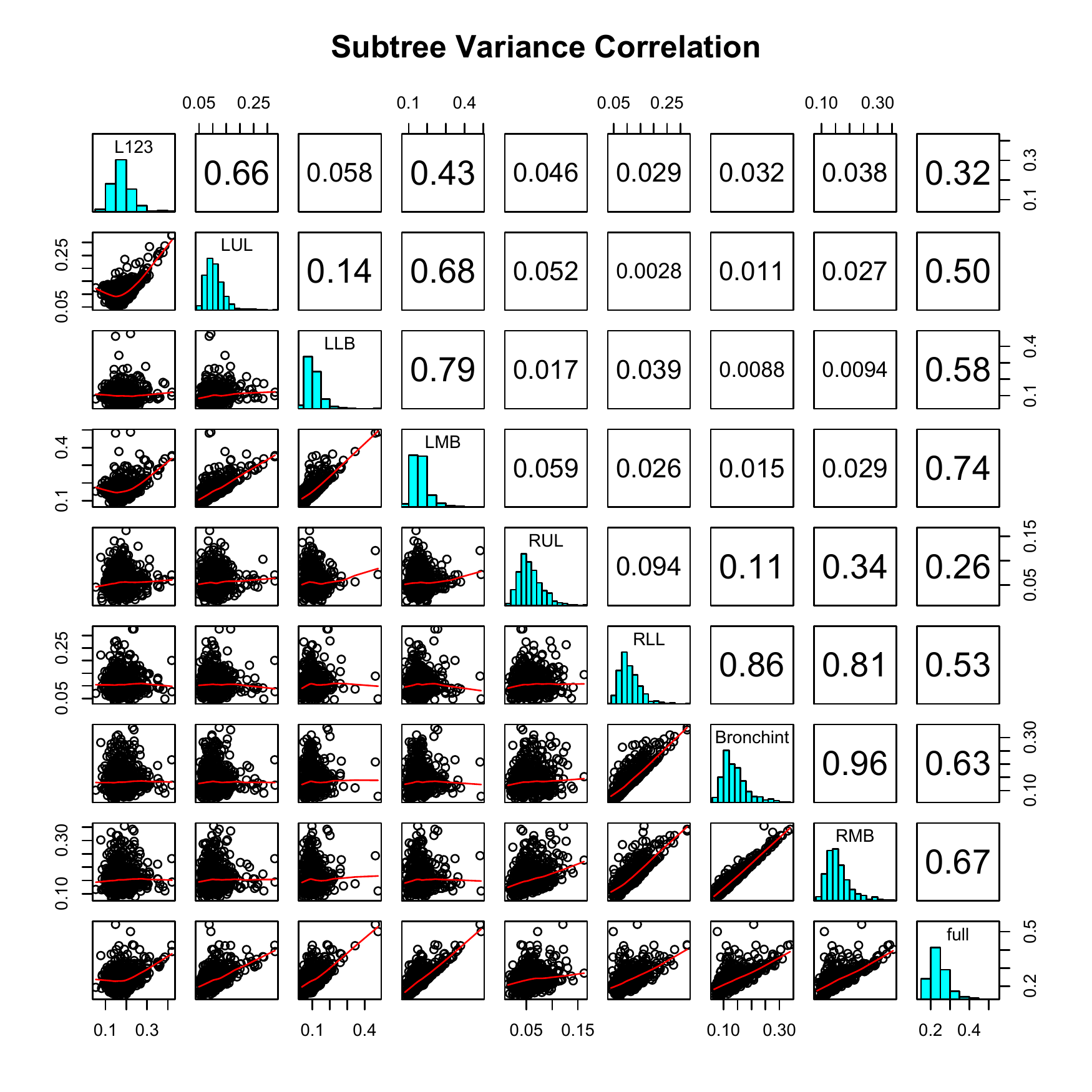}
    \caption{The plots in the lower triangle compare the distances between subtrees in the same tree to their corresponding mean subtree of healthy patients.  The plots along the diagonal are the histograms of these distances when the subtree is fixed.  The upper diagonal gives the correlation of the distances plotted in the corresponding plot in the lower triangle.}
    \label{fig:subtree_cor}
\end{figure}

The results of applying this test to the airway data set is shown in Fig.~\ref{fig:subtree_cor}.  Many of the subtree pairs exhibiting correlation in their variance are nested, as expected.  For example, variation is very correlated between the three subtrees RMB, Bronchint, and RLL, where RLL is a subtree of Bronchint, which is itself a subtree of RMB.  Similarly, there is a high correlation in the variation between the nested pairs LMB and LUL, and LMB and LLB.  However, not all nested subtrees are highly correlated.  In particular, RUL is also a subtree of RMB, but variation in it is not very correlated with that in RMB -- in fact, variance in RUL is not strongly correlated with any other subtree.  The trees are not separated by class as there was no significant difference in the behavior of the two classes.

\section{Visualization of NPC information spaces via Low-Distortion Embedding into the Hyperbolic Disc} \label{visualization}

High-dimensional data is often embedded into lower dimensional spaces in order to
improve the efficiency of computations, or, with a two- or three-dimensional embedding space,
for visualization. 
Besides being high-dimensional and stratified, the  
tree space $\mathcal{T}_n$ has negative curvature.  
While it is exactly this property that gives it unique geodesics, 
it also means that the number of trees within a given 
neighborhood can grow exponentially with the radius of the neighborhood. 
Our hypothesis was that embedding point sets in $\mathcal{T}_n$ into 
lower-dimensional hyperbolic space,
which also has negative curvature, would allow embeddings with 
lower distortion and/or lower total error. 
We explore the use of two different visualization techniques for general metric distance matrices,
Multi-Dimensional Scaling and Isomap.

\emph{Multidimensional Scaling} (MDS) is a classical approach that maps the original metric dataset to a target Euclidean space, usually of low dimension.  It transforms the input metric distance matrix into a set of coordinate positions for the data points - in our case, each tree is a data point - such that the  Euclidean distances between the coordinates approximate the input distances as well as possible. Using the new embedded coordinates, one can visualize dataset structure through the embedded dataset, where inter-point distances have been preserved as well as possible.  Different definitions for what it means to 
preserve the distances ``as well as possible" produce different computational problems.  Principle Components Analysis (PCA) can be seen as a version of MDS, for which the problem has a global solution, but other
definitions of optimal distance preservation often lead to better visualizations. 
These versions are all non-linear, so both the optimization criterion and the method of optimization can
lead to different results. 

\emph{IsoMap} ~\cite{isomap} is a more recent method
intended for points which lie on a lower-dimensional surface in the high-dimensional space.
It begins by constructing a neighborhood graph connecting nearby points in the input space. 
Then, using this graph, it  
approximates geodesic distances on the surface. 
Finally it applies MDS to the matrix of geodesic distances. 

While the standard approach in both of these methods
is to use a Euclidean target space for the embedding, in the past decades, 
hyperbolic multidimensional scaling has also been proposed. 
In a nutshell, the original Euclidean distance in the target space is replaced by the hyperbolic distance:
\begin{align}
	d (z_i,z_j)=2 \tanh^{-1} \frac{|z_i-z_j|}{|1-z_i\bar{z_j}|},
	\label{eqn:hyperdist}
\end{align}
where $z_i$ and $z_j$ denote two points in the target space. The modification in the distance metric makes the computation of gradients non-trivial. 
Our goal was to explore the question of whether hyperbolic space 
would be a more successful target space for the visualization of distributions of trees,
since tree space and hypoerbolic space are both non-positive curved. 

Recently, Cvetkovski and Crovella\cite{Cvetkovski} introduced a 
method MDS-PD (metric multidimensional scaling algorithm using the Poincar{\'e} disk model) which 
is based on a steepest decent method with hyperbolic line search. 
We adapted this software for our experiments with hyperbolic space, and we review the method here;
more details can be found in  \cite{Cvetkovski}.
Complex coordinates are used to present the points of the hyperbolic plane, making the Poincar{\'e} disk model a subset of the complex plane $\mathbf{C}$: $\mathbf{D}=\left\{ z\in \mathbf{C}| |z|<1\right\}$. 
The objective function to be minimized is the total embedding error 
\begin{align*}
	E=c\sum_{j=1}^{n}\sum_{k=j+1}^{n} c_{jk}\left( d_{jk} - \delta_{jk}\right)^2.
\end{align*}
where $c$ and $c_{jk}$ are constants,  $d_{jk}$ is the hyperbolic distance between points $z_j$ and $z_k$ (Equation~\ref{eqn:hyperdist}), and $\delta_{jk}$ denotes the dissimilarity/distance between 
points $z_j$ and $z_k$ in the input dissimilarity/distance matrix. 
More specifically, we use the Sammon Stress Criterion, in which $c$ and $c_{jk}$
are fixed based on $\delta_{jk}$ as follows:
\begin{align}
	E=\frac{1}{\sum_{j=1}^{n}\sum_{k=j+1}^{n}  \delta_{jk} }\sum_{j=1}^{n}\sum_{k=j+1}^{n} \frac{\left( d_{jk} - \delta_{jk}\right)^2}{\delta_{jk}}
\label{eqn:sam_error}
\end{align}
This criterion does not favor preserving large distances over small ones.  
The algorithms starts with a set of random points in the Poincar{\'e} disk. 
In each iteration, it moves each of 
the points along the gradient direction of the energy function shown in Equation~\ref{eqn:sam_error} 
with a Mobius transform 
until one of the stopping tolerances is met or the maximum iteration number is reached. 

%
%

\subsection{Experiments on real and synthetic data}

While much of tree-space looks locally like a Euclidean space, there are two local features which are decidedly not Euclidean:  corners and open books.  A corner is point concentration of negative curvature (see Fig.~\ref{fig:corner_ex}), while an open book is a set of Euclidean half-space attached together along their axes, or "spine" (see Fig.~\ref{fig:open book}).  These two features, as well as the high dimension of the local Euclidean space, are the sources of error for the low-distortion embedding.  We generate synthetic datasets that isolate the two features to determine how hyperbolic MDS (HMDS) and hyperbolic isomap (HIsomap) treat them.  We compare the results both qualitatively and quantitatively with embeddings done with classical MDS and isomap.  More specifically, the datasets are CORNER, in which 250 points are generated by sampling the distance from the origin from a Gaussian distribution $\mathcal{N}(0, 1)$ and sampling an angle with one of the orthant boundaries uniformly from the interval $[0, \frac{5\pi}{2}]$; 3SHEETS\_2D, in which 50 points are generated in each of 3 2-dimensional sheets; 3SHEETS\_3D, in which 50 points are generated in each of 3 3-dimensional sheets;  5SHEETS\_2D, in which 50 points are generated in each of 5 2-dimensional sheets; 5SHEETS\_3D, in which 50 points are  in each of 5 3-dimensional sheets; and COPD, in which the lung airway trees of 125 healthy patients and 125 patients with COPD are randomly selected.  Within each sheet, the 50 points were generated by sampling from a symmetric normal distribution in the underlying Euclidean space that is centered at the origin.

The multiplicative distortion for each embedding approach is summarized in Table~\ref{tab:errors}.  The multiplicative distortion for a single distance between two points in the dataset is $original\_distance/embedded\_distance$. The distortion for the whole dataset is $max\_distortion/min\_distortion$, where $max\_distortion$ is the maximum distortion of any two points and $min\_distortion$ is the minimum distortion for any two points.  HMDS and HIsomap perform the best for almost all of the datasets.  The embedded visualizations and the histograms for each dataset are found in Figures~\ref{fig:all_embeddings} and \ref{fig:all_error_hists}.  

\begin{table} 
   \centering
    \caption{Multiplicative distortion of the embeddings.}
    \begin{tabular}{l | rrrr }
         \hline
             & MDS & Isomap & HMDS & HIsomap  \\ \hline
             CORNER  & 1.4 & 5.0 & 18.3 & 2.96 \\ 
             3SHEETS\_2D & 71.4 & 98.9 & 76.3 & 44.0 \\ 
             3SHEETS\_3D & 44.0 & 189.54 & 54.1 & 68.2 \\ 
             5SHEETS\_2D & 551.9 & 567.4 & 87.8 & 76.1 \\
             5SHEETS\_3D & 2097.8 & 470.5 & 393.6 & 123.4 \\
             COPD\_250  & 253.9 & 952.3  & 62.0 & 64.3 \\
         \hline
    \end{tabular}
\label{tab:errors}
\end{table}


Qualitatively, for CORNER, all methods were qualitatively able to group the points in the same quadrant, and MDS performs best qualitatively, while the hyperbolic Isomap performs better than the Isomap. For the two-dimensional open books 3SHEETS\_2D and 5SHEETS\_2D, all methods also grouped the points by 
their respective sheets.  The two Euclidean methods overlaid all but two of the sheets, while the two hyperbolic methods kept the sheets distinct, particularly in 3SHEETS\_2D, better representing the true geometry.  
Despite increasing the dimension only by one, for 3SHEETS\_3D it was much harder for the methods to separate the distinct sheets.  While MDS performed the best quantitatively, this was not the case qualitatively, where the two hyperbolic methods gave better sheet separation.  All methods had trouble representing the more complex datasets 5SHEETS\_3D and COPD, although quantitatively, the hyperbolic methods did a far better job of reducing distortion. 

The ideal histogram would place all of the distances in the column corresponding to zero error. 
Although the embedded datasets in Fig.~\ref{fig:all_embeddings} do not provide much qualitative insight for the more complex datasets, the histograms in Fig.~\ref{fig:all_error_hists} show that the hyperbolic methods generally give the most accurate reduction to two dimensions.

\section{Discussion and conclusion}

We have considered two different approaches for quantifying and visualizing variance in datasets of trees. In Sec.~\ref{quantification} the dataset trees were divided into nested subtrees, in order to quantify the contribution of different subtrees in distinguishing two populations of trees through either hypothesis testing or classification. These approaches were applied to populations of airway trees from COPD patients and healthy individuals, where the most discriminative subtrees were extracted for the different tasks. In Sec.~\ref{visualization} visualization of population structure for datasets of trees was studied through multidimensional scaling and isomap in a hyperbolic disc as opposed to in the Euclidean plane. The choice of a hyperbolic visualization space was motivated by the fact that tree-space itself has singular points which are hyperbolic, and it thus seems likely that a hyperbolic visualization space can give a more truthful rendering of the structure of the population of trees than a Euclidean space. We demonstrate a quantitative and visual improvement in dataset visualization on a set of synthetic datasets sampled from singular spaces representing the types of singularities found in tree-space, as well as on a set of airway trees.

These approaches supply a new set of tools, and give insight into new potential solutions, for analysis of tree-structured data. Future work includes development of structured sparsity methods using subtrees where the correlation between different subtrees is explicitly taken into account, as well as low-distortion embedding into more complex non-Euclidean visualization spaces whose geometry is similar to that of tree-space.

\begin{table}[!ht] 
	\caption{The embedded datasets.  For the CORNER, 3SHEETS\_2D, 3SHEETS\_3D, 5SHEETS\_2D, and 5SHEETS\_3D dataset embeddings, points have the same color if they are located in the same quadrant or sheet.  For the COPD dataset embeddings, the class of healthy patients is colored in red, and the class of patients with COPD are colored in blue. } 	
	\tabcolsep=0.11cm
	\begin{tabular}
		{ccccc}
		& Classical MDS & Classical Isomap & HyperMDS & HyperIsomap \\
		\rotatebox{90}{~~~~\parbox{2mm}{CORNER}~}&
		\includegraphics[trim =4.6cm 8cm 4cm 7cm,clip=true,scale=0.21]{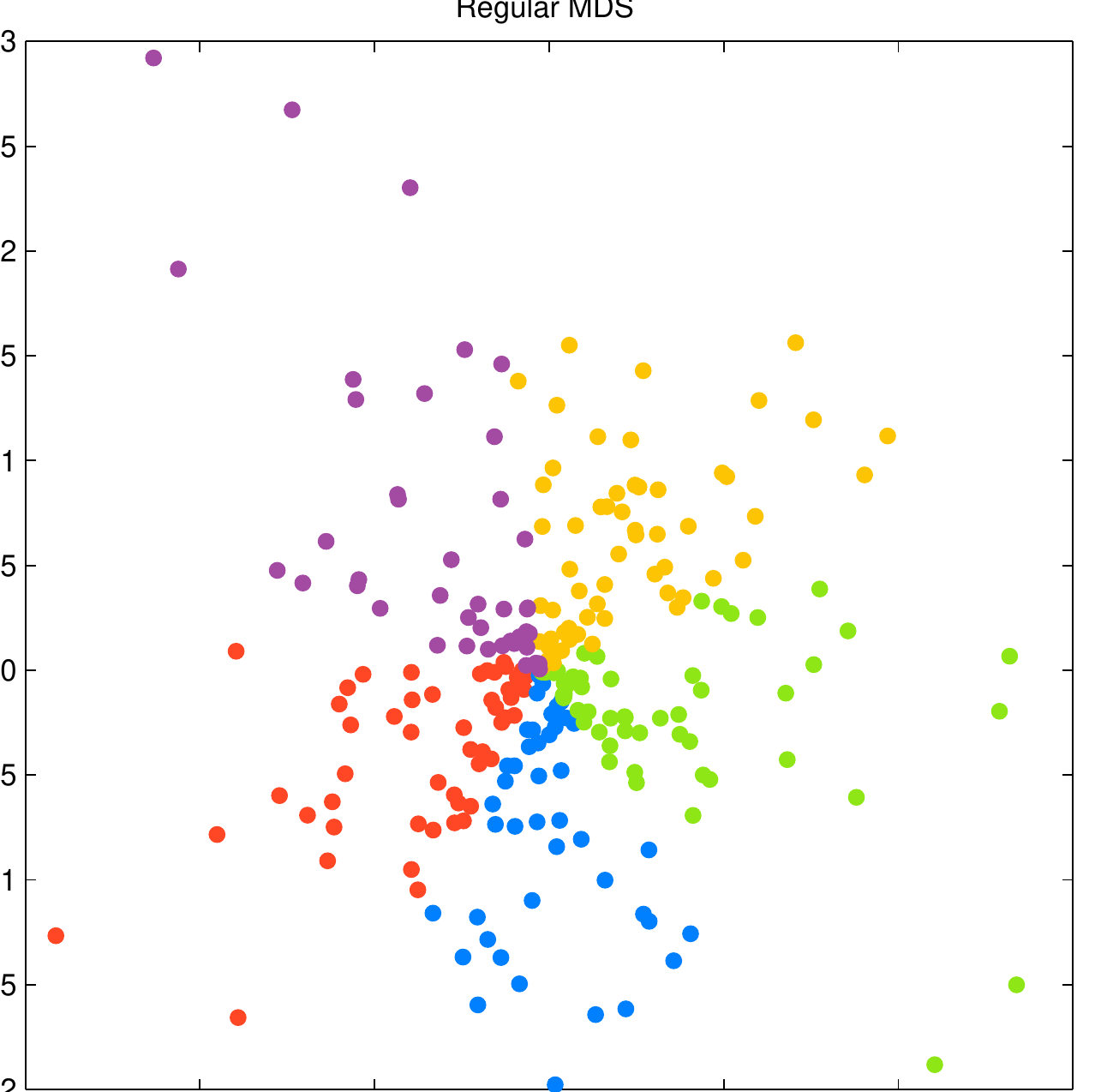} &
		\includegraphics[trim =4.6cm 8cm 4cm 7cm,clip=true,scale=0.21]{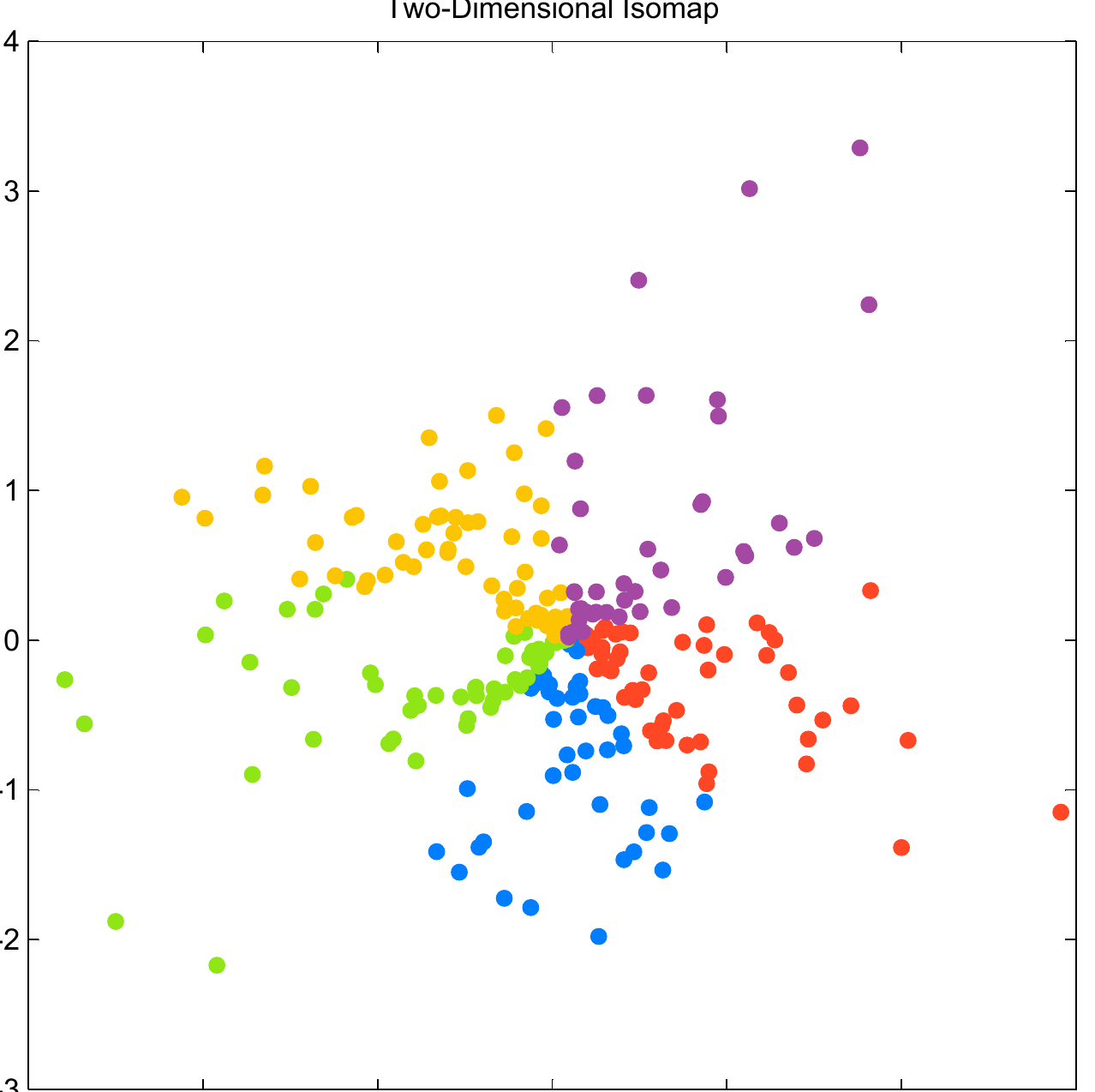} &
		\includegraphics[trim=4.6cm 8cm 4cm 7cm,clip=true,scale=0.21]{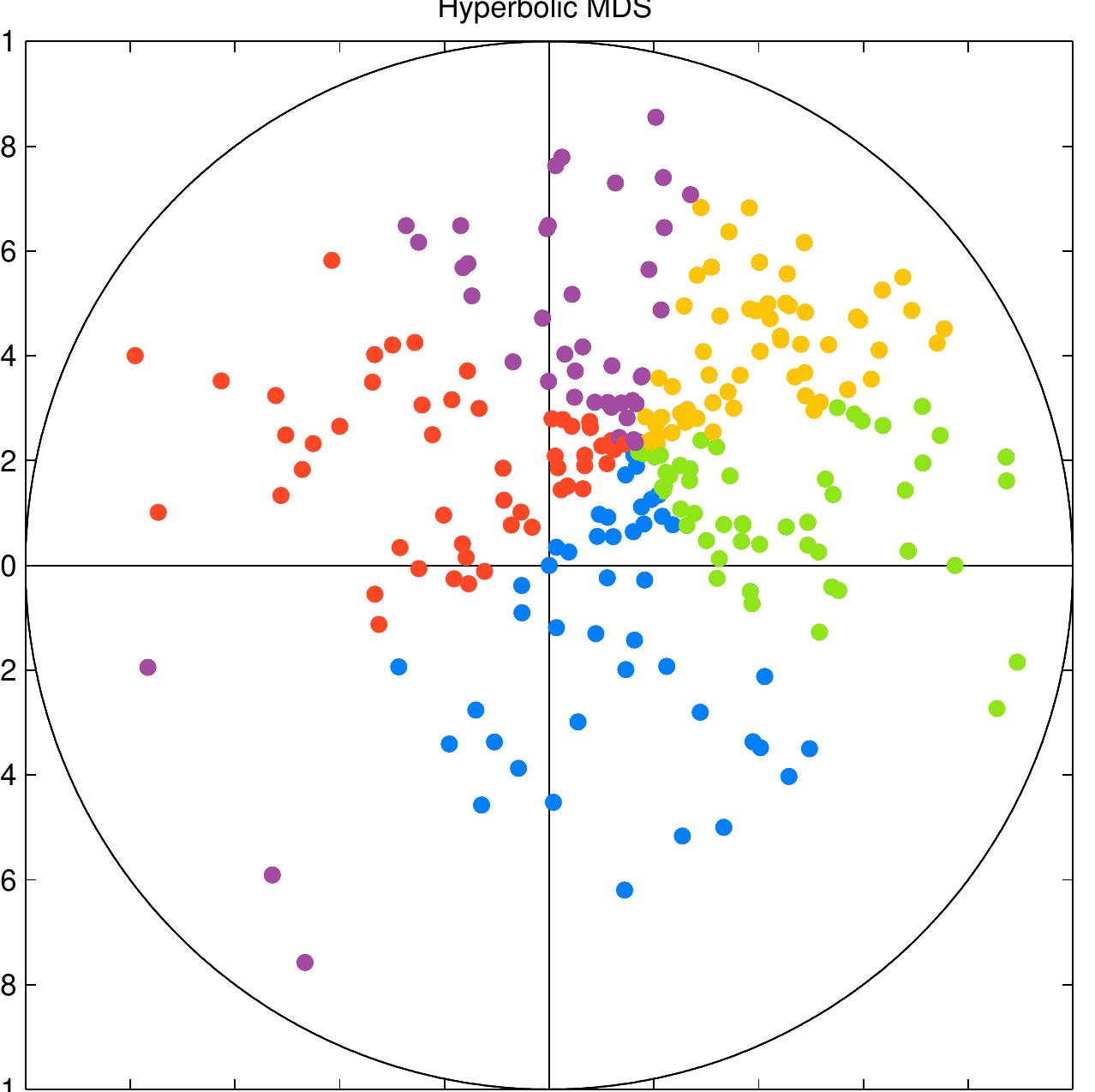} &
		\includegraphics[trim=4.6cm 8cm 4cm 7cm,clip=true,scale=0.21]{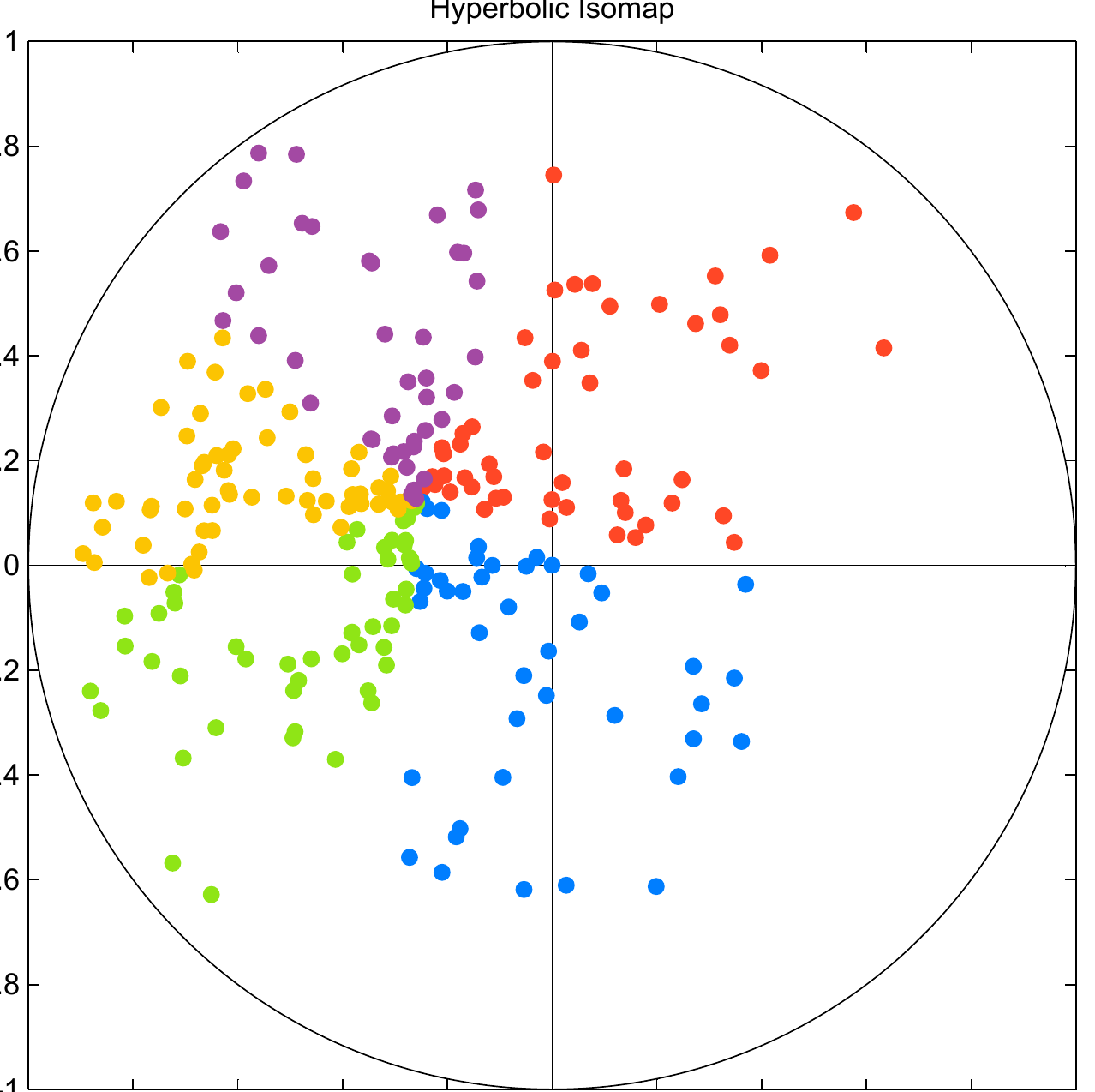} \\
		\rotatebox{90}{~~~~\parbox{2mm}{3SHEETS\_2D}~}&
		 \includegraphics[trim =4.6cm 8cm 4cm 7cm,clip=true,scale=0.21]{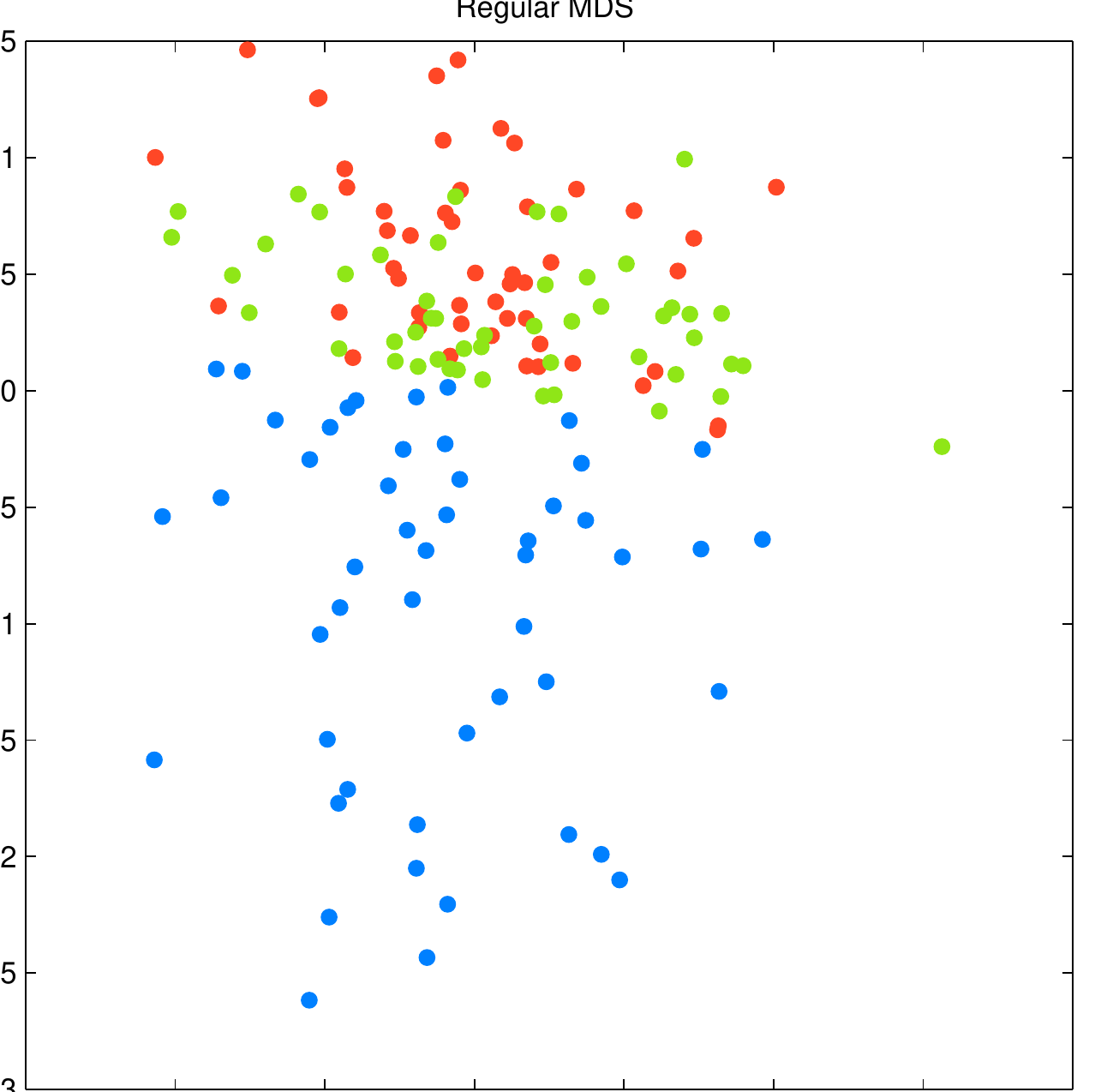} &
		\includegraphics[trim =4.6cm 8cm 4cm 7cm,clip=true,scale=0.21]{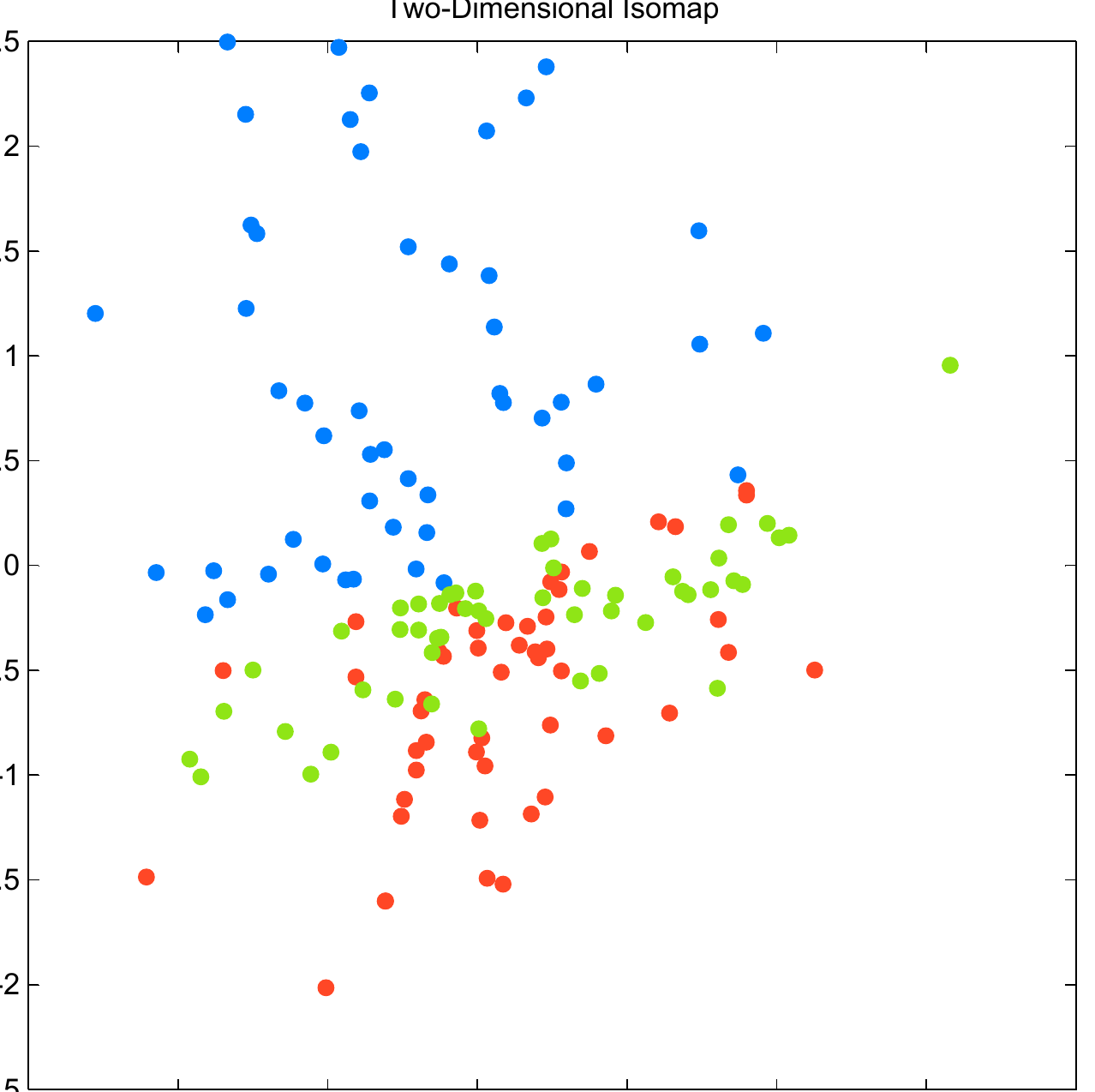} &
		\includegraphics[trim=4.6cm 8cm 4cm 7cm,clip=true,scale=0.21]{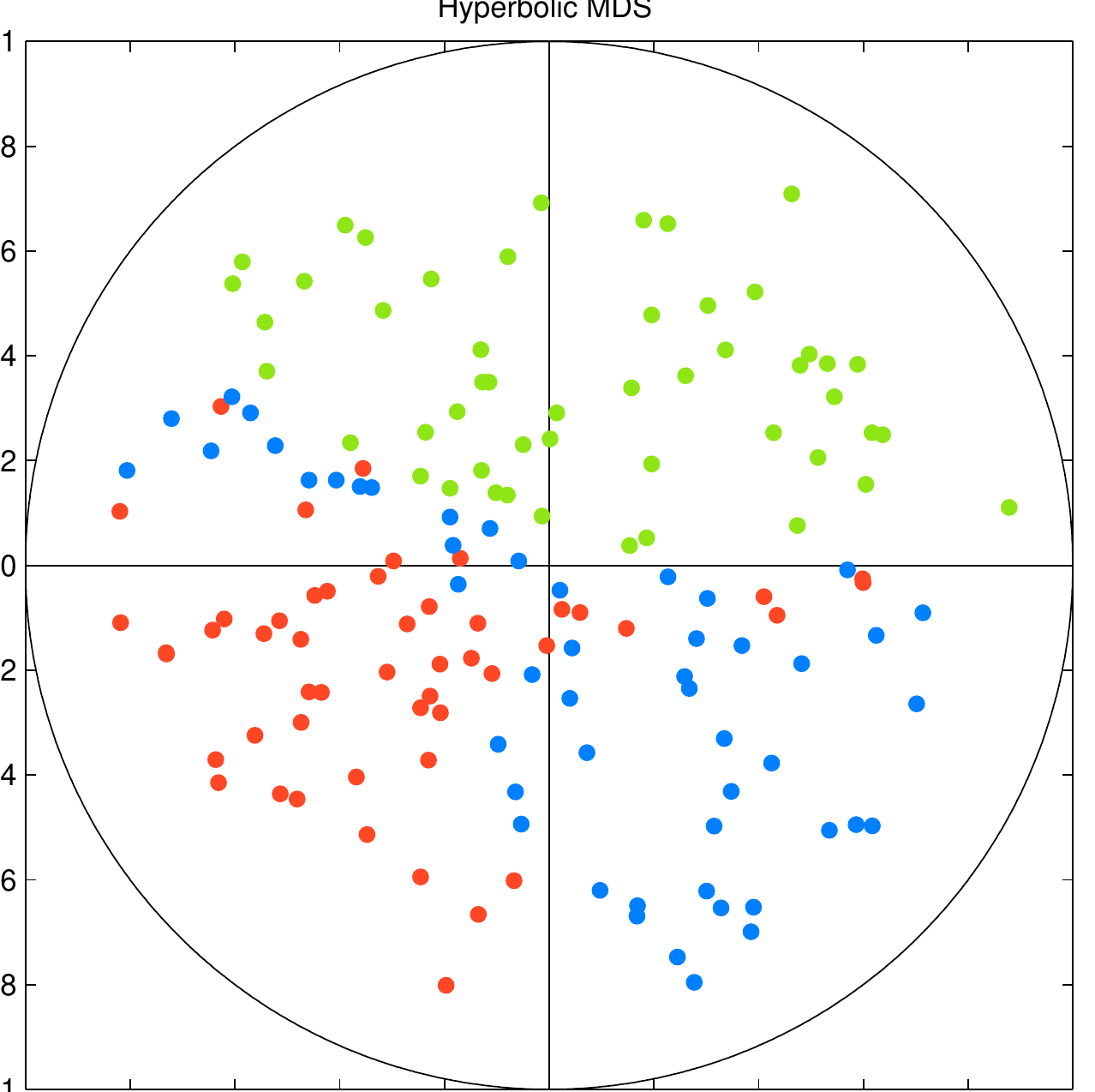} &
		\includegraphics[trim=4.6cm 8cm 4cm 7cm,clip=true,scale=0.21]{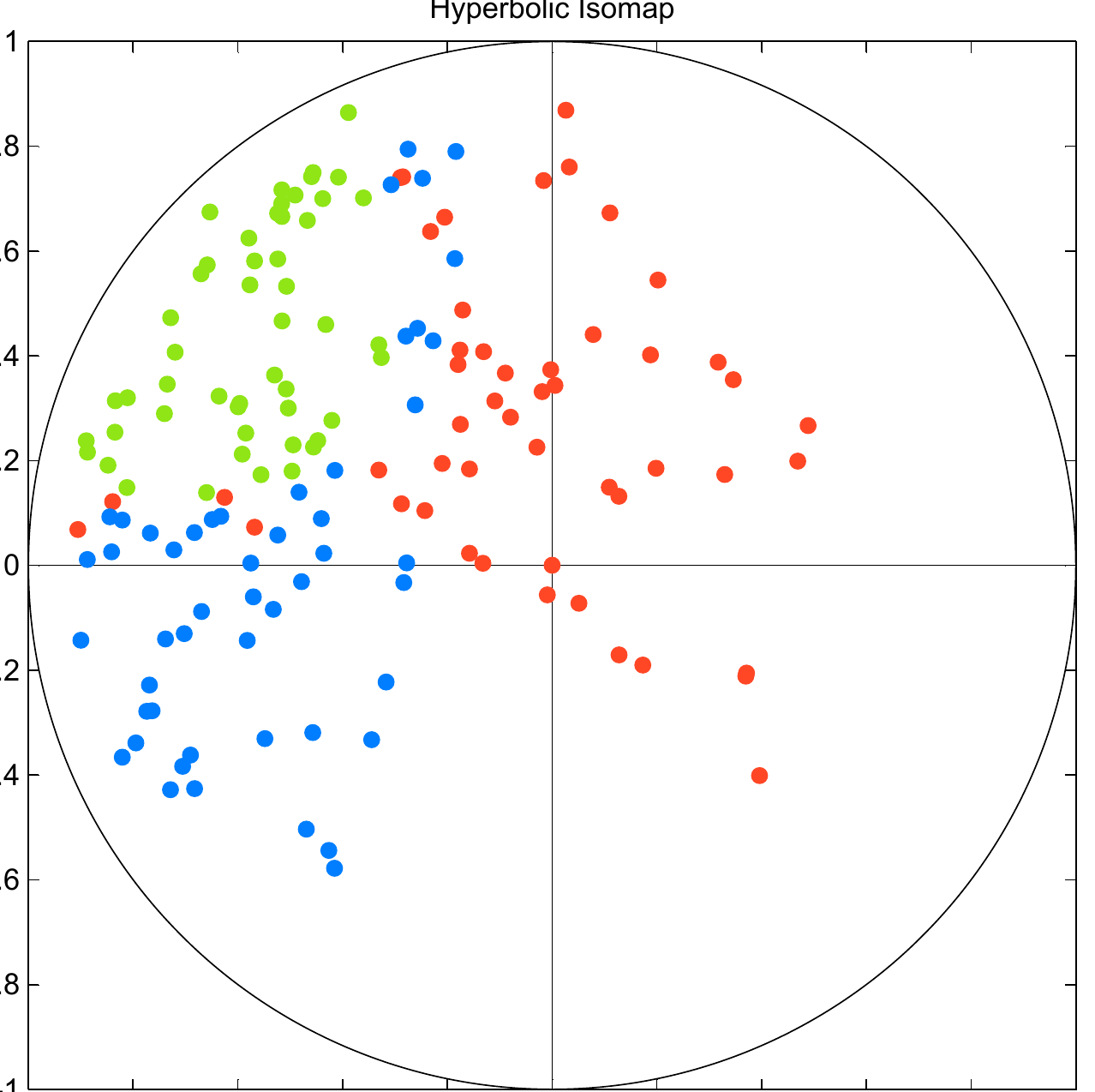} \\
		\rotatebox{90}{~~~~\parbox{2mm}{3SHEETS\_3D}~}&
		 \includegraphics[trim =4.6cm 8cm 4cm 7cm,clip=true,scale=0.21]{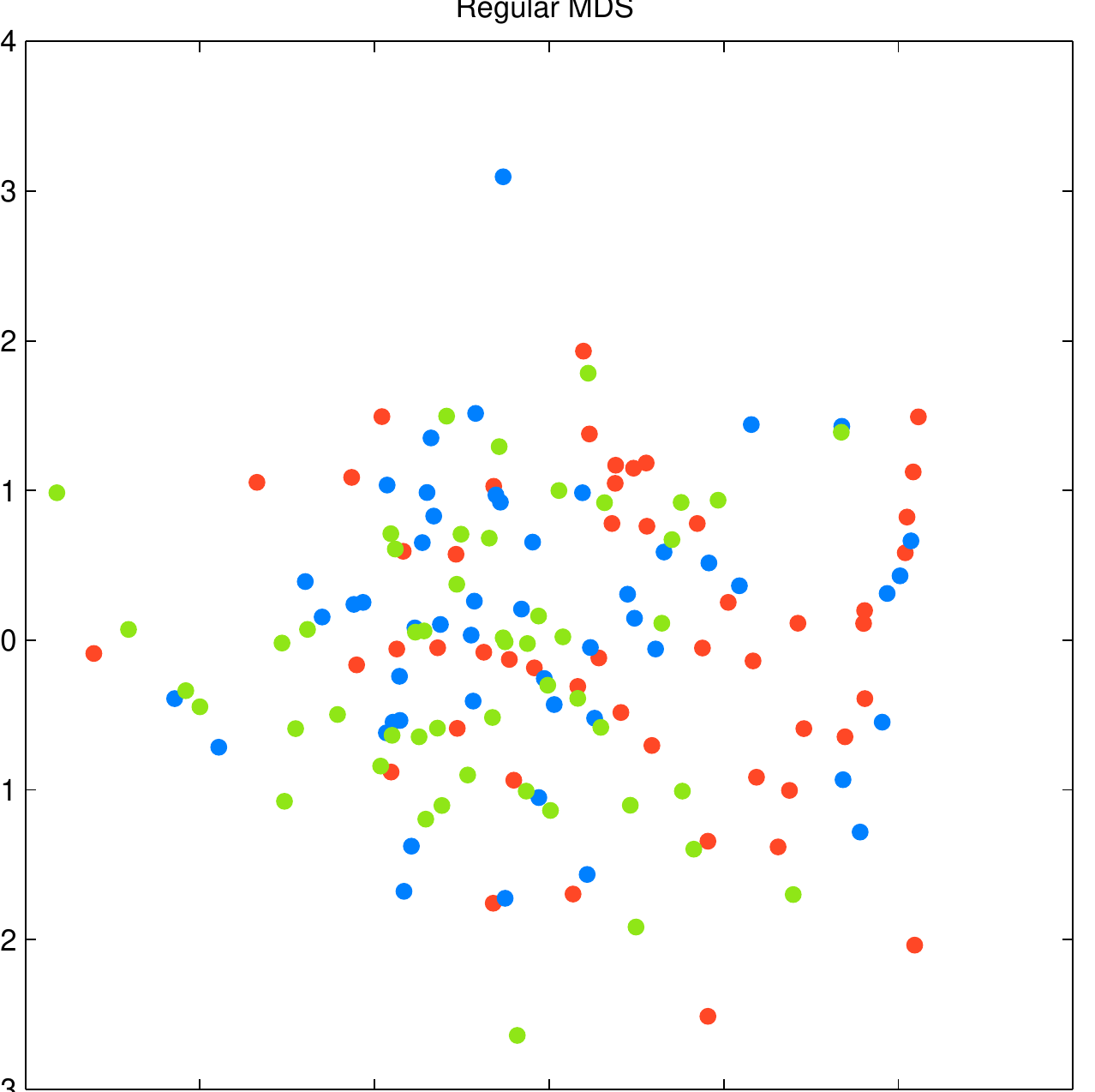} &
		\includegraphics[trim =4.6cm 8cm 4cm 7cm,clip=true,scale=0.21]{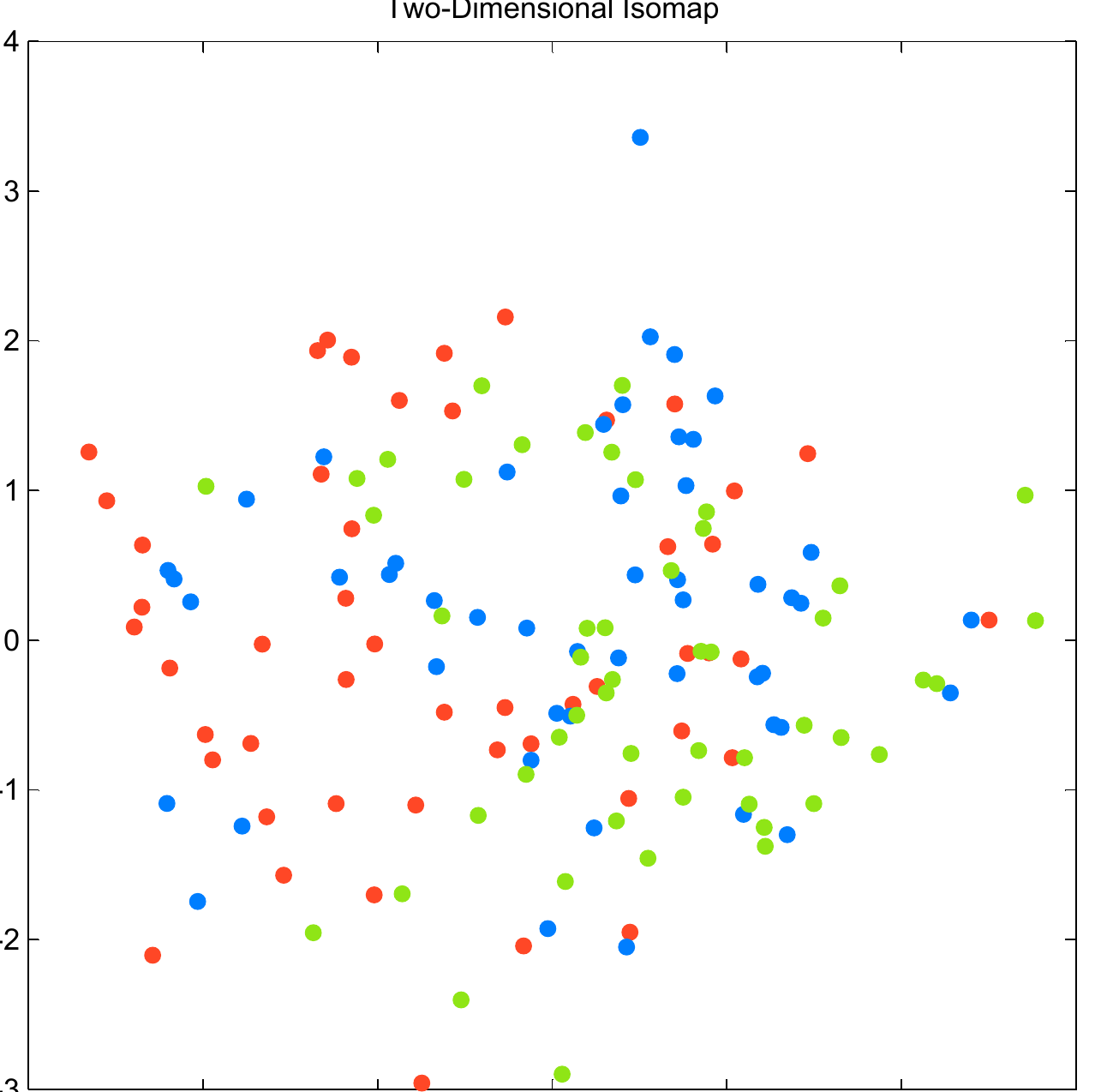} &
		\includegraphics[trim=4.6cm 8cm 4cm 7cm,clip=true,scale=0.21]{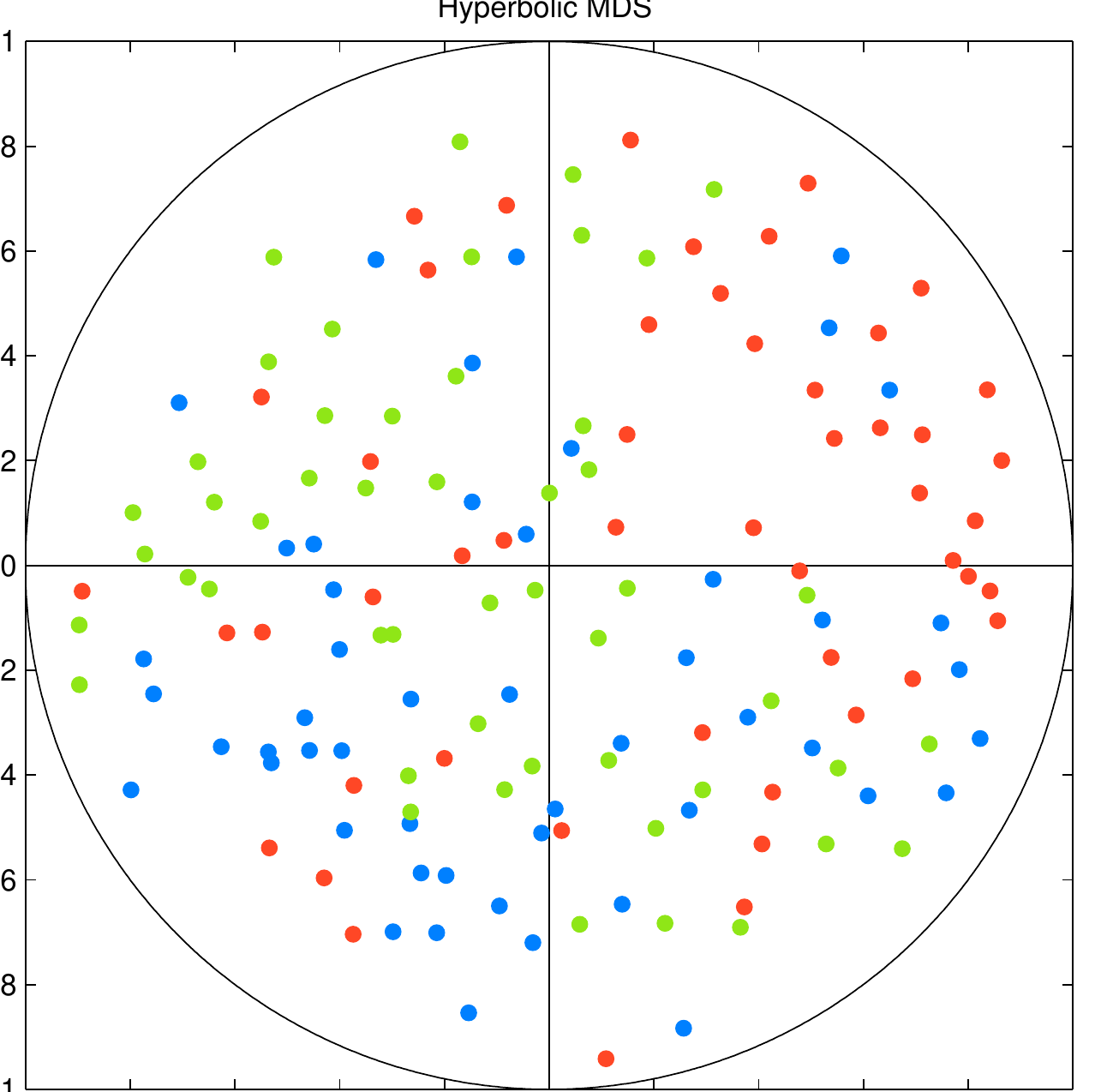} &
		\includegraphics[trim=4.6cm 8cm 4cm 7cm,clip=true,scale=0.21]{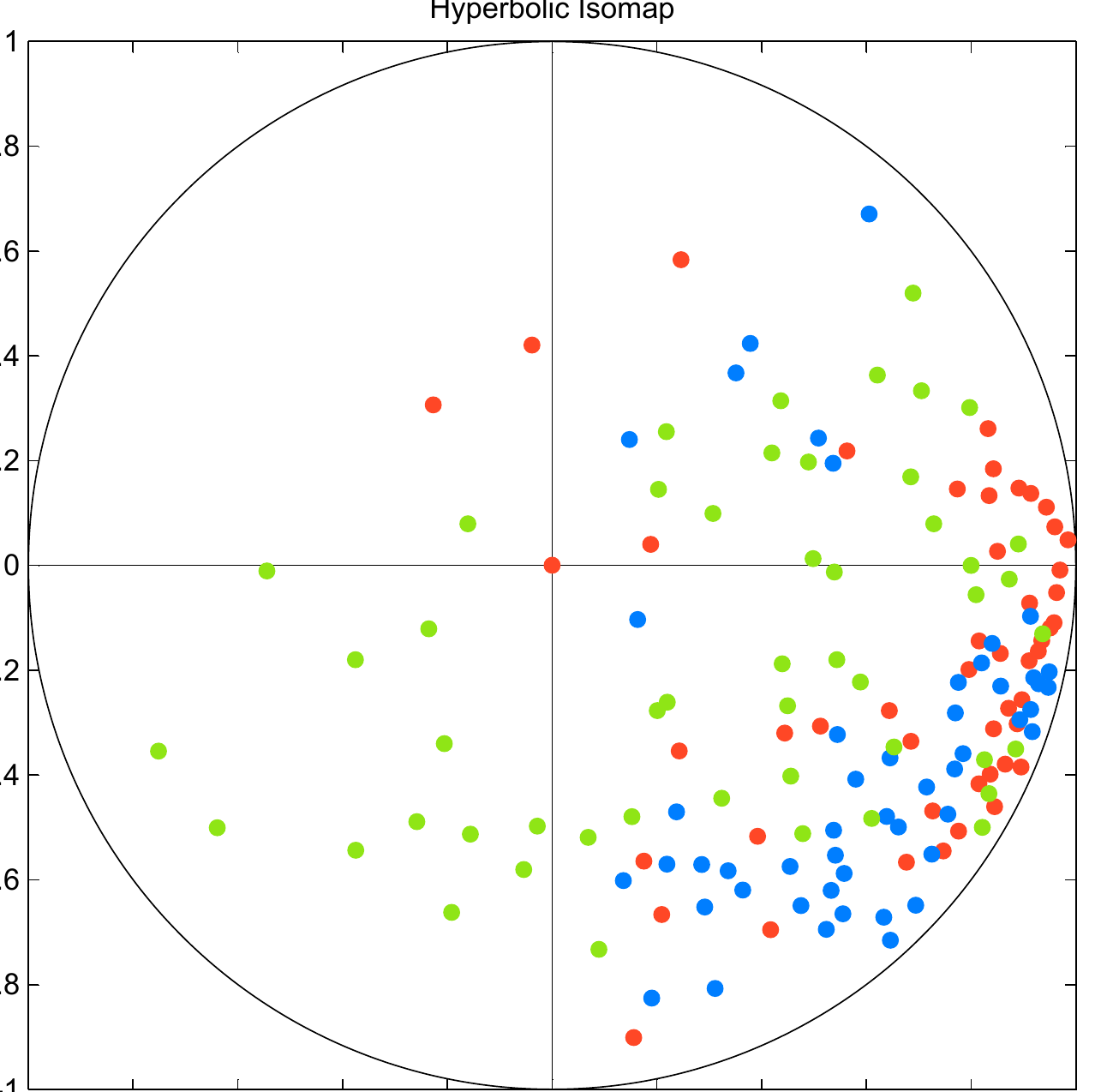} \\
		\rotatebox{90}{~~~~\parbox{2mm}{5SHEETS\_2D}~}&
		 \includegraphics[trim =4.6cm 8cm 4cm 7cm,clip=true,scale=0.21]{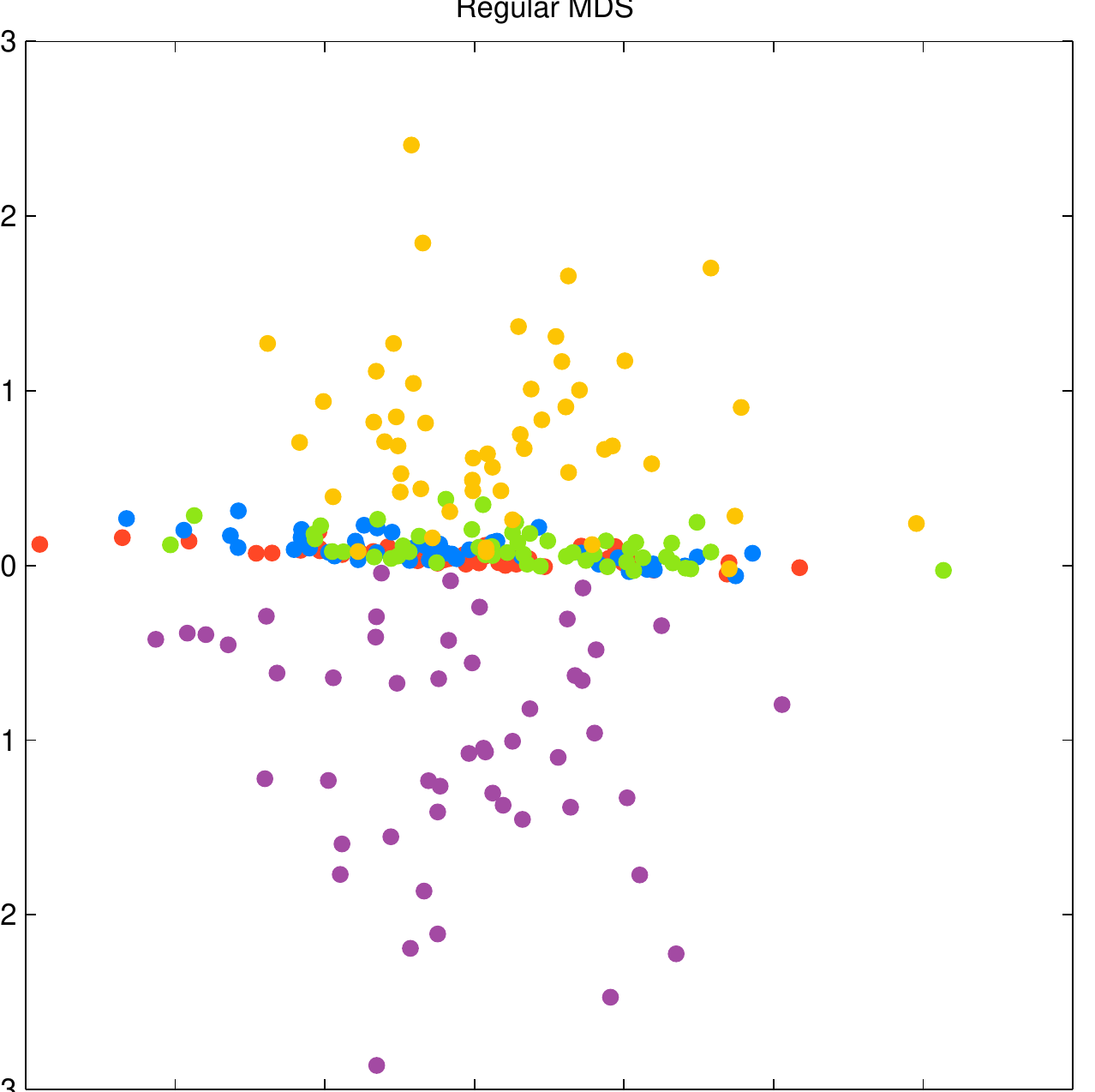} &
		\includegraphics[trim =4.6cm 8cm 4cm 7cm,clip=true,scale=0.21]{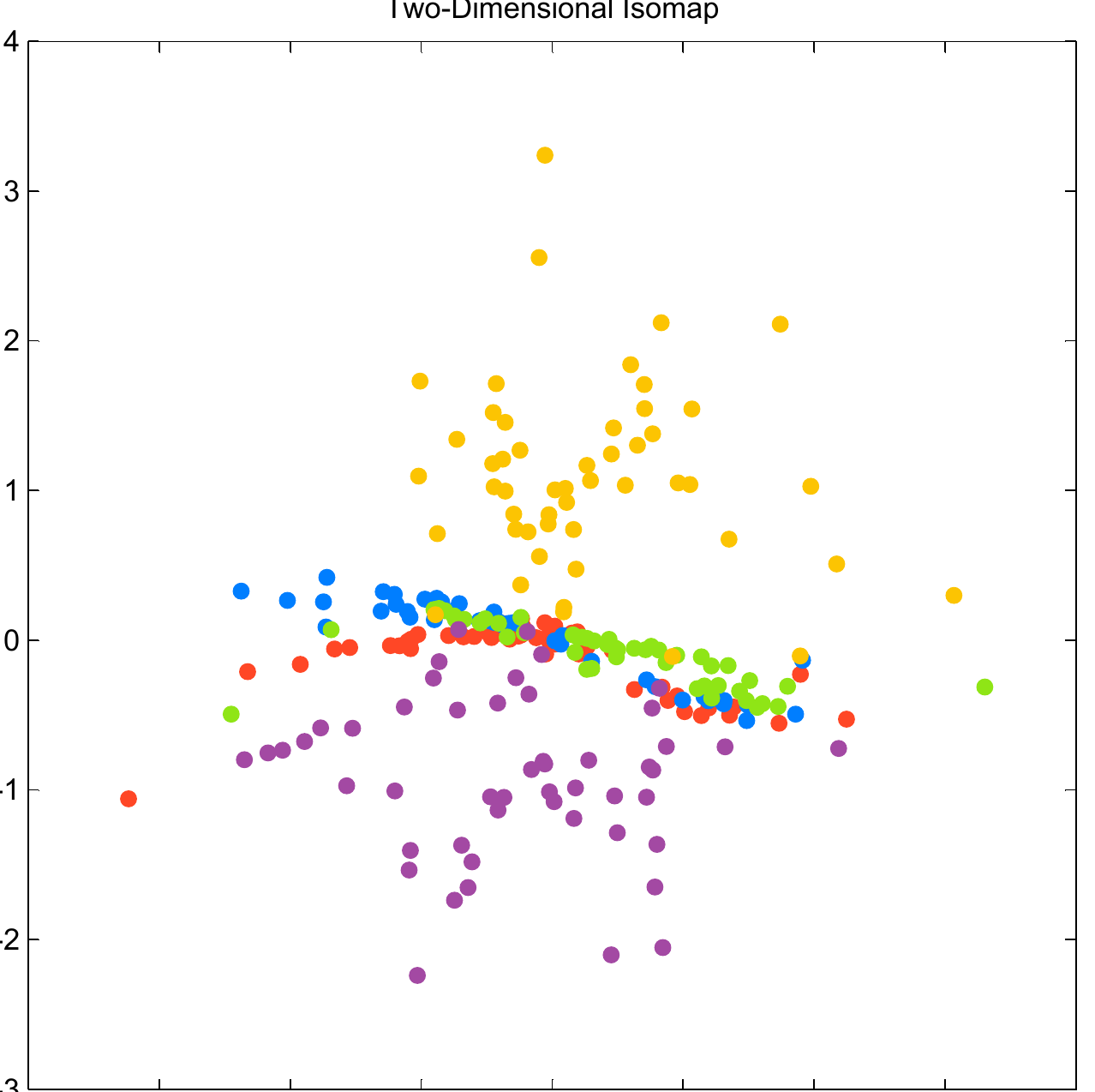} &
		\includegraphics[trim=4.6cm 8cm 4cm 7cm,clip=true,scale=0.21]{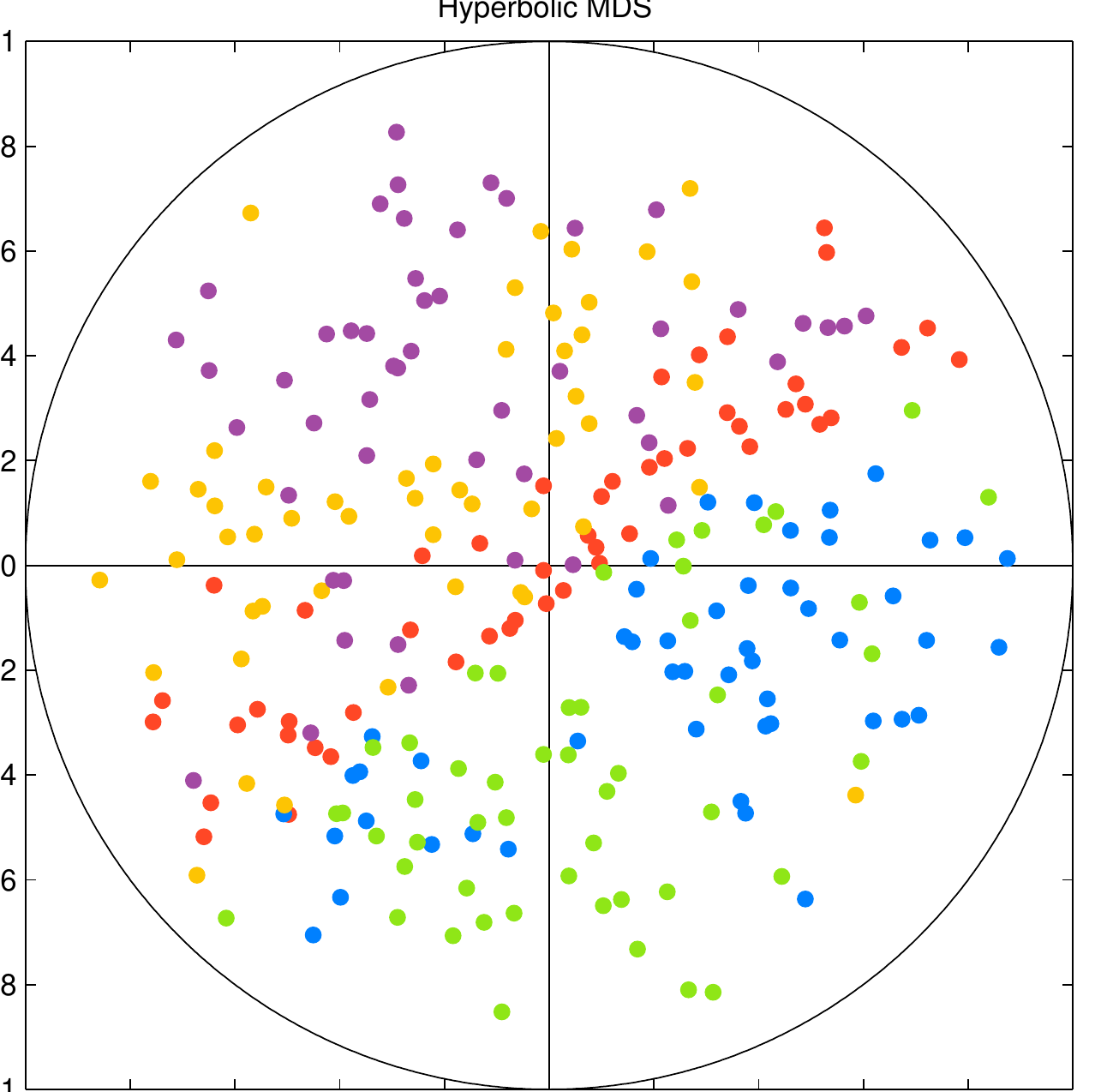} &
		\includegraphics[trim=4.6cm 8cm 4cm 7cm,clip=true,scale=0.21]{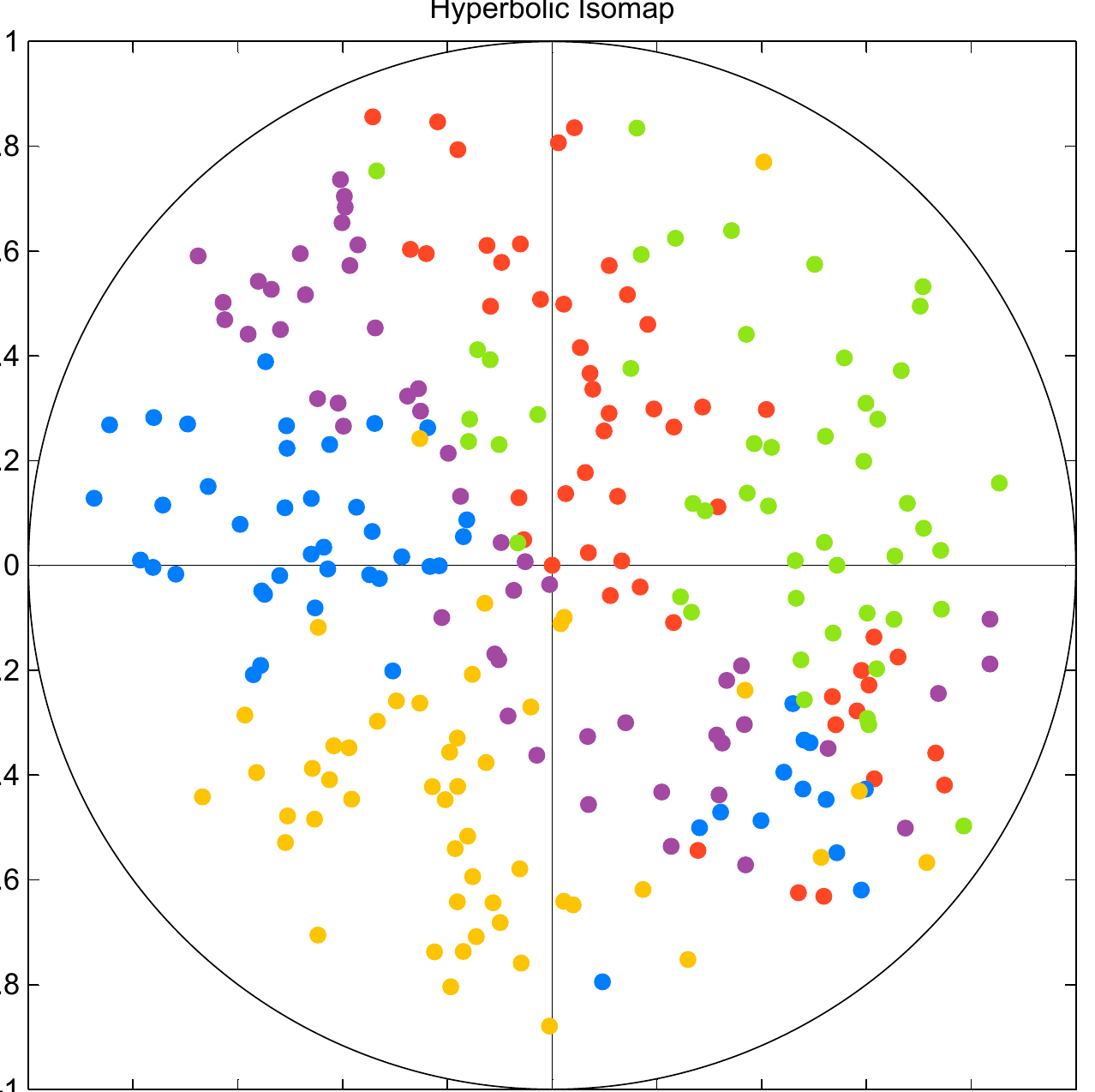} \\
		\rotatebox{90}{~~~~\parbox{2mm}{5SHEETS\_3D}~}&
		 \includegraphics[trim =4.6cm 8cm 4cm 7cm,clip=true,scale=0.21]{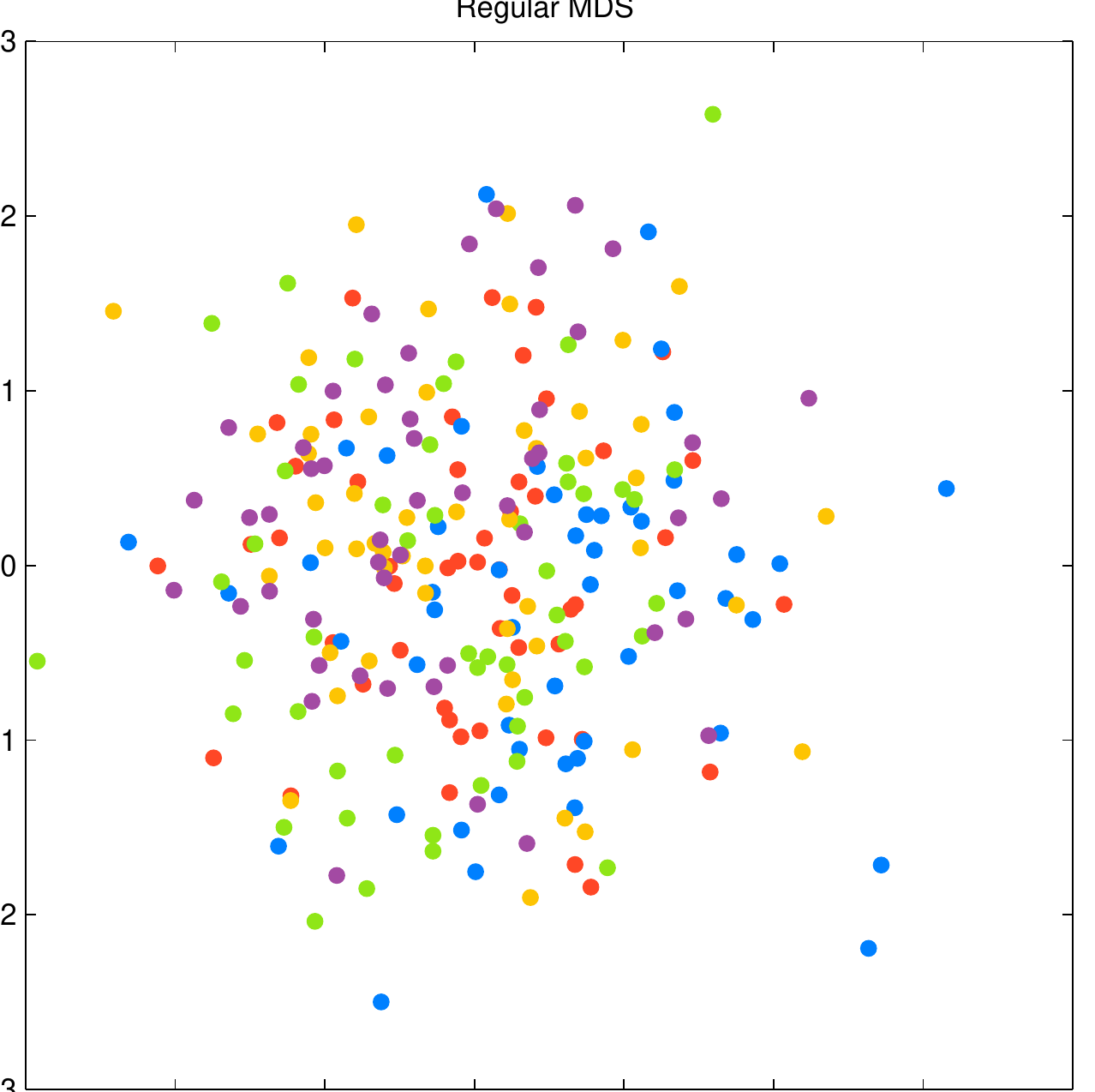} &
		\includegraphics[trim =4.6cm 8cm 4cm 7cm,clip=true,scale=0.21]{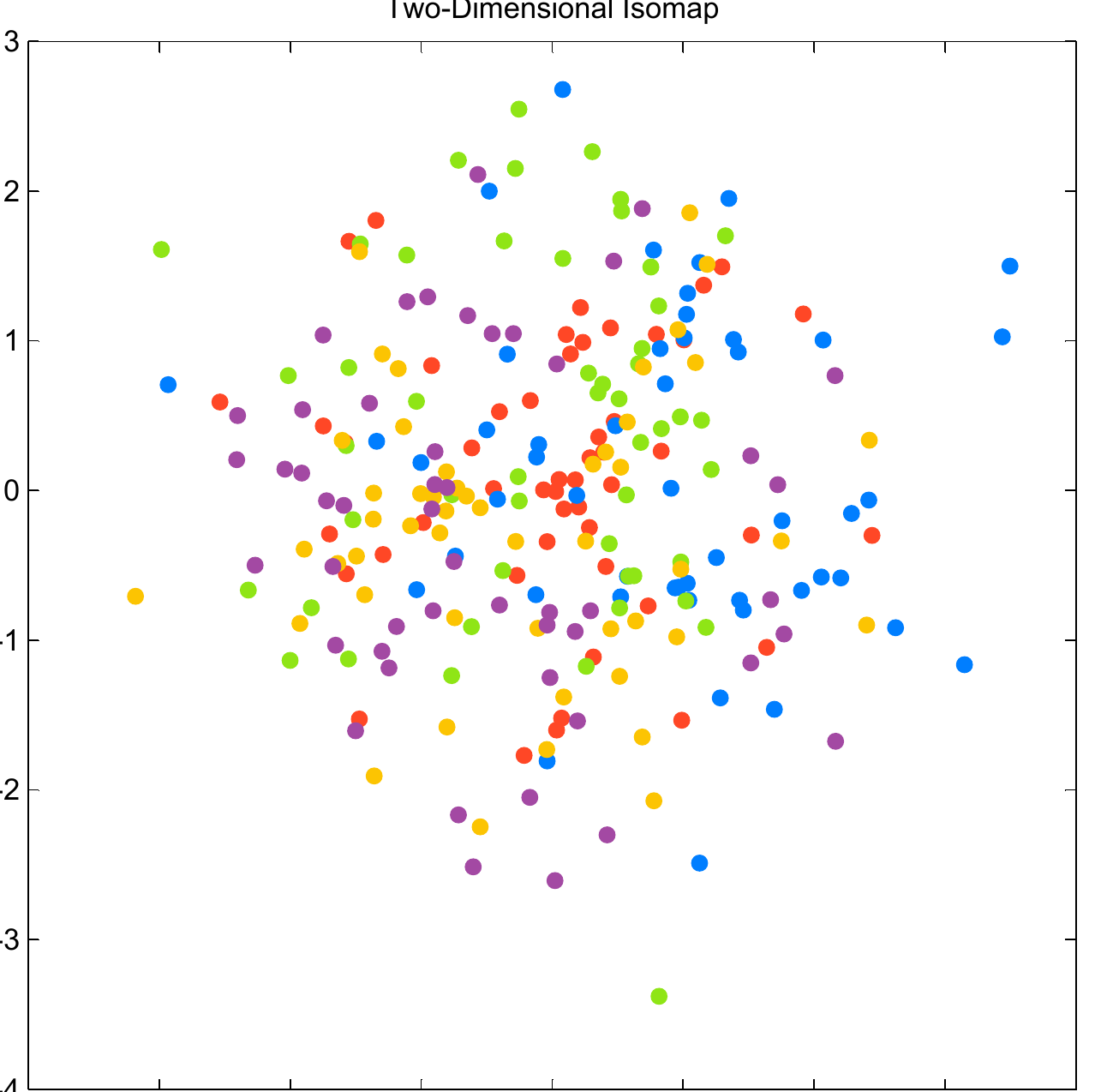} &
		\includegraphics[trim=4.6cm 8cm 4cm 7cm,clip=true,scale=0.21]{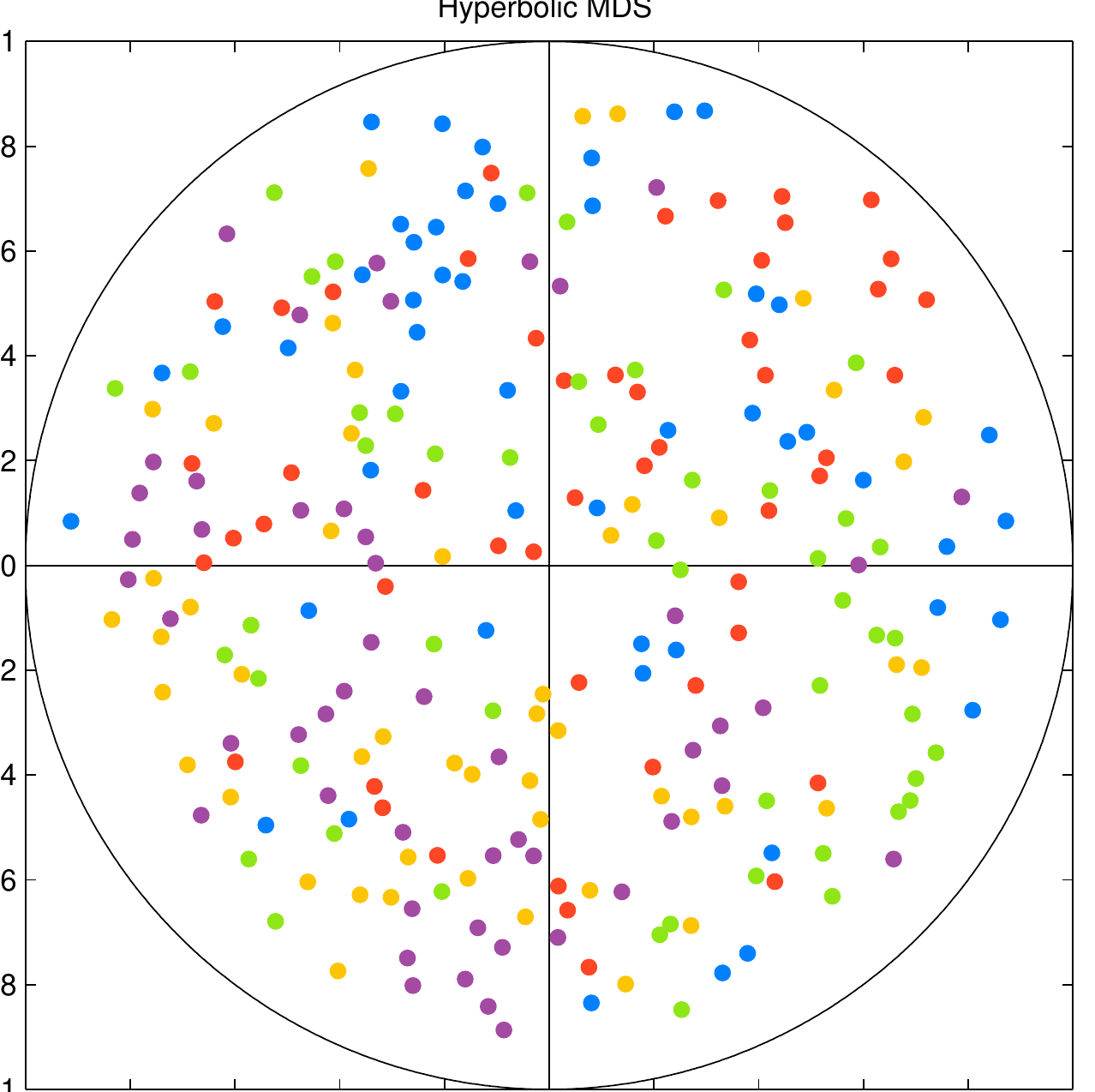} &
		\includegraphics[trim=4.6cm 8cm 4cm 7cm,clip=true,scale=0.21]{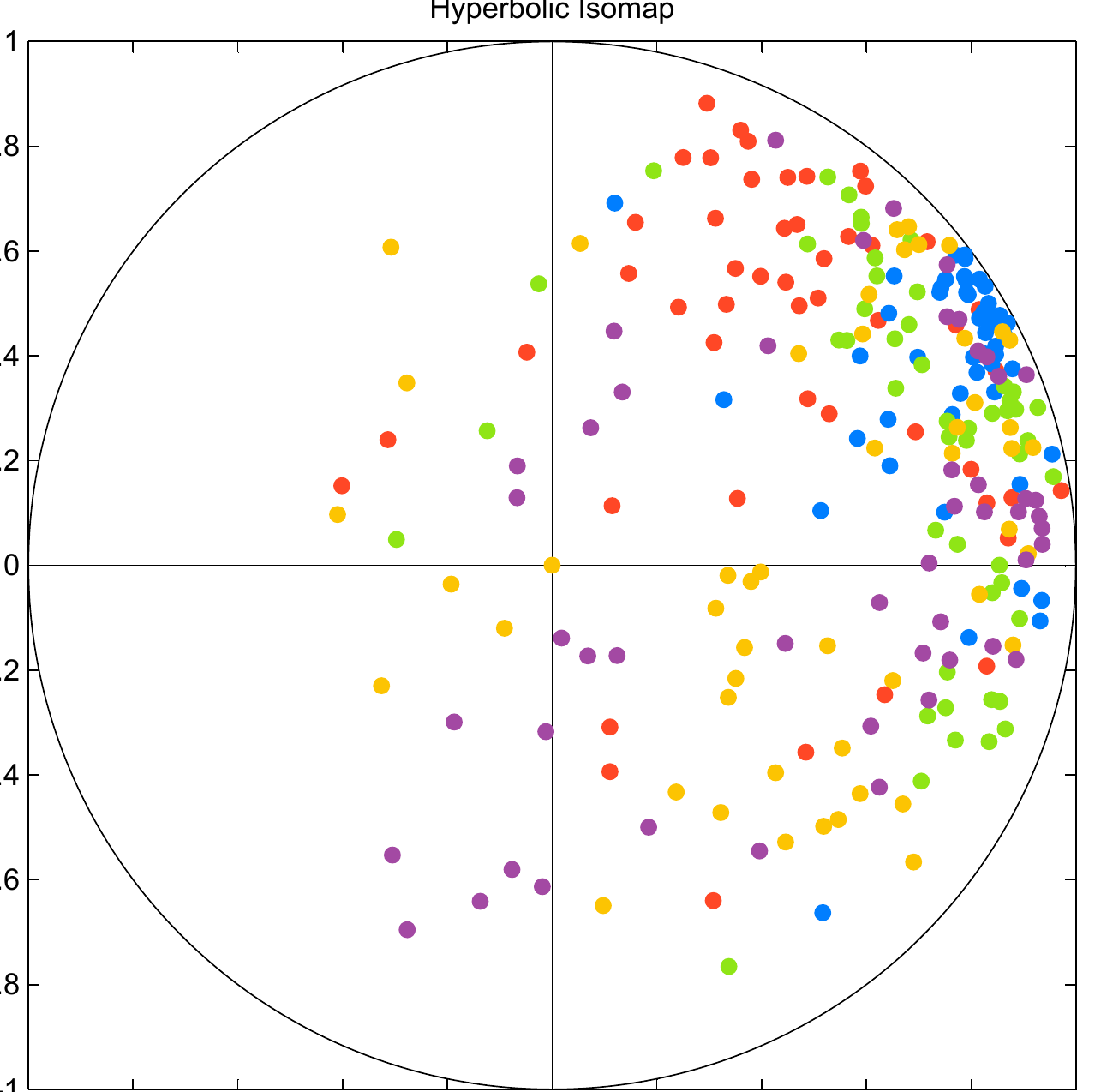} \\
		\rotatebox{90}{~~~~\parbox{2mm}{COPD}~}&
		 \includegraphics[trim =4.6cm 8cm 4cm 7cm,clip=true,scale=0.21]{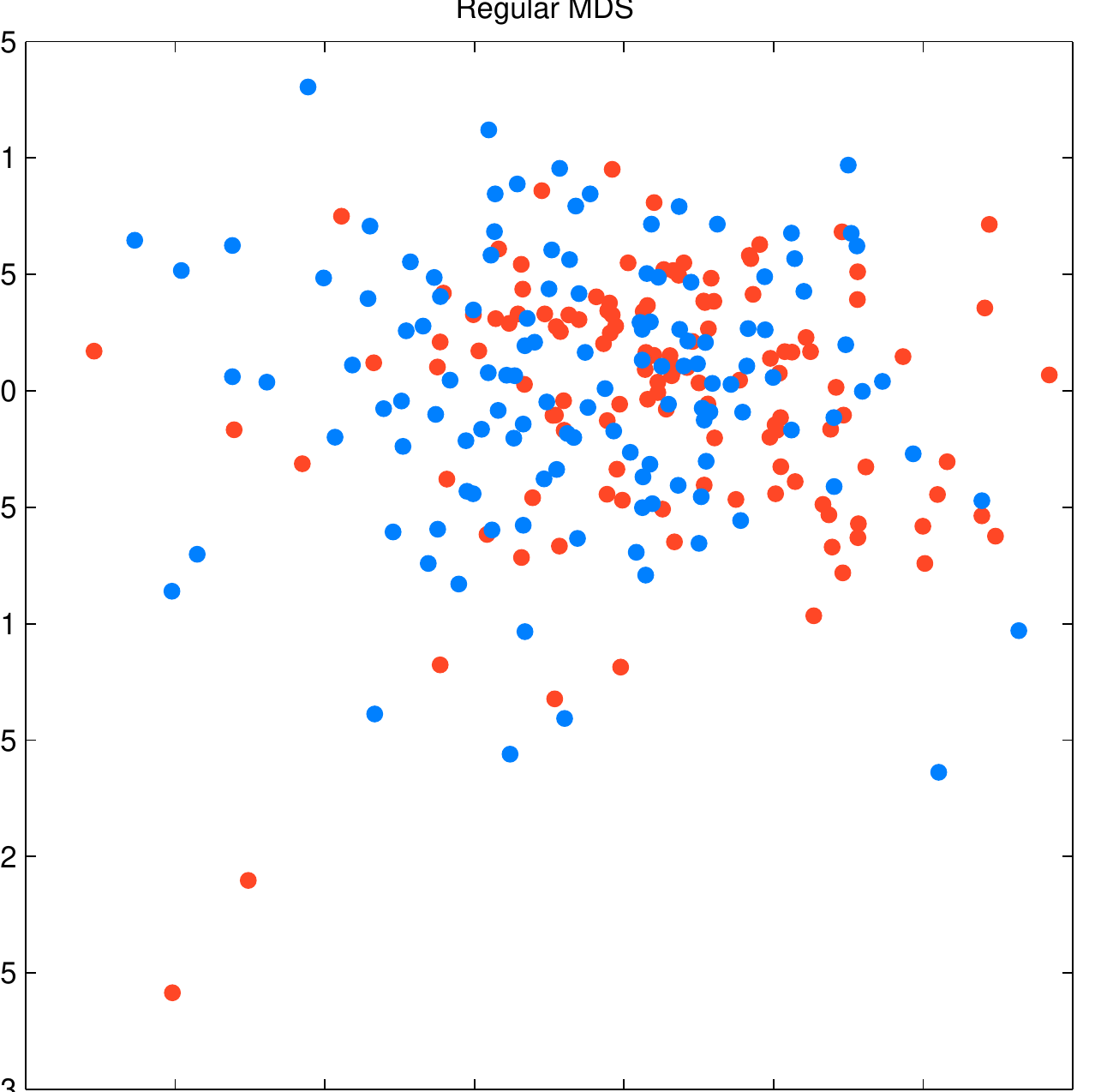} &
		\includegraphics[trim =4.6cm 8cm 4cm 7cm,clip=true,scale=0.21]{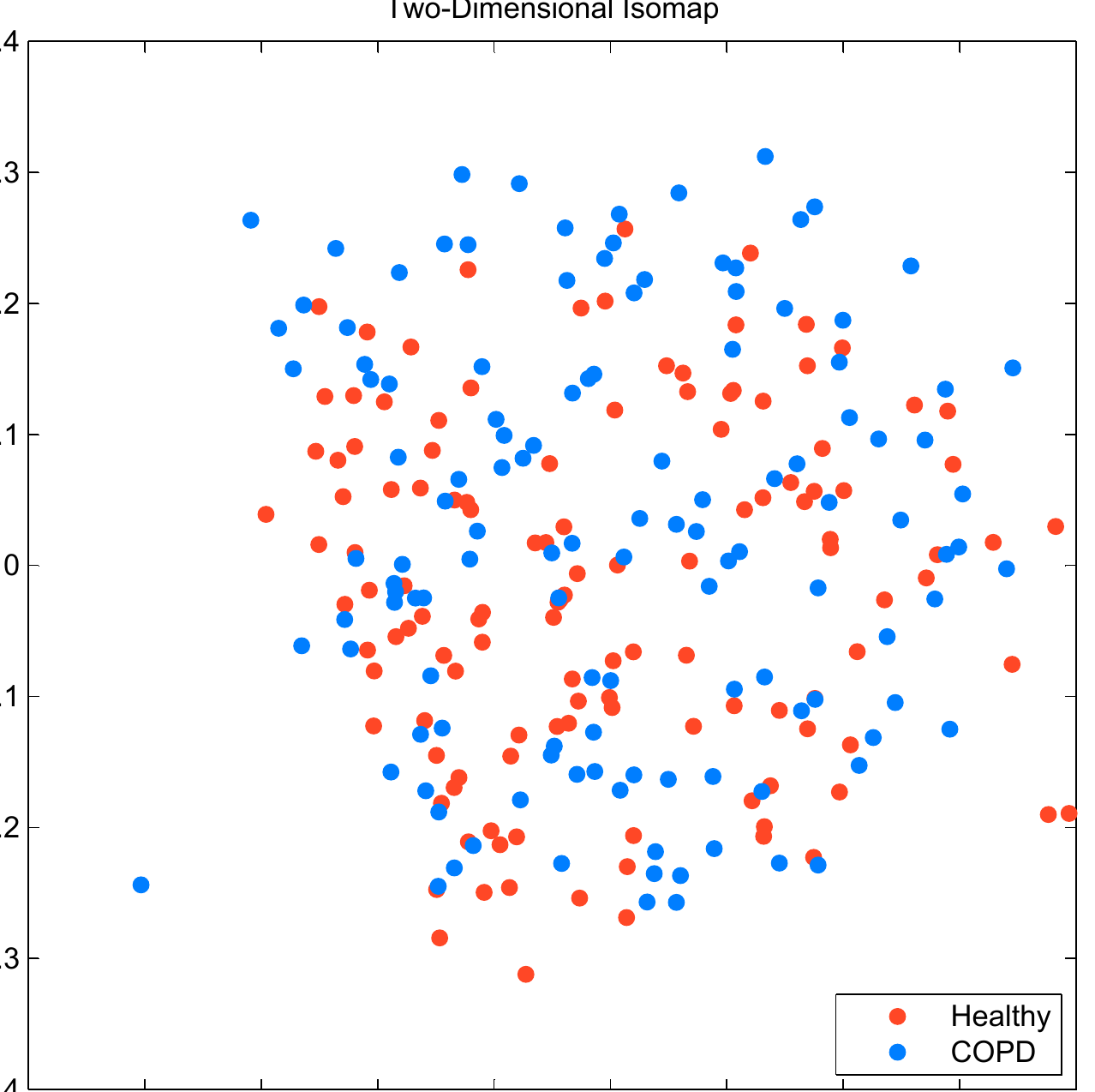} &
		\includegraphics[trim=4.6cm 8cm 4cm 7cm,clip=true,scale=0.21]{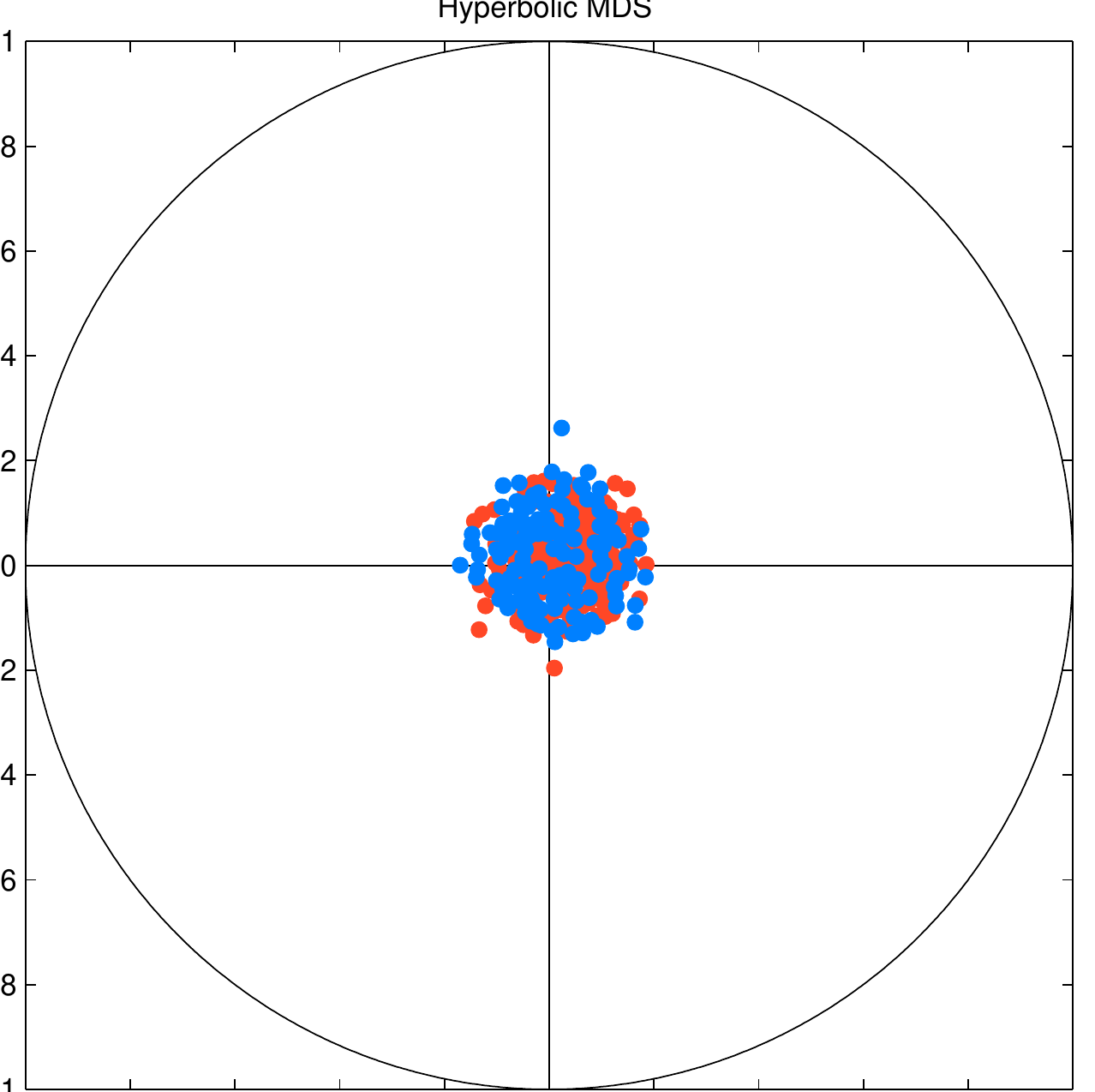} &
		\includegraphics[trim=4.6cm 8cm 4cm 7cm,clip=true,scale=0.21]{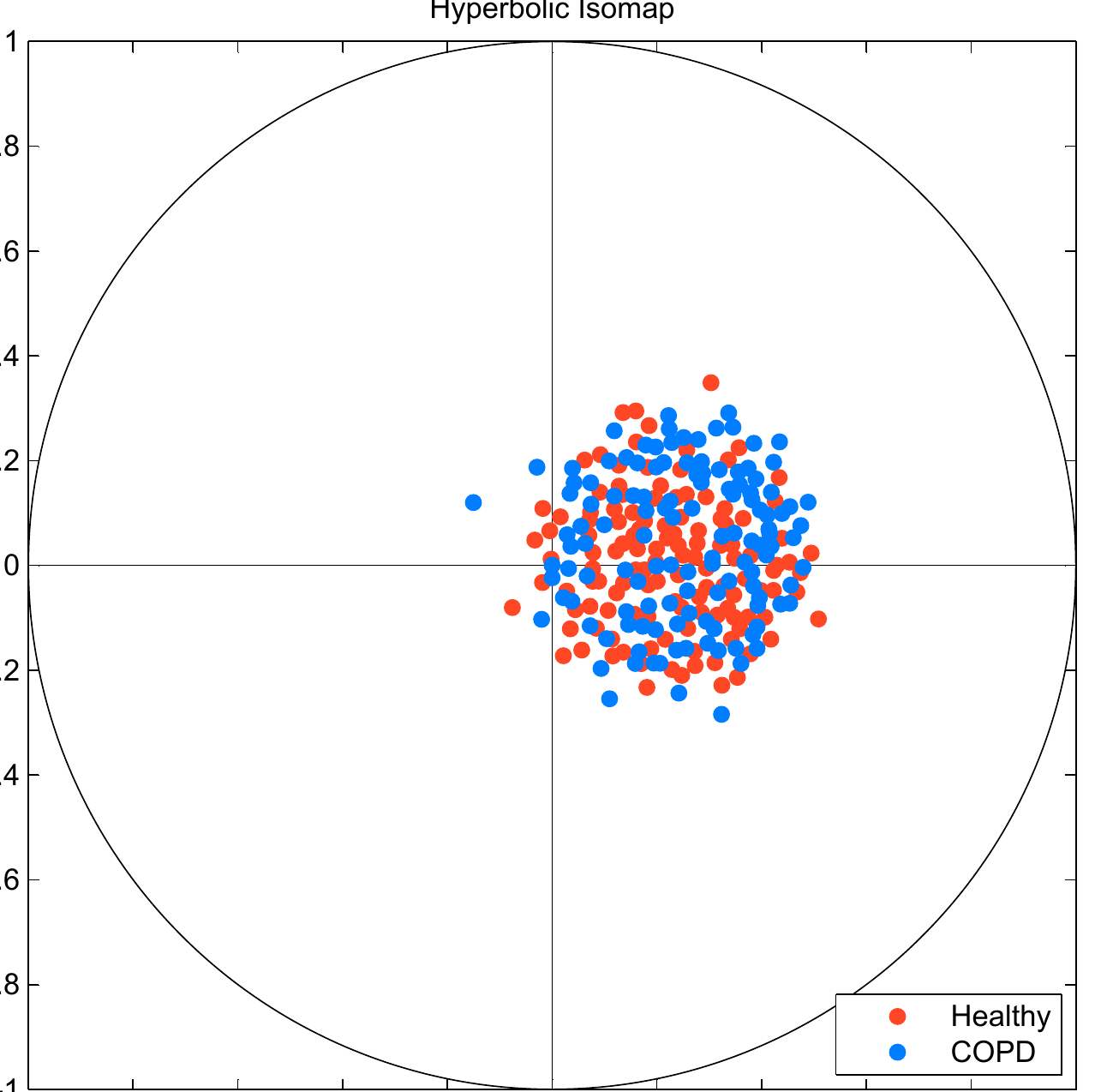}
	\end{tabular}

	\label{fig:all_embeddings}
\end{table}

\begin{table}
	[!ht] 
		\caption{The error histograms of the embedded datasets in Figure~\ref{fig:all_embeddings}.  Every pair of points is binned according to the error in the embedding, which is the difference between the original distance between the pair of points, and the distance between them in the embedding.}
	\tabcolsep=0.01cm
	\begin{tabular}
		{ccccc}
		& Classical MDS & Classical Isomap & HyperMDS & HyperIsomap \\
		\rotatebox{90}{~~~~\parbox{2mm}{CORNER}~}&
		\includegraphics[trim=1.2cm 1.5cm 1.5cm 2cm,clip=true,scale=0.18]{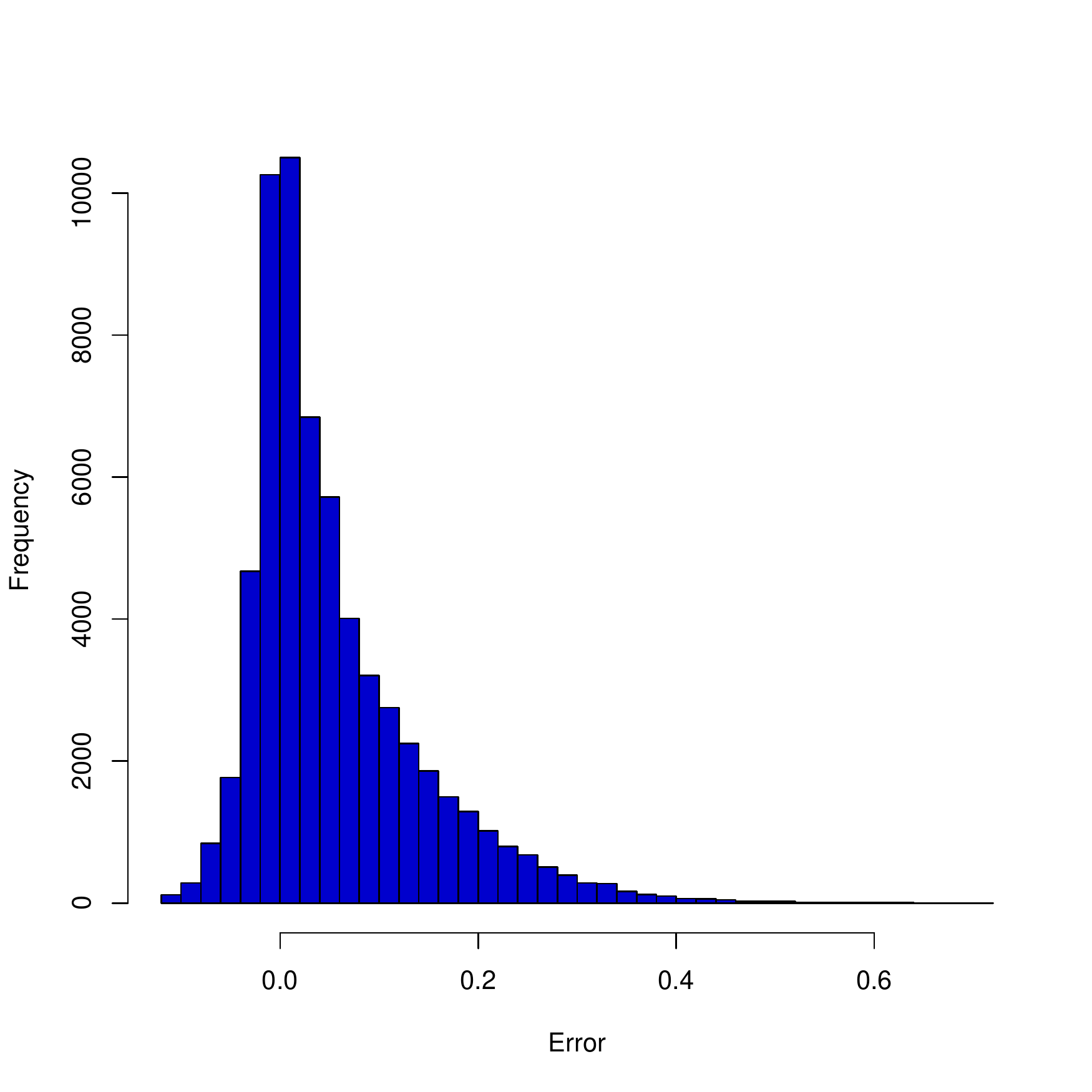} &
		\includegraphics[trim=1.2cm 1.5cm 1.5cm 2cm,clip=true,scale=0.18]{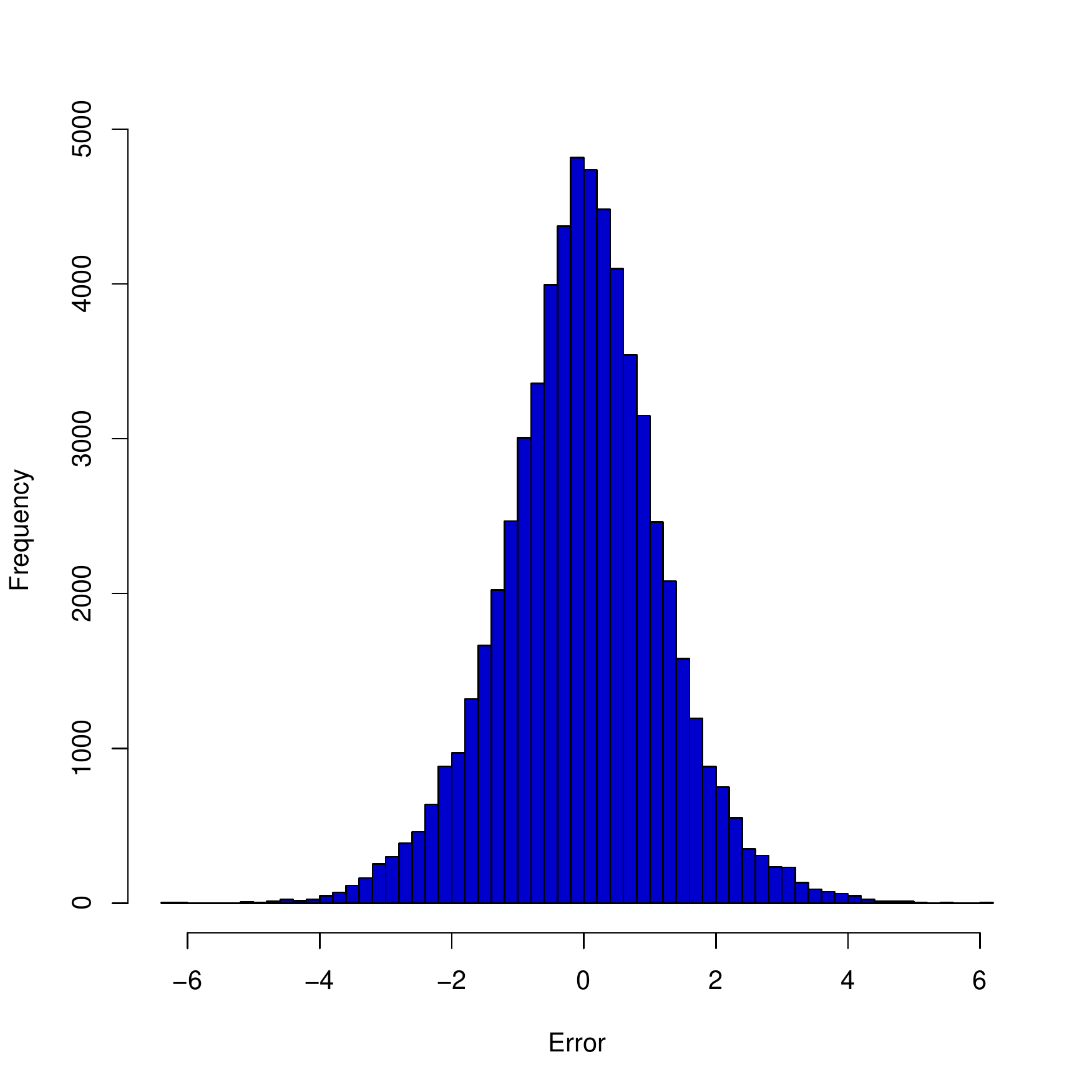} &
		\includegraphics[trim=1.2cm 1.5cm 1.5cm 2cm,clip=true,scale=0.18]{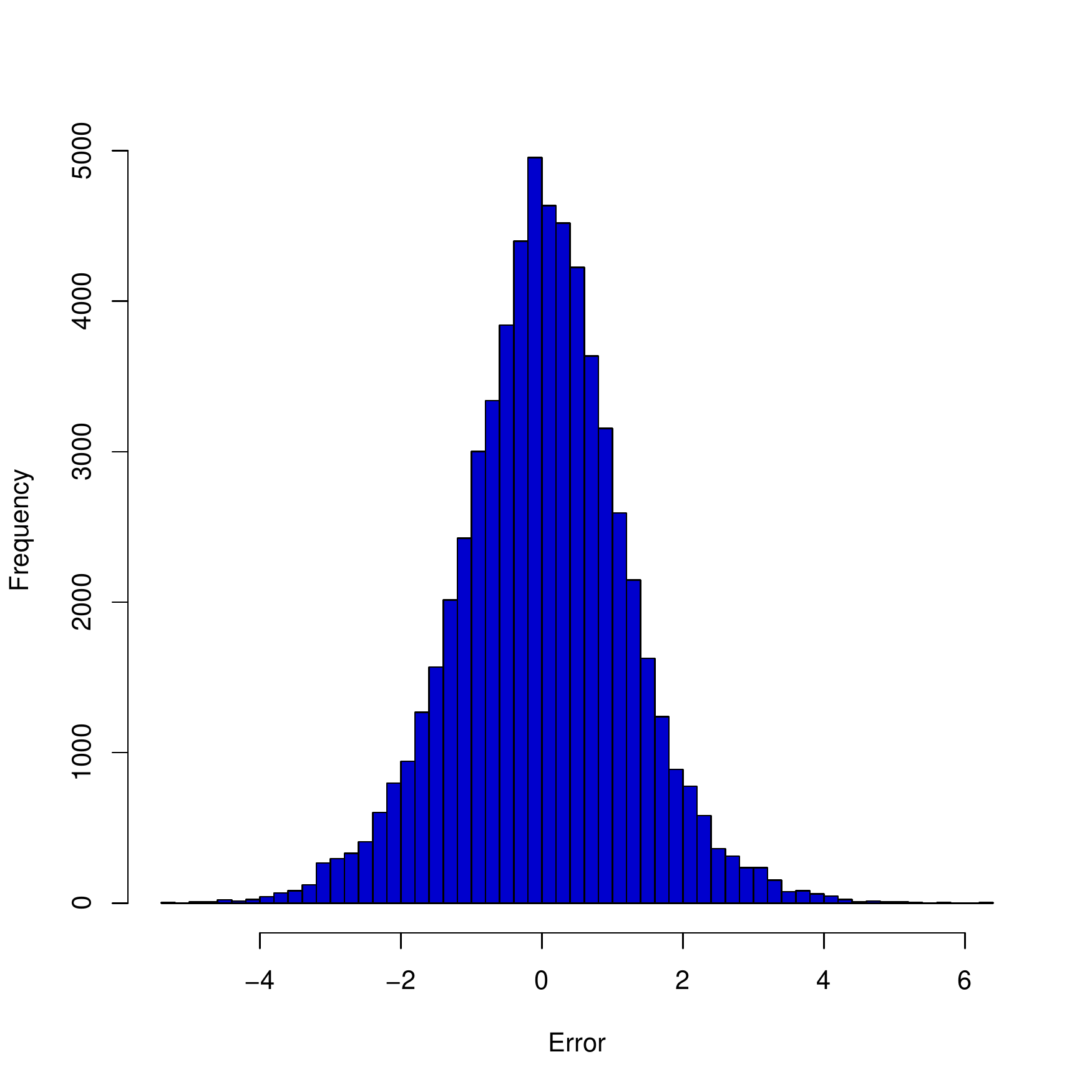} &
		\includegraphics[trim=1.2cm 1.5cm 1.5cm 2cm,clip=true,scale=0.18]{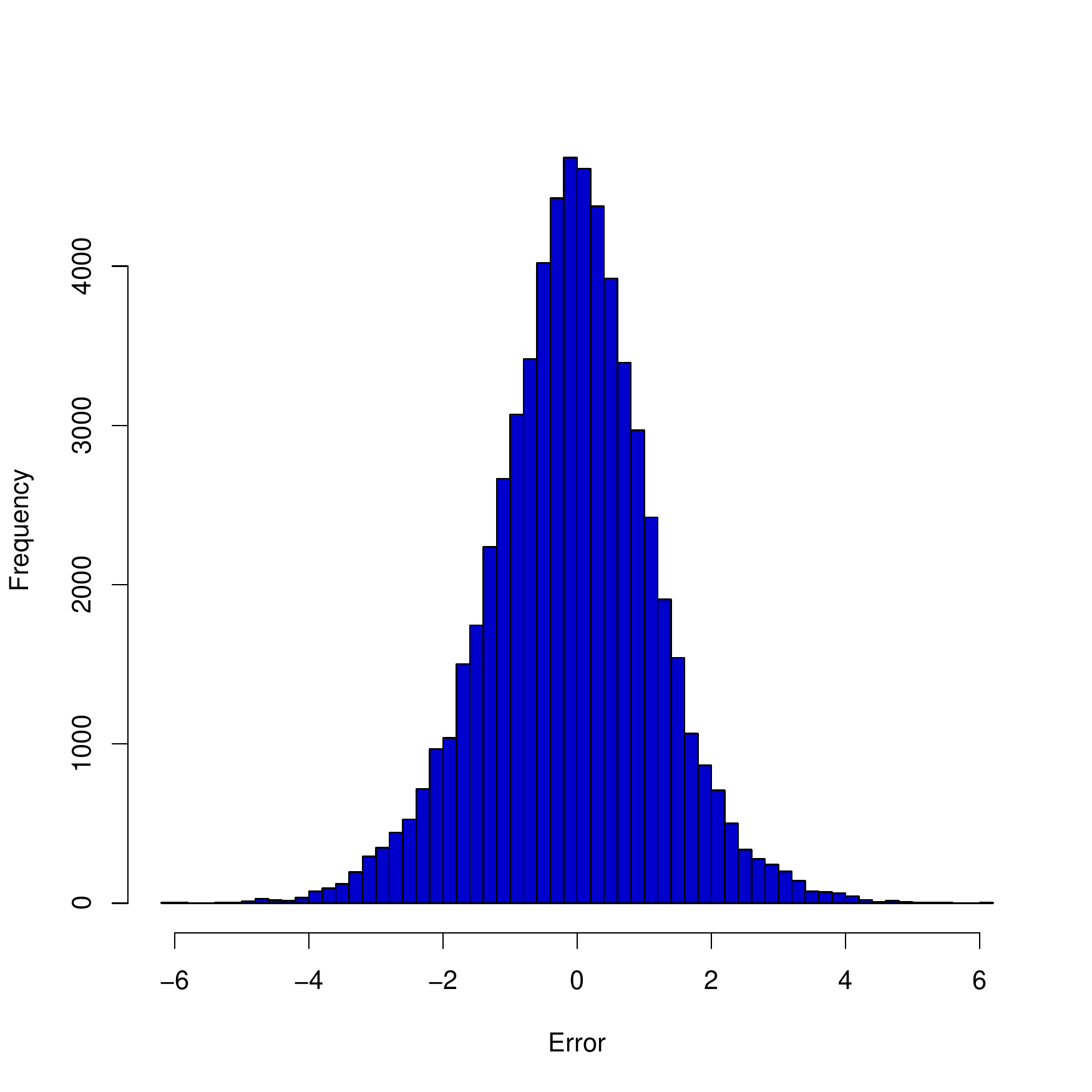} \\
		\rotatebox{90}{~~~~\parbox{2mm}{3SHEETS\_2D}~}&
		\includegraphics[trim=1.2cm 1.5cm 1.5cm 2cm,clip=true,scale=0.18]{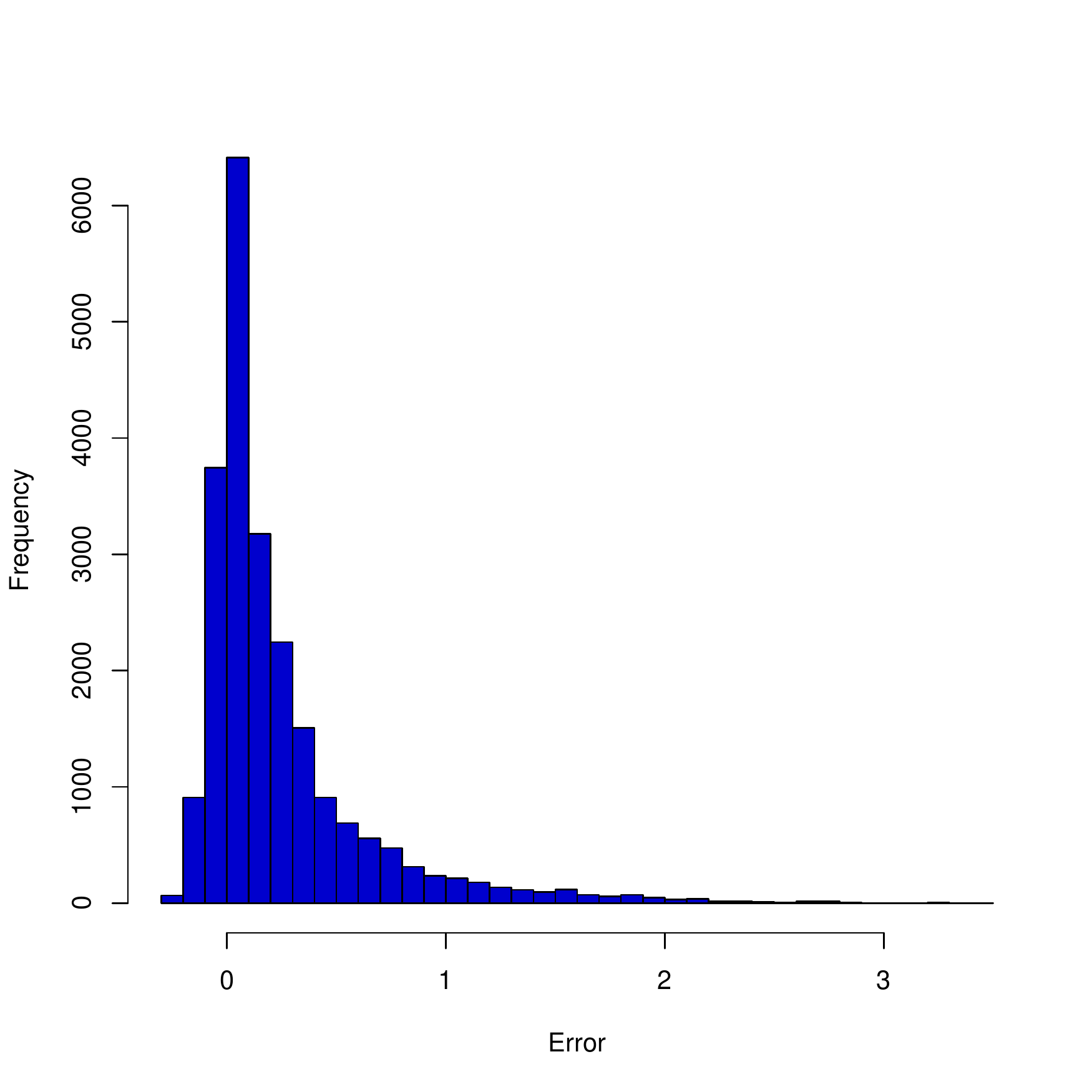} &
		\includegraphics[trim=1.2cm 1.5cm 1.5cm 2cm,clip=true,scale=0.18]{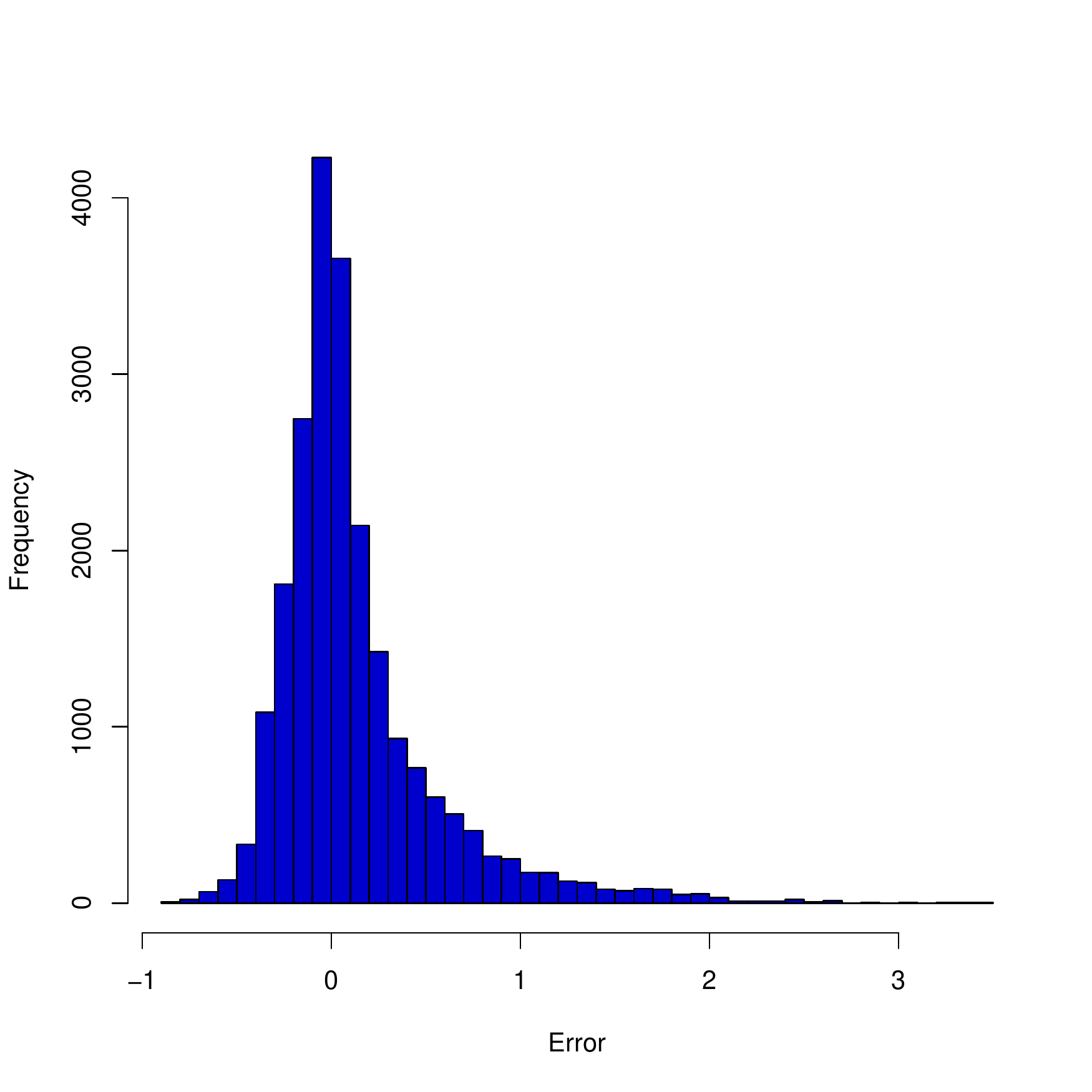} &
		\includegraphics[trim=1.2cm 1.5cm 1.5cm 2cm,clip=true,scale=0.18]{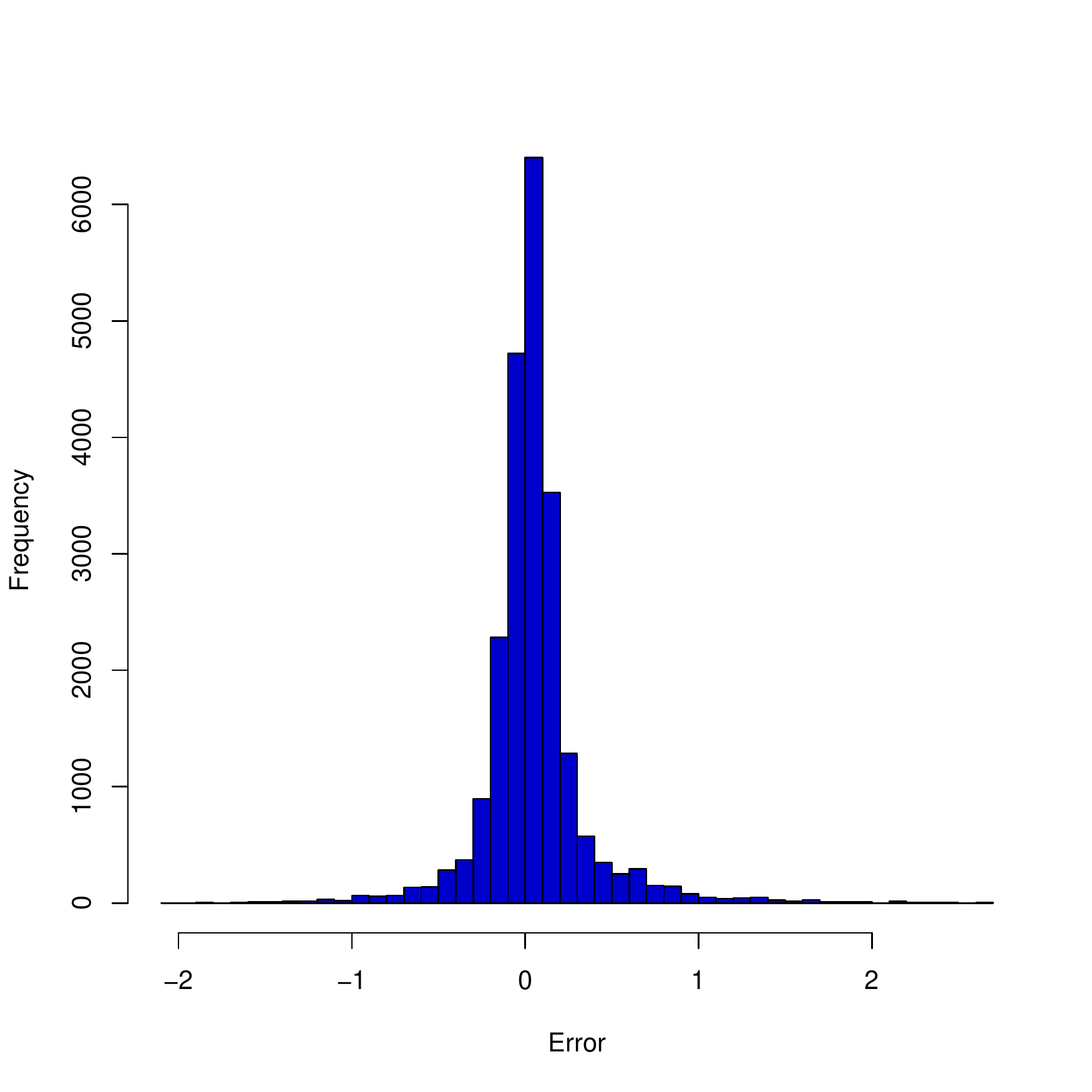} &
		\includegraphics[trim=1.2cm 1.5cm 1.5cm 2cm,clip=true,scale=0.18]{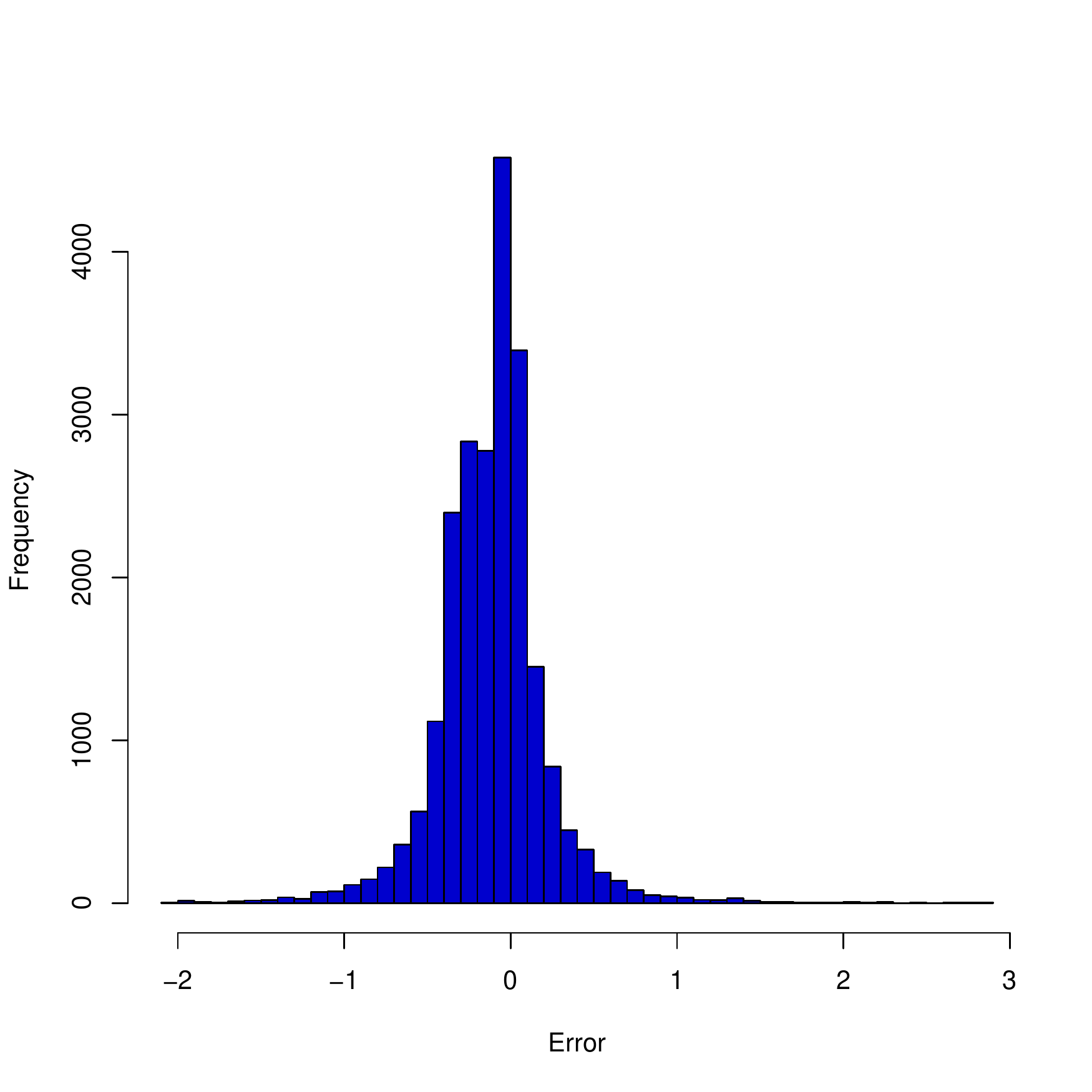} \\
		\rotatebox{90}{~~~~\parbox{2mm}{3SHEETS\_3D}~}&
		\includegraphics[trim=1.2cm 1.5cm 1.5cm 2cm,clip=true,scale=0.18]{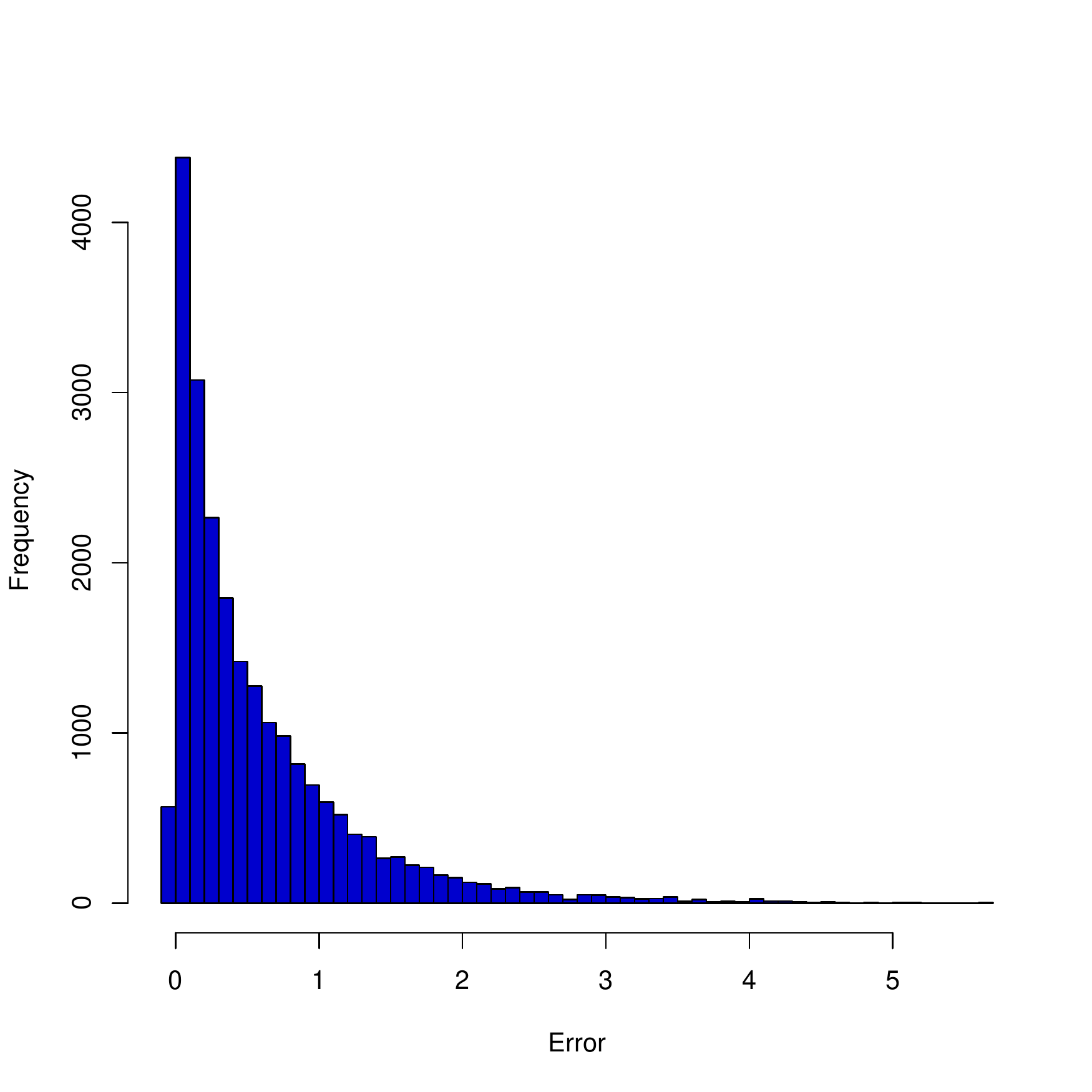} &
		\includegraphics[trim=1.2cm 1.5cm 1.5cm 2cm,clip=true,scale=0.18]{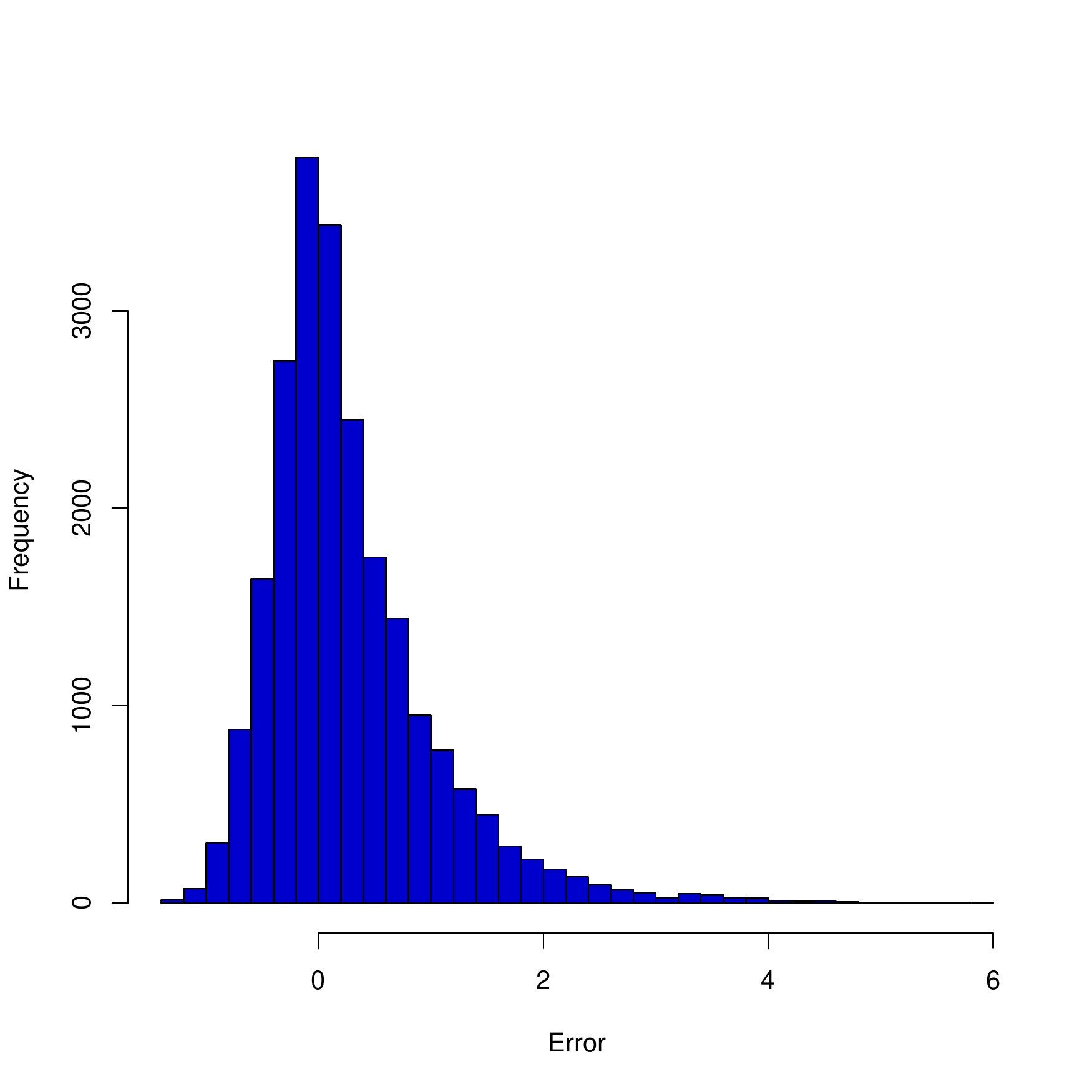} &
		\includegraphics[trim=1.2cm 1.5cm 1.5cm 2cm,clip=true,scale=0.18]{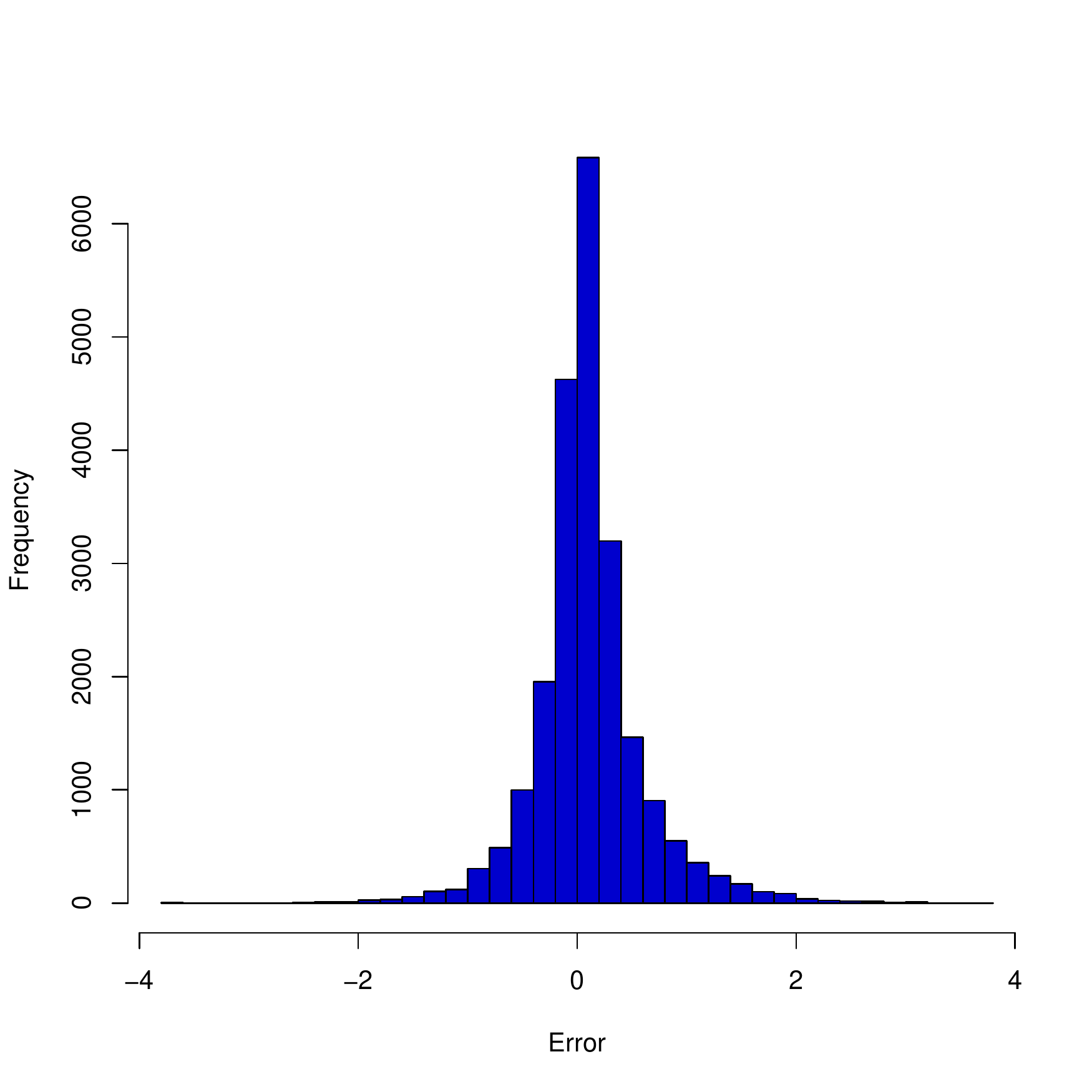} &
		\includegraphics[trim=1.2cm 1.5cm 1.5cm 2cm,clip=true,scale=0.18]{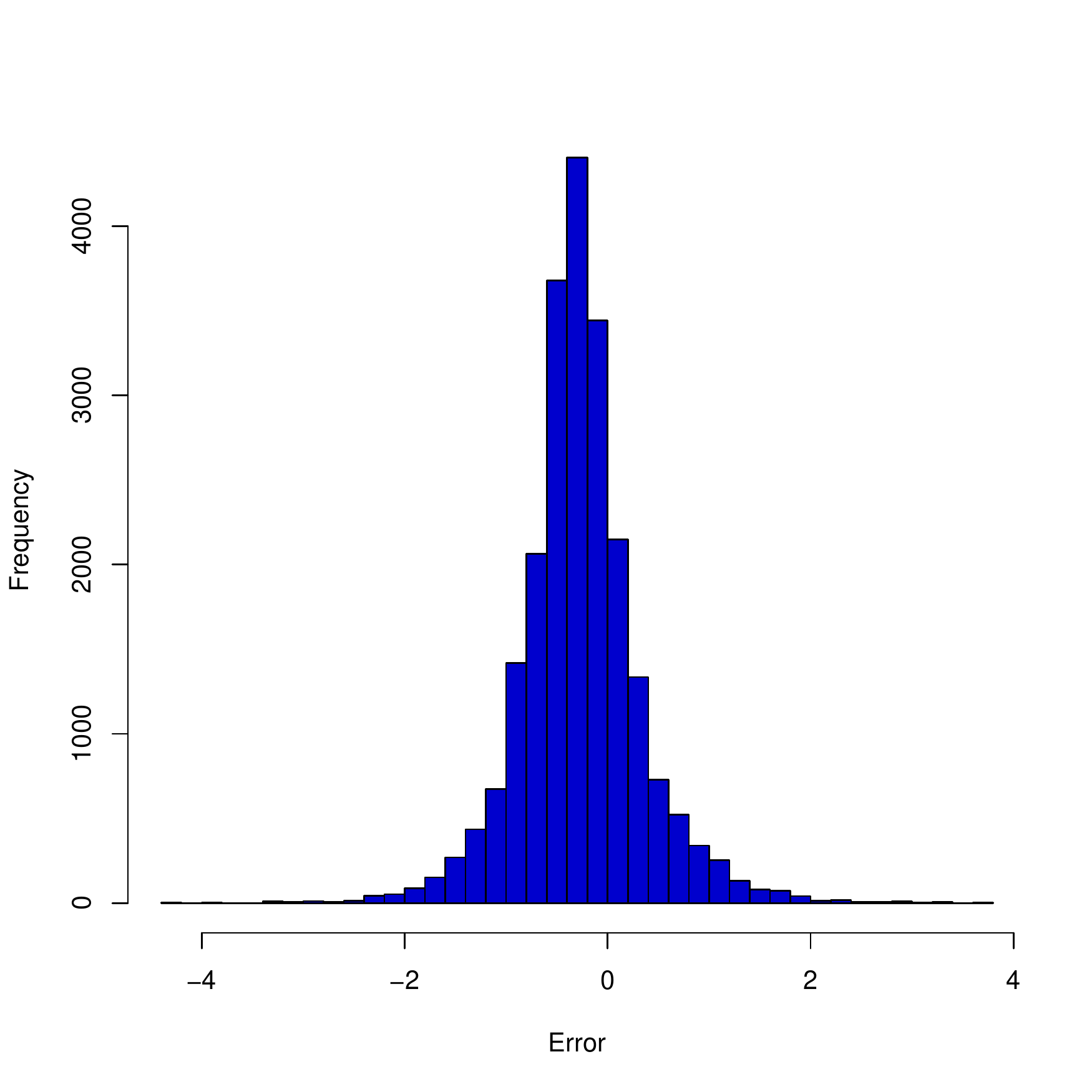} \\
		\rotatebox{90}{~~~~\parbox{2mm}{5SHEETS\_2D}~}&
		\includegraphics[trim=1.2cm 1.5cm 1.5cm 2cm,clip=true,scale=0.18]{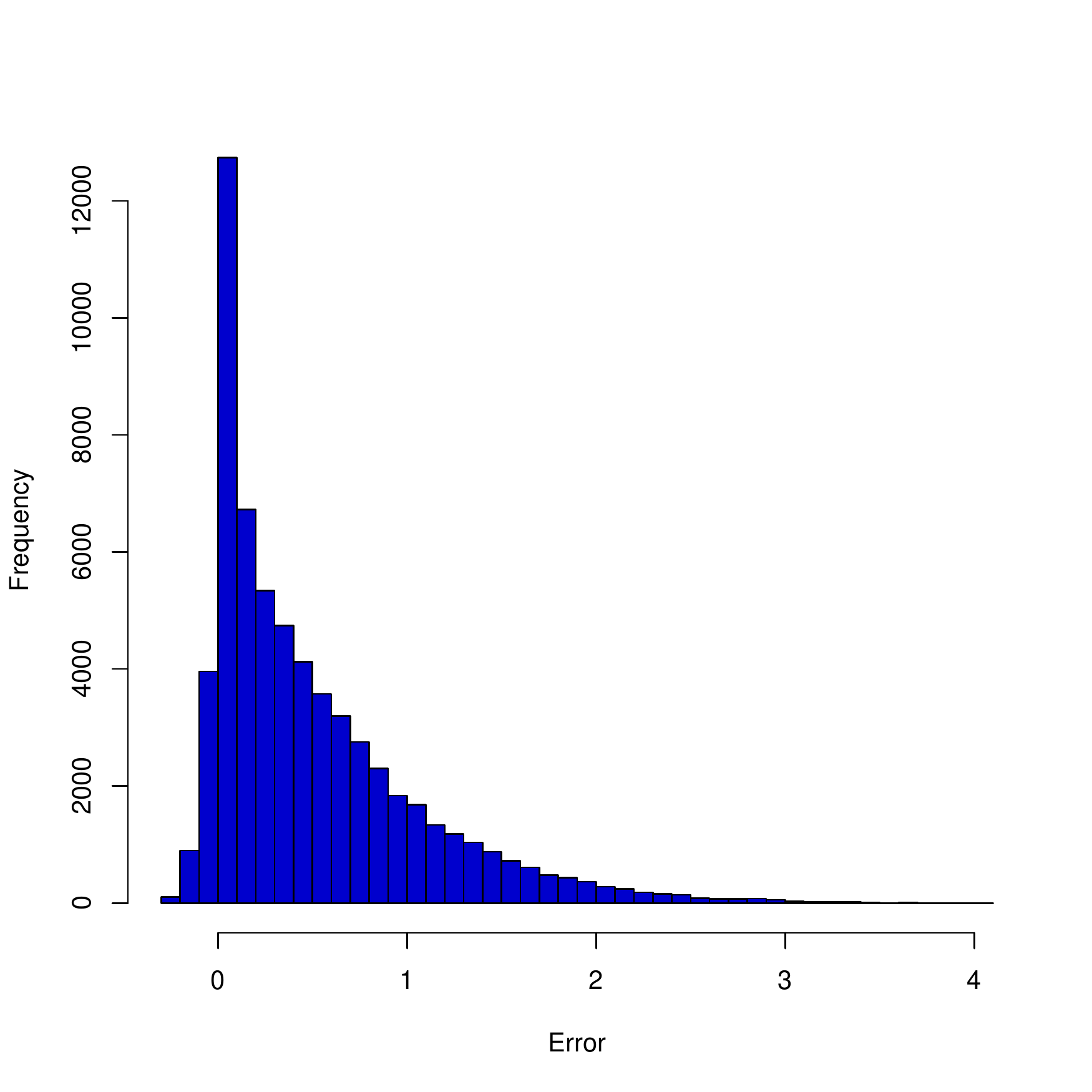} &
		\includegraphics[trim=1.2cm 1.5cm 1.5cm 2cm,clip=true,scale=0.18]{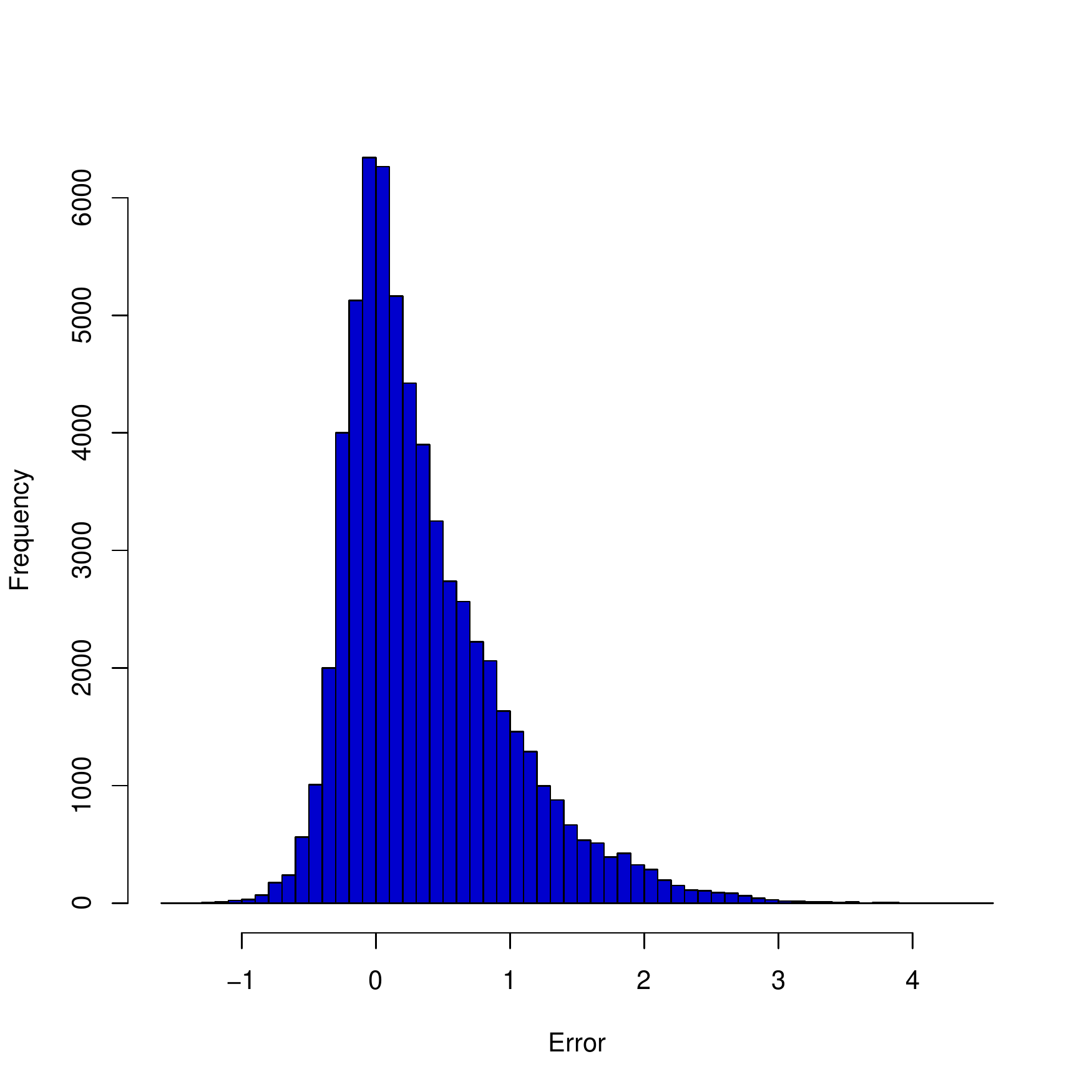} &
		\includegraphics[trim=1.2cm 1.5cm 1.5cm 2cm,clip=true,scale=0.18]{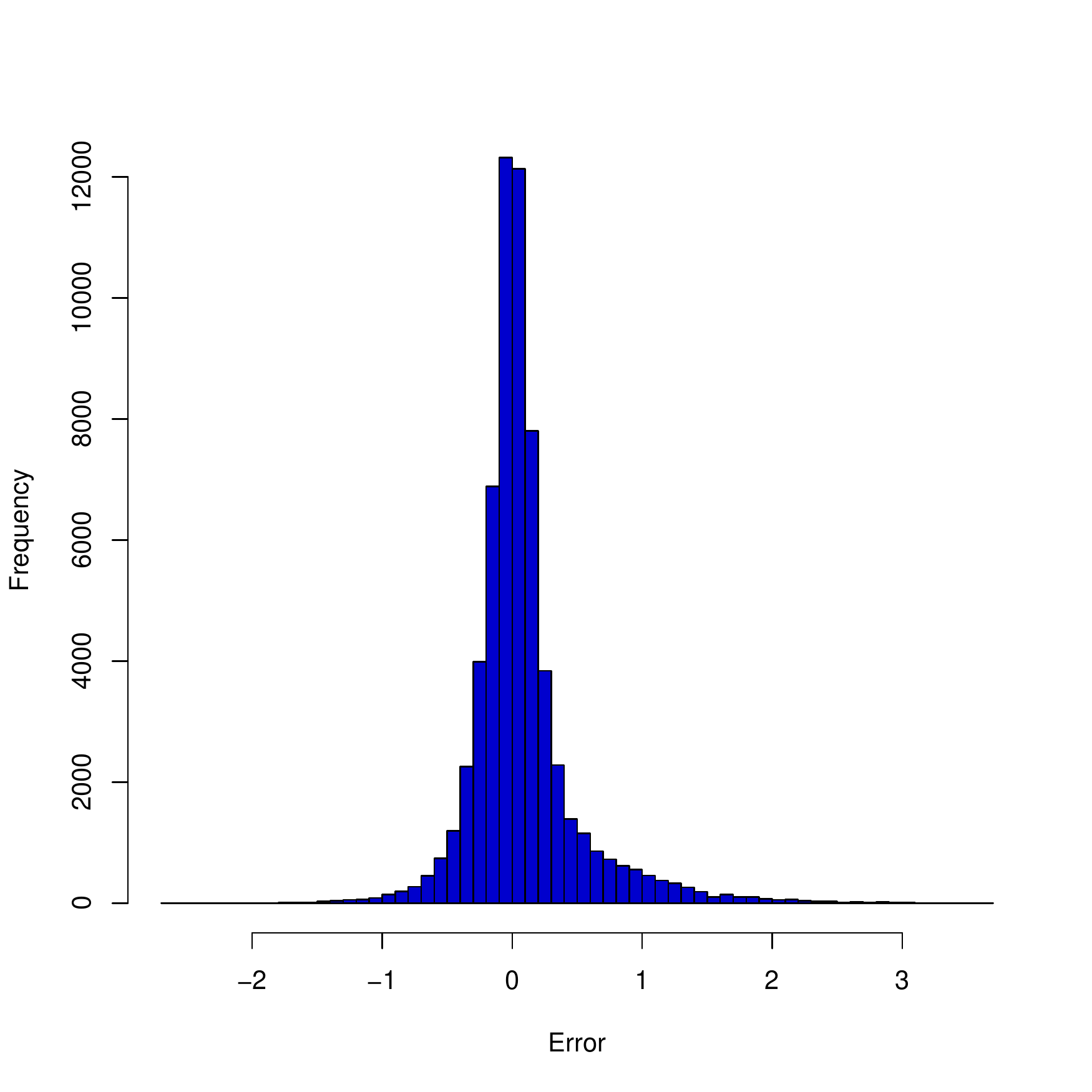} &
		\includegraphics[trim=1.2cm 1.5cm 1.5cm 2cm,clip=true,scale=0.18]{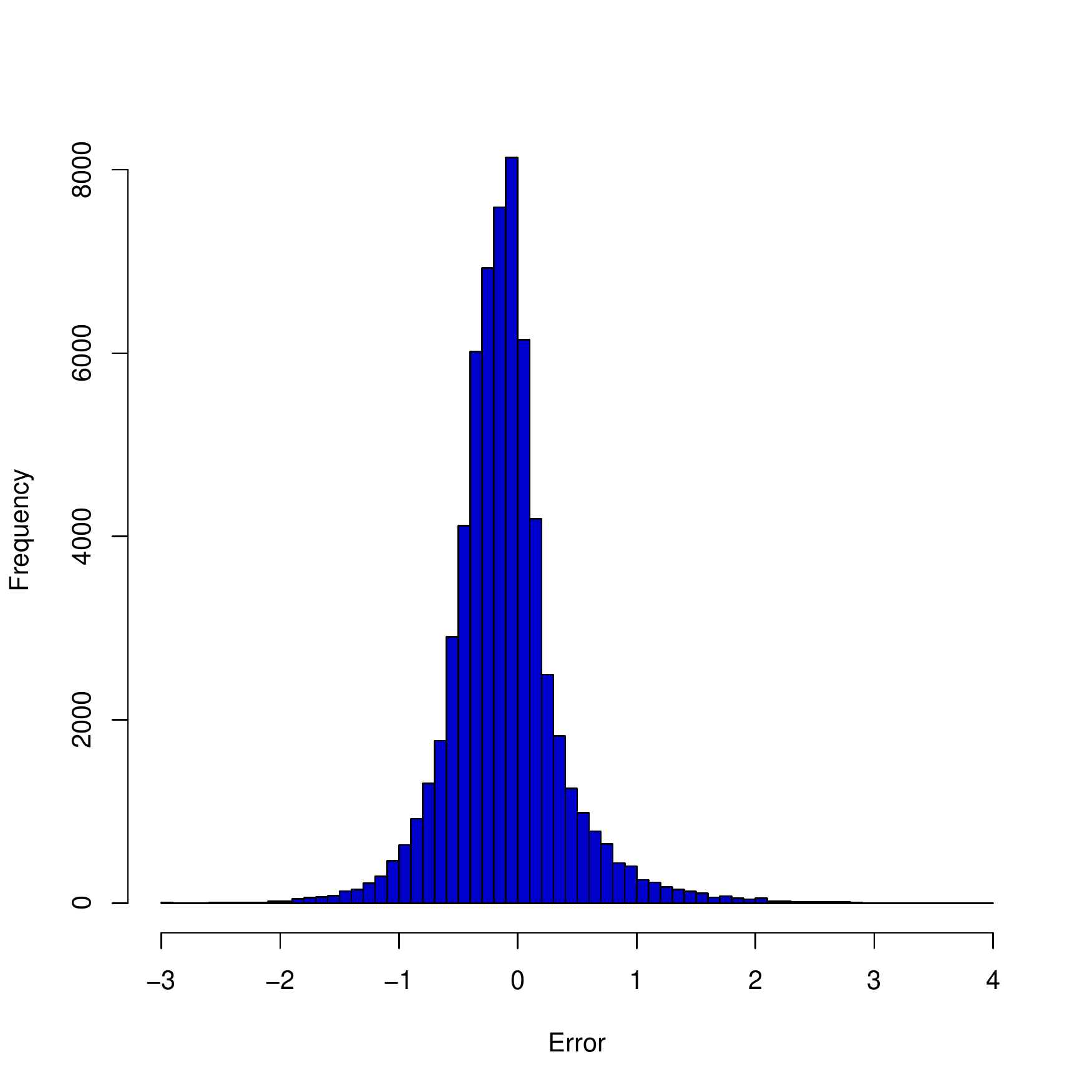} \\
		\rotatebox{90}{~~~~\parbox{2mm}{5SHEETS\_3D}~}&
		\includegraphics[trim=1.2cm 1.5cm 1.5cm 2cm,clip=true,scale=0.18]{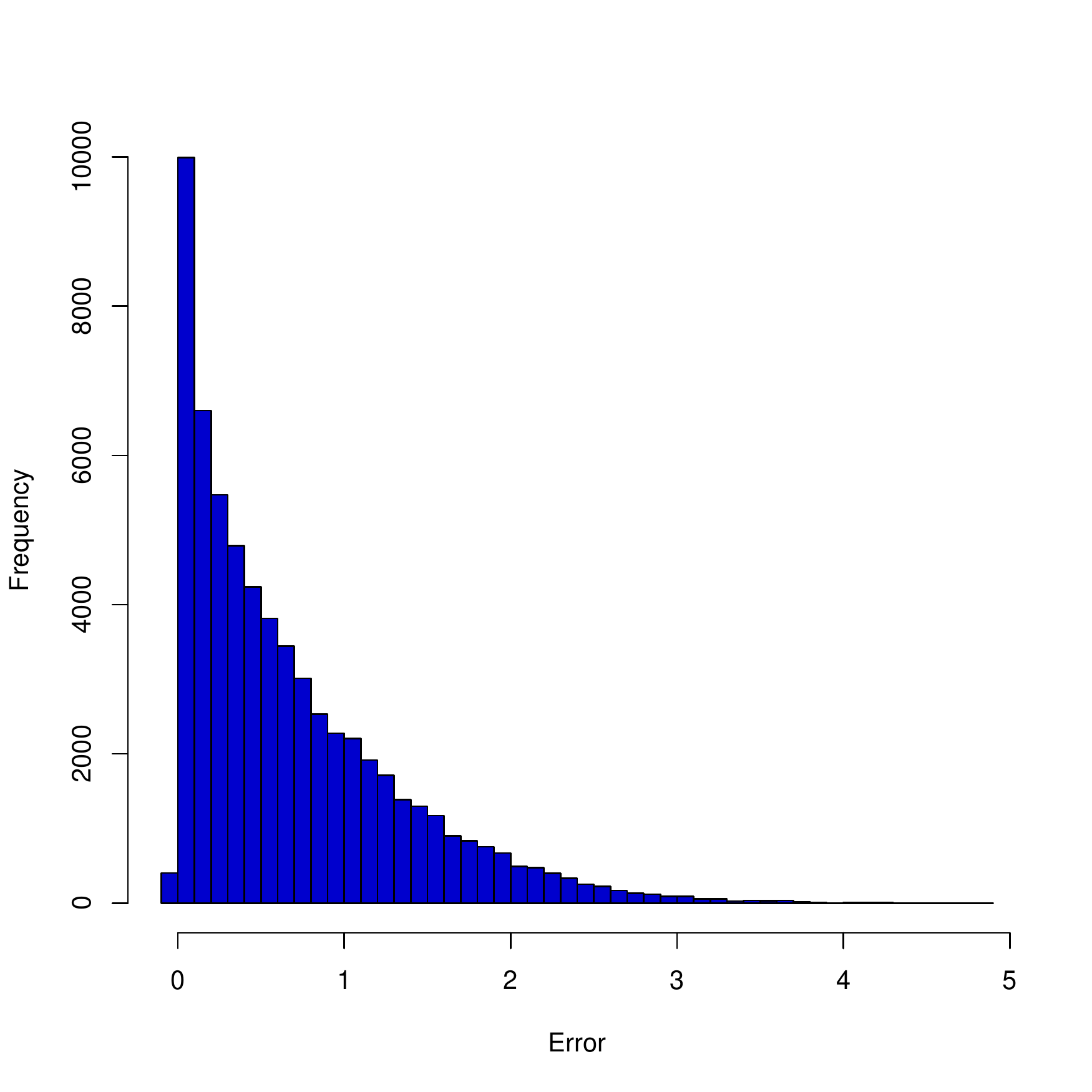} &
		\includegraphics[trim=1.2cm 1.5cm 1.5cm 2cm,clip=true,scale=0.18]{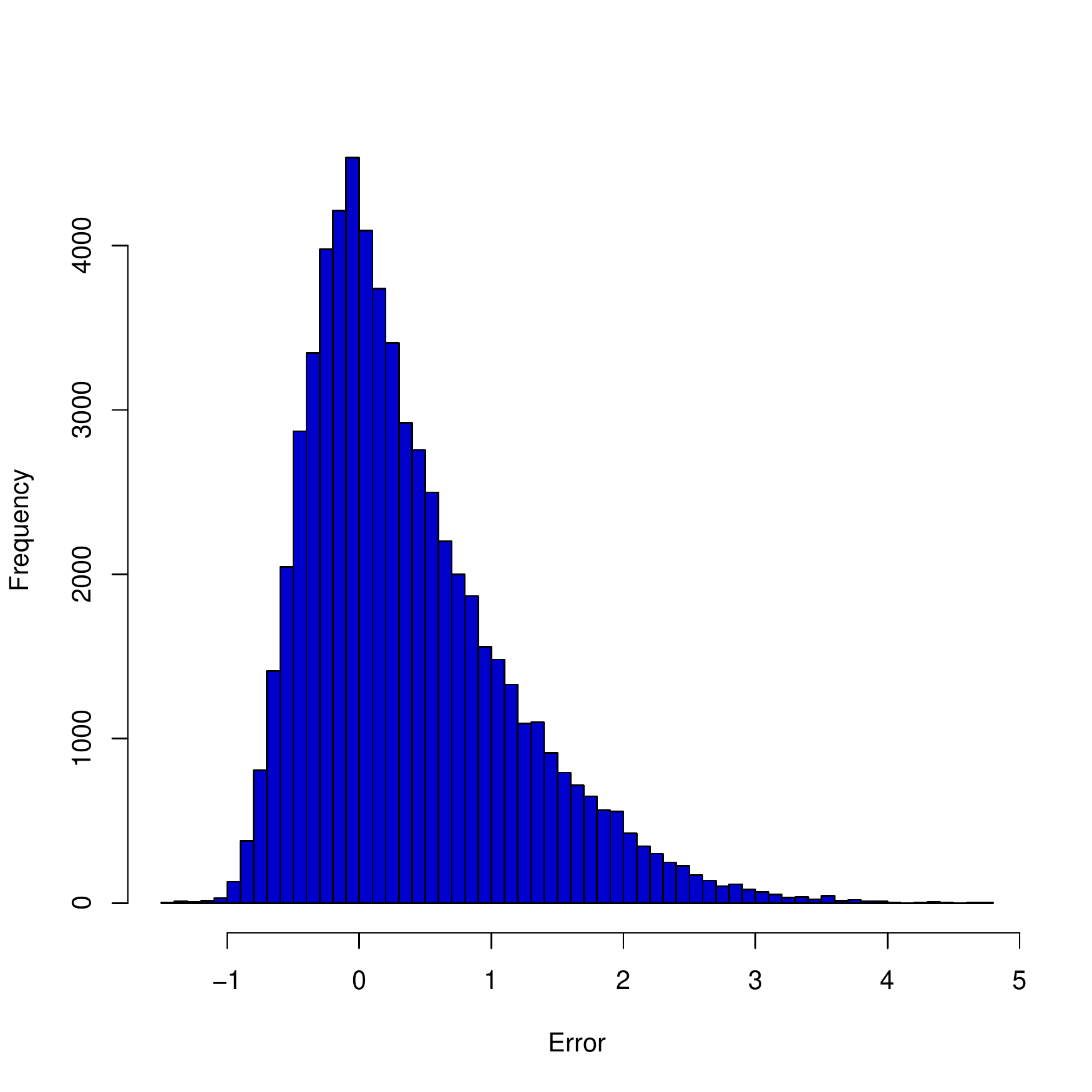} &
		\includegraphics[trim=1.2cm 1.5cm 1.5cm 2cm,clip=true,scale=0.18]{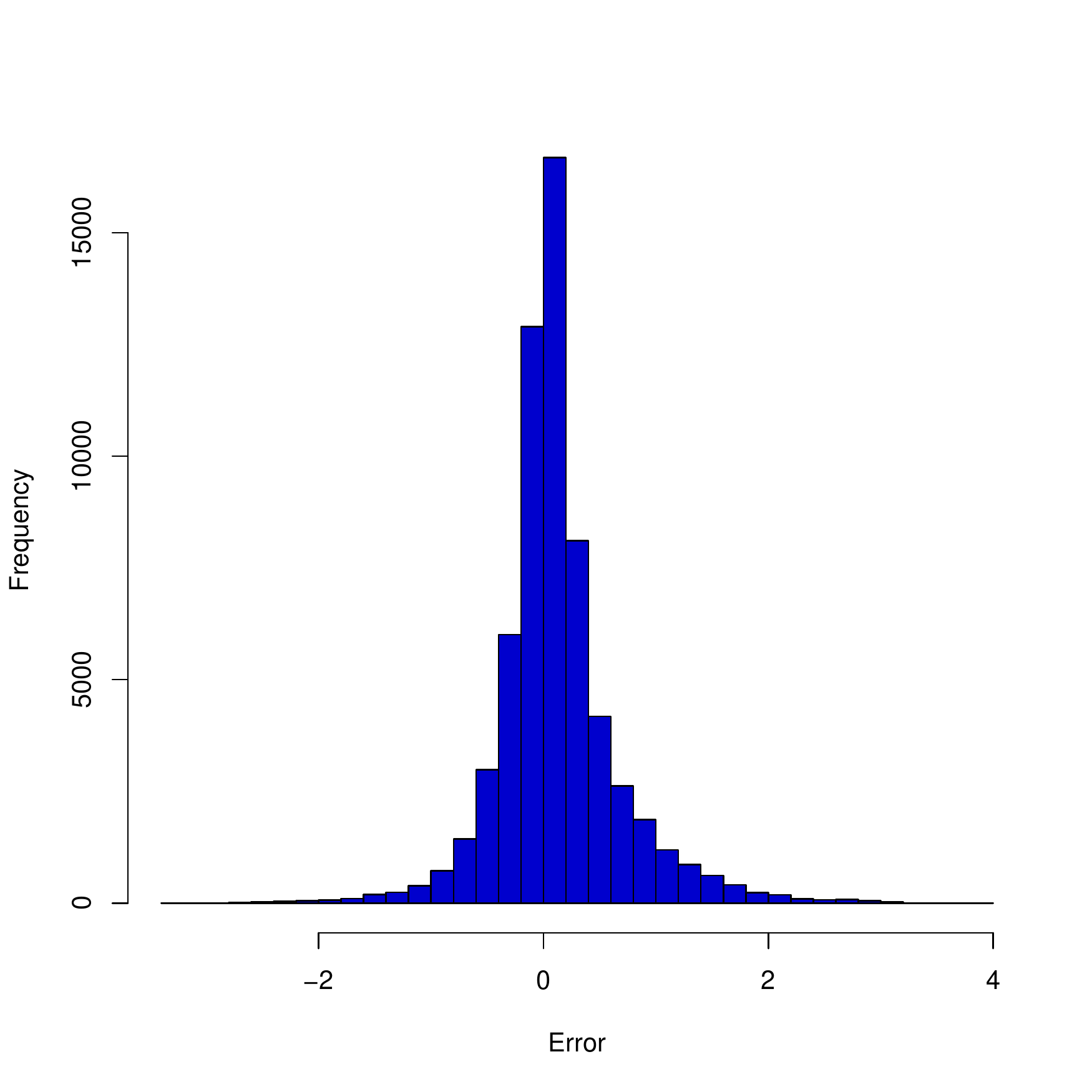} &
		\includegraphics[trim=1.2cm 1.5cm 1.5cm 2cm,clip=true,scale=0.18]{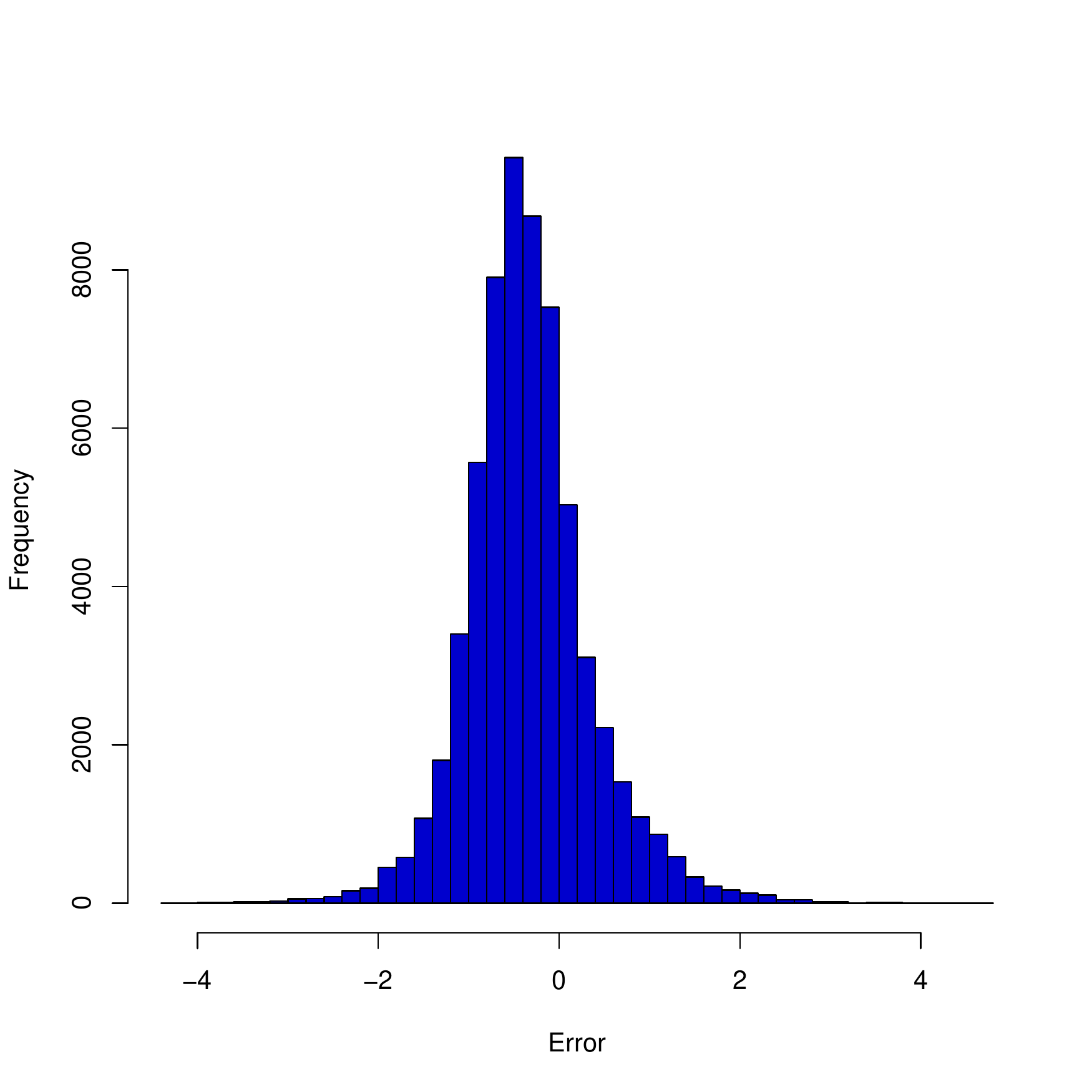} \\
		\rotatebox{90}{~~~~\parbox{2mm}{COPD}~}&
		\includegraphics[trim=1.2cm 1.5cm 1.5cm 2cm,clip=true,scale=0.18]{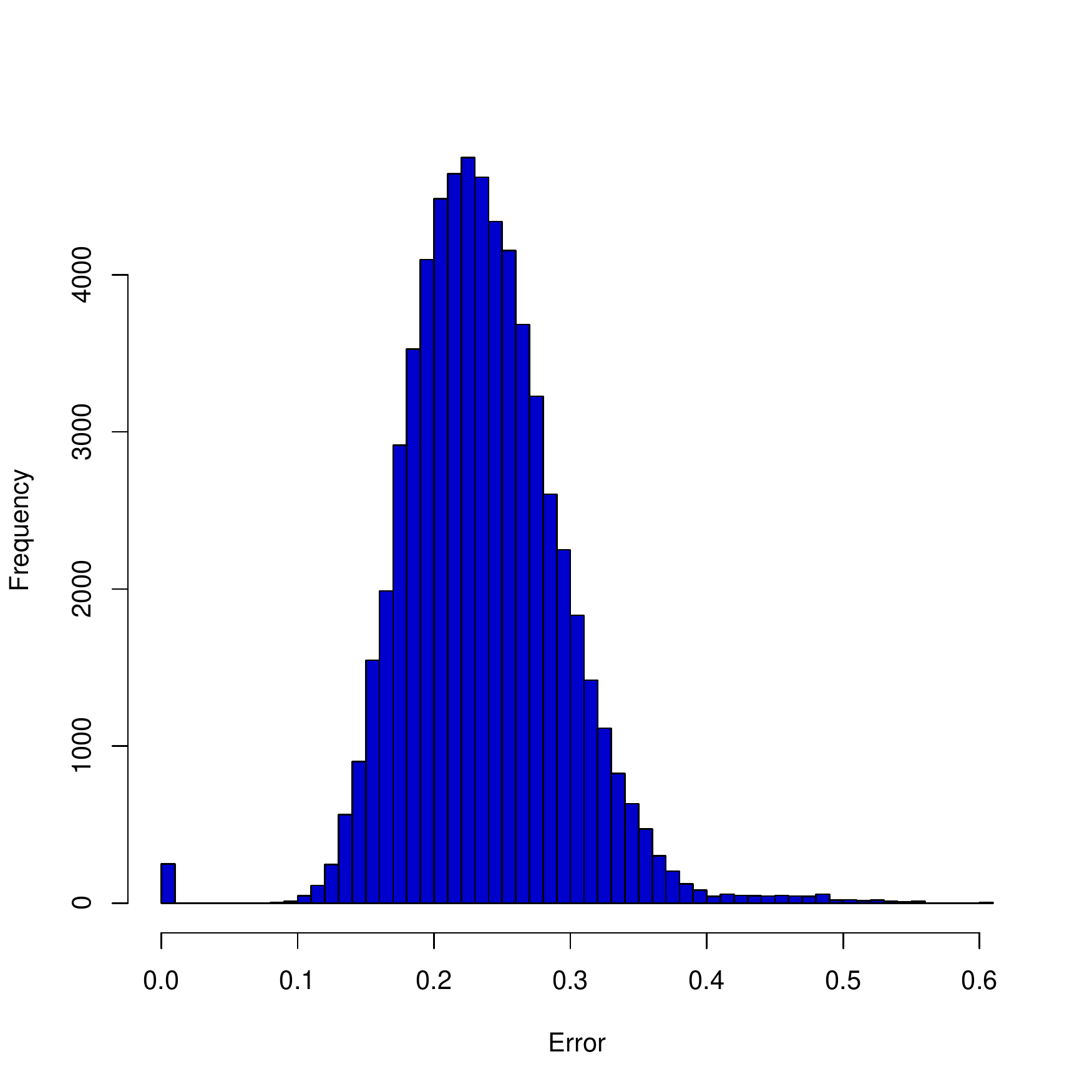} &
		\includegraphics[trim=1.2cm 1.5cm 1.5cm 2cm,clip=true,scale=0.18]{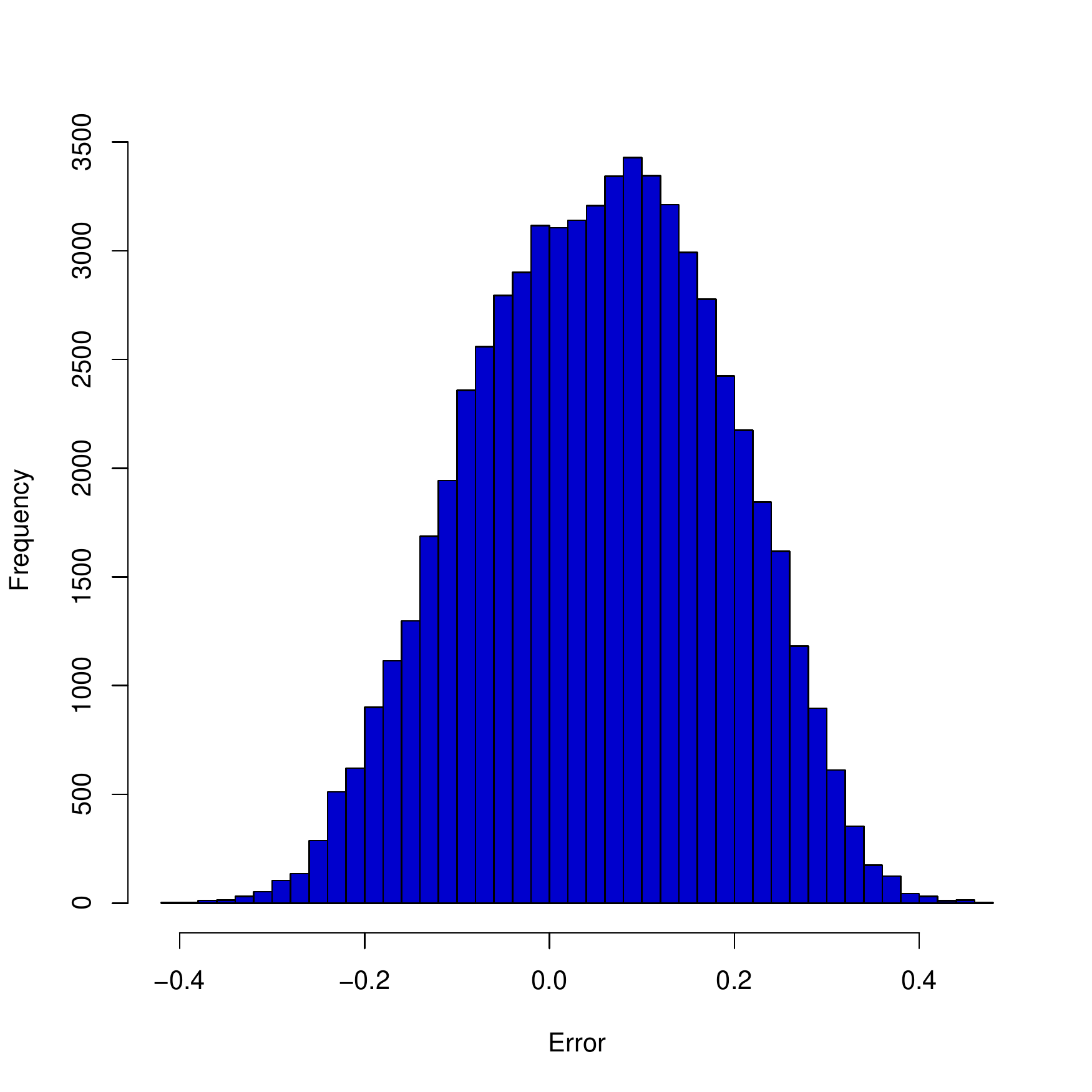} &
		\includegraphics[trim=1.2cm 1.5cm 1.5cm 2cm,clip=true,scale=0.18]{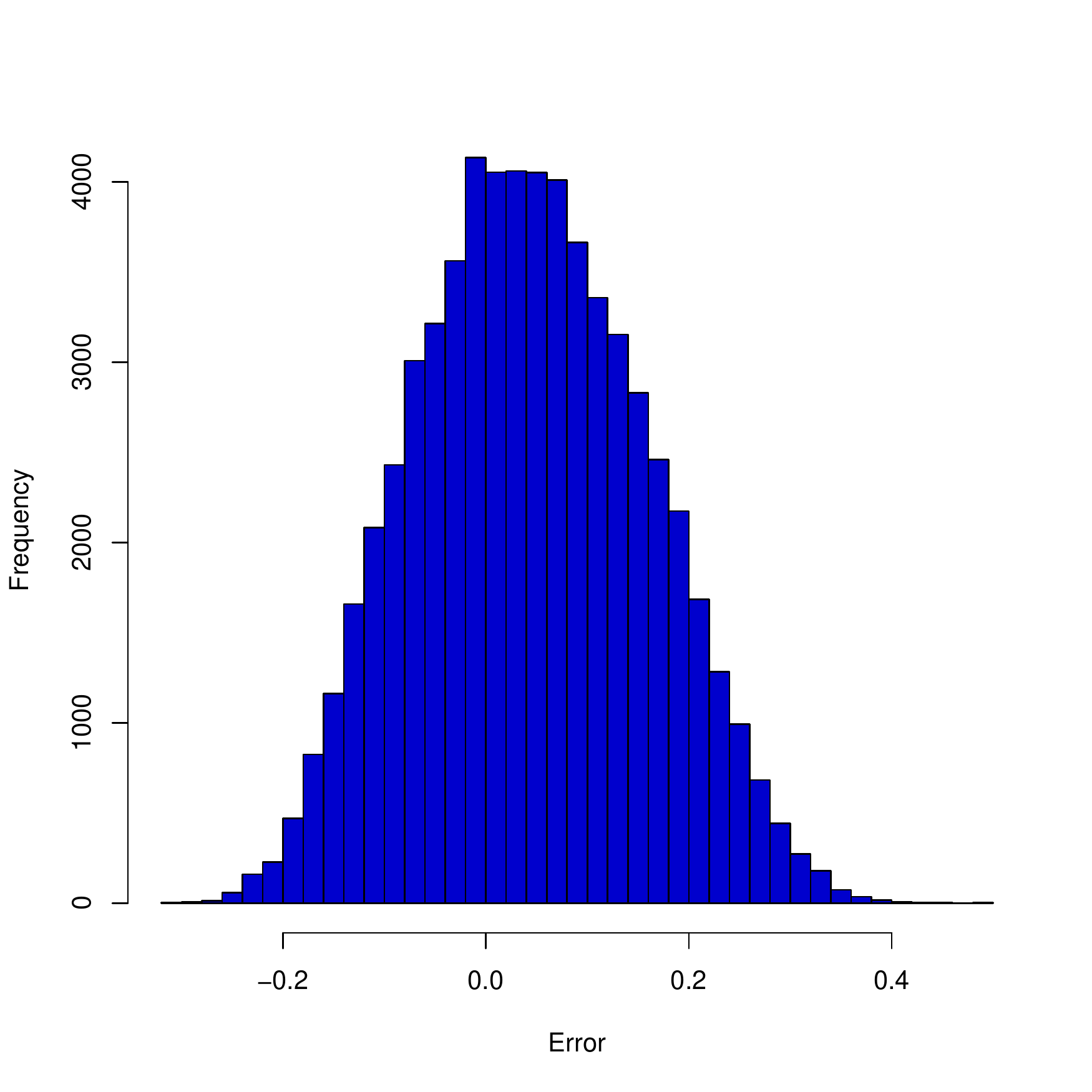} &
		\includegraphics[trim=1.2cm 1.5cm 1.5cm 2cm,clip=true,scale=0.18]{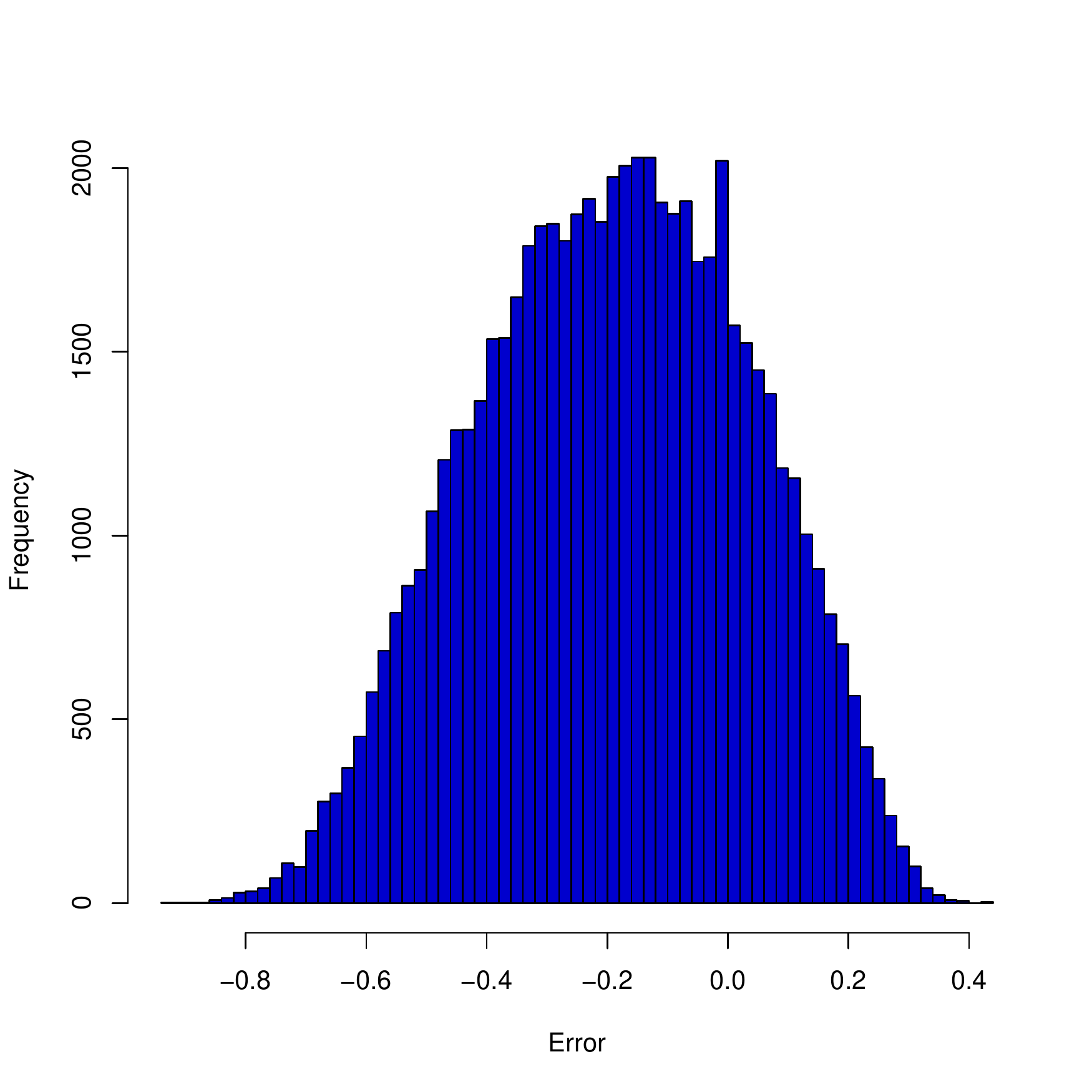} \\
	\end{tabular}
	\label{fig:all_error_hists}
	\end{table}

\section*{Acknowledgements}

This work was supported by NSF IIS  111766; MO was supported through a Fields-Ontario Postdoctoral Fellowship; AF was supported by the Danish Council for Independent Research $|$ Technology and Production. MH works for National Security Technologies, LLC, under Contract No. DE-AC52-06NA25946 with the U.S. Department of Energy/National Nuclear Security Administration, DOE/NV/25946{-}{-}2016.

We thank the organizers, Kathryn Leonard and Luminita Vese, and the sponsors of the collaboration workshop Women in Shape(WiSh): Modeling Boundaries of Objects in 2- and 3- Dimensions that was held at the Institute for Pure and Applied Mathematics (IPAM) at UCLA July 15-19 2013, where the collaboration leading to this paper was established.

\clearpage
\bibliographystyle{plain}
\bibliography{treestuff}

%

\end{document}